\documentclass[12pt]{article}
\usepackage{a4wide}
\usepackage[ngerman,english]{babel}
\usepackage{amsmath, amssymb, latexsym}
\usepackage{graphicx}
\usepackage{braket}
\usepackage{bm}
\usepackage{hyperref}
\usepackage{color}
\usepackage{cite}
\usepackage{wrapfig}
\date{}

\begin{document}
\begin{titlepage}

\begin{center}
\vspace{2mm}

\par\end{center}

\begin{center}
\textbf{\Large{}Polarimetric signatures of the photon ring of a black hole that is pierced by a cosmic axion string}\textbf{ }
\par\end{center}

\begin{center}
Alexander Gu{\ss}mann
\\
 
\par\end{center}

\begin{center}
\textsl{\small{}Stanford Institute for Theoretical Physics, Stanford University}\\
\textsl{\small{} Stanford, California 94305, USA}\\
\texttt{\small{} gussmann@stanford.edu }\\
\par\end{center}{\small \par}

\vskip 0.3 cm
\begin{abstract}
A black hole image contains a bright ring of photons that have closely circled the black hole on their way from the source to the detector. Here, we study the photon ring of a rotating black hole which is pierced by a global hyper-light axion-type cosmic string. We show that the coupling $\phi F \tilde{F}$ between the axion $\phi$ and the photon can give rise to a unique polarimetric structure of the photon ring. The structure emerges due to an Aharonov-Bohm type effect that leads to a change of the polarization directions of linear polarized photons when they circle the black hole. For several parameter choices, we determine concrete polarization patterns in the ring. Measuring these patterns can provide us with a way of determining the value of the coefficient of the mixed anomaly between electromagnetism and the symmetry that gave rise to the cosmic string. Finally, we briefly review a possible formation mechanism of black holes that are pierced by cosmic strings and discuss under which conditions we can expect such objects to be present as supermassive black holes in the center of galaxies.
\end{abstract}
\end{titlepage}

\section{Introduction}

General relativity predicts the existence of unstable bound photon orbits around Kerr black holes (see e.g. \cite{Chandrasekhar:1985kt, Teo:2020sey}). Unperturbed photons on such orbits circle the black hole forever. Photons on orbits that deviate slightly from the bound orbits first circle the black hole but eventually escape to infinity (or fall into the black hole). When these photons reach a distant observer, they produce an image on the screen of the observer that is known as the ``photon ring" \cite{Bardeen:1973tla, Luminet:1979nyg}. Using very large baseline interferometry, the event horizon telescope collaboration has recently obtained the first time-averaged horizon-scale image of a black hole that includes its photon ring \cite{Akiyama:2019cqa, Akiyama:2019brx, Akiyama:2019sww, Akiyama:2019bqs, Akiyama:2019fyp, Akiyama:2019eap}. Although not resolved by current experiments, we expect from predictions of general relativity that the photon ring itself is composed of an infinite sequence of nested subrings (see e.g. \cite{Gralla:2019xty, Johnson:2019ljv, Gralla:2019drh}).\footnote{Note that there are different terminologies used in the literature: Whereas \cite{Gralla:2019xty} uses the terms ``photon ring" and ``lensing ring", \cite{Johnson:2019ljv, Gralla:2019drh} refer to ``subrings of the photon ring". We use the terminology of \cite{Johnson:2019ljv, Gralla:2019drh}.} The photons that form the $n^{{\rm th}}$ subring have completed $n$ half-orbits around the black hole on their way from the source to the detector. It is anticipated in the literature that some subrings can be resolved by potential future interferometry experiments that use very large baselines with a detector placed in space (or on the moon) \cite{Johnson:2019ljv}. Such experiments also have the potential of measuring the polarization direction of linearly polarized light in the subrings \cite{Himwich:2020msm}.
\newline

Very large baseline interferometry therefore allows us to test aspects of fundamental physics that affect the structure and properties of the photon ring in the regime close to a black hole. This includes testing gravity, the black hole no hair conjecture and various aspects of particle physics in that regime (works along these lines include \cite{Bambi:2008jg, Johannsen:2010ru, Loeb:2013lfa, Psaltis:2014mca, Cunha:2015yba, Johannsen:2015hib, Johannsen:2016uoh, Johannsen:2016vqy, Lacroix:2016qpq, Bambi:2019tjh, Psaltis:2018xkc, Cunha:2018gql, Banerjee:2019xds, Vagnozzi:2019apd, Chen:2019fsq, Gralla:2020srx, Volkel:2020xlc, Psaltis:2020ctj, Glampedakis:2021oie}). In this work we shall study one new aspect of this sort, namely the impact that certain cosmic strings can have on the black hole photon ring when they pierce the black hole.\footnote{See \cite{Tinchev:2013nba} for another work on aspects of the photon ring of a black hole that is pierced by a cosmic string. In that work, the authors study effects that are generated by the gravitational backreaction of the cosmic string. The strength of this backreaction increases with the string tension. In this work we shall not take into account gravitational backreaction of the cosmic string but study polarimetric signatures of the photon ring which are independent of the value of the string tension. Neglecting gravitational backreaction is justified when the tension of the cosmic string that pierces the black hole is sufficiently small. As we shall argue, we can expect this to be the case for the strings of our interest.}
\newline

Cosmic strings are predicted to be formed in phase transitions in the early universe in many particle physics models (see e.g. \cite{Hindmarsh:1994re, Vilenkin:2000jqa} for two reviews). Depending on the particular model, there can be Nielsen-Olesen type local cosmic strings \cite{Nielsen:1973cs} (in case the spontaneously broken symmetry is local) and/or global cosmic strings (in case the spontaneously broken symmetry is global). The latter typically appear in models of the QCD axion \cite{Peccei:1977hh, Weinberg:1977ma, Wilczek:1977pj} and in models of more general axion-like particles \cite{Arvanitaki:2009fg}. Under certain conditions such strings can pierce black holes. For example, black holes pierced by cosmic strings are inevitably formed in the early universe in case primordial overdensities collapse to (primordial) black holes at a time when a string network has reached a scaling regime \cite{Vilenkin:2018zol}. If formed in this way, many of these pierced primordial black holes will loose their strings over time. A nonzero density of such black holes of primordial origin that are pierced by cosmic strings will however remain present at any time during the evolution of the universe \cite{Vilenkin:2018zol}.\footnote{This, of course, assumes that the cosmic strings themselves do not all decay. In this work we shall focus on such stable strings. Below, in footnote \ref{qcdaxion}, we shall comment on this point in a bit more detail.}  As argued in \cite{Vilenkin:2018zol}, such black holes can be seeds of supermassive astrophysical black holes of present time if the string tension is sufficiently small. From the observational point of view, there have been speculations that the black hole in the center of our milky way is pierced by a cosmic string \cite{Morris_2017}.
\newline

We shall argue that global hyper-light axion-type cosmic strings that pierce a black hole can give rise to characteristic polarization patterns in the photon ring of the black hole. These patterns emerge due to the coupling $\phi F \tilde{F}$ between the Goldstone boson (axion) $\phi$ and the photon as this coupling leads to a change of the polarization direction of linearly polarized photons when they circle the black hole (and the string) \cite{Agrawal:2019lkr}. The change of the polarization direction is generated by a topological Aharonov-Bohm type effect. It only depends on the total angle that a photon has passed when circling the black hole (and the string) and on the coefficient $C$ in the coupling $\frac{C}{2}\frac{\phi}{v} F \tilde{F}$ \cite{Agrawal:2019lkr}.\footnote{Here $v$ is the vacuum expectation value of the spontaneously broken symmetry that gave rise to the cosmic string, see section \ref{section:section2}.} In models with charged fermions and Yukawa interactions between $\phi$ and the fermions, the value of $C$ is set by the coefficient of the mixed anomaly of electromagnetism and the (spontaneously broken) symmetry that gave rise to the cosmic string. Measuring the polarization patterns can thus provide us with a new way of determining the value of this anomaly coefficient. Recently, a potential way to measure the anomaly coefficient via the same method was proposed in different setups in \cite{Agrawal:2019lkr}.\footnote{Knowing the value of the anomaly coefficient can teach us a lot about the UV theory realized in nature that gave rise to the anomaly \cite{Agrawal:2019lkr}.}
\newline

We shall work out polarization patterns in the photon ring of a pierced black hole both in time-averaged setups and, in a simple example, in setups with time dependent light sources. The detailed structures of the patterns are determined by several parameters: The black hole parameters (mass and spin), the inclination of the distant observer from the black hole spin axis, the coefficient $C$ in the coupling $\frac{C}{2}\frac{\phi}{v} F \tilde{F}$ and the light source distribution. We study examples with various black hole parameters and observer inclinations. In the time-averaged setups we shall look for polarization patterns that are independent of the light source distribution. We shall argue that, although the polarization directions of linearly polarized photons in any individual subring of the black hole photon ring generally depend on the distribution of the light sources, the relative polarization directions between linear polarized photons of different even/odd subrings are universal (independent of the light source distribution). Promising universal signatures to consider in time-averaged setups are therefore the relative polarizations of linear polarized photons from different even/odd subrings. In setups with time dependent sources, we shall consider the simple example of point-like sources that are located close to the black hole and study corresponding time dependent signatures in the photon ring. All our proposed signatures can be considered as complementary to signatures that a cosmic string itself can generate irrespective of the presence of the black hole (works on some potential signatures include \cite{Vilenkin:1984ea, Chudnovsky:1986hc, Gruzinov:2016hqs, Agrawal:2020euj}). 
\newline

We shall briefly comment on the potential relevance of ``environmental effects" on the propagating photons, generated for example by interactions with the surrounding plasma. A detailed study on the potential impact of such effects on our results will however require further analysis that is beyond the scope of this work.
\newline

The work is organized as follows. First, we review both aspects of the theory and a potential formation mechanism of rotating black holes pierced by cosmic strings (section \ref{section:section2}). We then discuss photon orbits in these backgrounds. We show how the polarization direction of linear polarized photons gets changed once they move around the pierced black hole before they reach out to the distant observer (section \ref{section:section3}). Finally, we determine polarization patterns in the black hole photon ring both in time-averaged and in time dependent setups (section \ref{section:section4}). We conclude with a short summary and an outlook (section \ref{section:section5}). We focus on global $U(1)$ axion-type cosmic strings that pierce the black hole such that they coincide with the black hole spin axis. We work in units $c = \hbar = 1$. For the metric we use the signature $(-,+,+,+)$.

\section{Black hole pierced by a cosmic string} 
\label{section:section2}

In this section we shall introduce the objects which we are going to study in this work. First, we shall review some aspects of global cosmic strings in a $U(1)$ invariant model. We then review one mechanism of how black holes can become pierced by such strings in the early universe. We point out the existence of theoretical solutions that describe such pierced black holes restricting to the case of a cosmic string that coincides with the rotation axis of the black hole. Finally, we discuss how the pierced black holes evolve in time and under what conditions we can expect black holes pierced by cosmic strings to appear in the center of galaxies at present time. 

\subsection{Global cosmic strings}

As an example for a model that allows for solitonic string solutions, we consider the $U(1)$ invariant Lagrangian density
\begin{equation}
\mathcal{L} = |\partial_\mu \psi|^2 - \lambda \left(|\psi|^2 - v^2\right)^2 \, ,
\label{lag}
\end{equation}
with a complex scalar field $\psi$ and the two real constants $\lambda$ and $v$. The $U(1)$ symmetry gets spontaneously broken when $\psi$ develops a vacuum expectation value, $|\braket{\psi}|^2 = v^2$, or
\begin{equation}
\braket{\psi(x)} = v e^{i\frac{\phi(x)}{v}} \, .
\end{equation}
Here $\phi$ is a real scalar field which has a periodicity $\phi \rightarrow \phi + 2 \pi v$ and which can be identified with the Goldstone boson (axion) of the broken symmetry.
\newline

The classical equations of motion of the model (\ref{lag}) admit solitonic string solutions (see e.g. \cite{Hindmarsh:1994re, Vilenkin:2000jqa} for two reviews). The prototype example of such strings are infinitely long, straight, static strings with winding number $\pm 1$ of the form
\begin{equation}
\psi(x) = vh(r {\rm sin} \theta)e^{\pm i \varphi} \, ,
\label{strings}
\end{equation}
where we used spherical coordinates $x = (r, \theta, \varphi)$. The sign $\pm$ sets the orientation of the string. The profile function $h(r {\rm sin} \theta)$ is a real function that has boundary conditions $h(0) = 0$ (so that the field is regular) and $h(\infty) = 1$ (so that the field approaches a vacuum solution). The energy density of such a ``global string" is well localized around the one-dimensional line $r {\rm sin} \theta = 0$ but, due to the gradient term in (\ref{lag}), has a long tail (see e.g. \cite{Hindmarsh:1994re, Vilenkin:2000jqa}). Due to this tail, the energy per unit length of one string (string tension) diverges logarithmically for $r {\rm sin} \theta \rightarrow \infty$. Therefore, a single global string has infinite energy in an infinite space. In practice, when there is more than one string present, this divergence is however cut off at the distance $R$ of a neighboring string with opposite orientation. This leads to a string tension $\mu$ of
\begin{equation}
\mu \sim v^2 \mathrm{ln}\left(\frac{R}{d}\right) \, ,
\label{stringtension}
\end{equation}
where $d$ is the thickness of the string (see e.g. \cite{Hindmarsh:1994re, Vilenkin:2000jqa}). Around the string, the field $\phi$ has a profile of $\frac{\phi}{v} = \pm \varphi$ (\ref{strings}) ($\pm$ is set by the string orientation) and thus winds from $0$ to $\pm 2 \pi v$ when going once around a string. More intricate string configurations that go beyond the prototype case include curved strings and closed string loops. In these cases the infrared divergence in energy is cut off at the curvature radius of the string.
\newline

Cosmic strings of this kind are formed in the early universe when the $U(1)$ symmetry in the model (\ref{lag}) gets spontaneously broken as the universe cools below a temperature of $v$, $T \leq v$. In the corresponding phase transition the field $\phi$ randomly takes different values in different Hubble patches. Therefore, $\phi$ can change from $0$ to $\pm 2 \pi v$ at the boundaries of the Hubble patches. Cosmic strings are formed there. This formation mechanism is the famous Kibble mechanism \cite{Kibble:1976sj, Kibble:1980mv}. The evolution of the resulting string network has been studied numerically (works include \cite{Albrecht:1984xv, Bennett:1985qt, Bennett:1986zn, Bennett:1987vf, Allen:1990tv, Vincent:1996rb, Yamaguchi:1998gx, Yamaguchi:2002sh, Kawasaki:2018bzv, Vaquero:2018tib, Gorghetto:2018myk, Hindmarsh:2019csc, Hindmarsh:2021vih}). The simulations indicate that, for a phase transition that happens after inflation, a string network approaches a scaling regime as a result of frequent string interactions; that is a regime in which the energy density $\rho$ of the string network scales like the dominant form of the energy in the universe,\footnote{Note that the effective tension $\mu$ of all strings in the network generically also depends on time because the effective curvature scale in the string network (the parameter $R$ in (\ref{stringtension})) typically scales like the Hubble distance, see e.g. \cite{Vilenkin:2000jqa}. This however gives only rise to a logarithmic scaling.}
\begin{equation}
\rho \sim \mu H^2 \, ,
\label{scaling}
\end{equation}
with the Hubble parameter $H$.\footnote{There is some recent discussion on potential small (logarithmic) deviations from this scaling behavior (see e.g. \cite{Hindmarsh:2021vih, Gorghetto:2020qws} for two recent works).} This implies an approximately constant number of total string length (measured in Hubble units) per Hubble volume throughout the evolution of the universe after the scaling regime has reached. Simulations demonstrate that most of the total string length is contained in long strings that are larger than a Hubble volume whereas smaller string loops only give rise to some subcomponent of the total string length. Thus, as a good approximation, the total number of (long) strings per Hubble volume remains constant in the scaling regime.\footnote{In cases when the field $\phi$ acquires a mass, as is the case for example for the QCD axion \cite{Peccei:1977hh, Weinberg:1977ma, Wilczek:1977pj}, $N$ domain walls form when $H$ becomes of order this mass. The strings end on these domain walls. If $N = 1$, all strings will disappear in a short period of time because they are pulled together by the domain wall. In this work we are interested in strings that survive until present time and therefore restrict to cases where either $N \neq 1$ or to cases where $\phi$ does at most acquire a tiny mass. We have in mind global hyper-light axion-type cosmic strings \cite{Arvanitaki:2009fg}. \label{qcdaxion}}
\newline

Global cosmic strings have interesting properties generated by various couplings of $\psi$ to standard model particles. Later in this work we shall consider the coupling of the Goldstone boson $\phi$ to standard model photons of the form
\begin{equation}
\delta \mathcal{L} = \frac{C}{2} \frac{\phi}{v} \tilde{F}_{\mu \nu} F^{\mu \nu} \, ,
\label{coupling}
\end{equation}
where $\tilde{F}_{\mu \nu} \equiv \frac{1}{2} \epsilon_{\mu \nu \alpha \beta} F^{\alpha \beta}$. In the presence of charged fermions that couple to $\phi$ via Yukawa couplings, the coupling (\ref{coupling}) gets induced by the mixed anomaly of the $U(1)$ symmetry from above and electromagnetism (see e.g \cite{Callan:1984sa}). The coupling constant $C$ can be expressed in terms of the electromagnetic fine structure constant $\alpha$ and the anomaly coefficient $\mathcal{A}$ as
\begin{equation}
C = \frac{\alpha \mathcal{A}}{2 \pi} \, .
\label{anomaly}
\end{equation}
The coupling (\ref{coupling}) can lead to an inflow of charges onto the cosmic string that move on the string in the form of fermion zero modes \cite{Callan:1984sa, Lazarides:1984zq}. This can happen for example when the cosmic string passes through an external magnetic field. As a consequence, the cosmic string can become superconducting \cite{Witten:1984eb, Lazarides:1984zq}.

\subsection{Black holes piercing}
Under certain conditions cosmic strings, such as the ones just reviewed, can pierce black holes in our universe (see e.g. \cite{Aryal:1986sz, Lonsdale:1988xd, DeVilliers:1997nk, DeVilliers:1998nm, Snajdr:2002aa, Dubath:2006vs, Vilenkin:2018zol, Xing:2020ecz} for several possible formation mechanisms). Black holes pierced by cosmic strings can for example be formed in the early universe as primoridial black holes \cite{Vilenkin:2018zol}:\footnote{Note that the authors of \cite{Vilenkin:2018zol} refer to local Nielsen-Olesen type strings with magnetic flux. When taking the appropriate string tension into account, many of their arguments however also apply for the global strings (\ref{strings}) that we consider in this work.} When primordial overdensities become order one (as predicted to happen by some inflationary models), the entire corresponding Hubble volume collapses to a (primordial) black hole of mass $\frac{l_H}{G_N}$, where $l_H$ is the size of the horizon at formation time \cite{Zeldovich:1967lct, Hawking:1971ei, Carr:1974nx}. In this way a certain fraction of Hubble volumes can collapse into primordial black holes. If primordial black holes form via this mechanism at a time when a string network has already reached its scaling regime, all formed black holes will turn out to be pierced by cosmic strings at formation time (because, as reviewed above, in the scaling regime there is an approximately constant nonzero number of strings per Hubble volume).
\newline

On the theoretical side black holes pierced by cosmic strings have been studied a lot in the literature (works include \cite{Aryal:1986sz, Achucarro:1995nu, Frolov:1995vp, Bonjour:1998rf, Santos:1999if, Ghezelbash:2001pq, Gregory:2013xca, Kubiznak:2015hsg, Gregory:2014uca, Kinoshita:2016lqd, Igata:2018kry}). Here we consider Kerr black holes that are pierced by cosmic strings. In Boyer-Lindquist coordinates $(t, r, \theta, \varphi)$ the line element of the Kerr metric is given by \cite{Boyer:1966qh}
\begin{equation}
ds^2 = -\frac{\Delta}{\rho^2}\left(dt-a {\rm sin}^2\theta d\varphi\right)^2 + \frac{\rho^2}{\Delta}dr^2 + \rho^2 d\theta^2 + \frac{{\rm sin}^2\theta}{\rho^2}\left(\left(r^2 + a^2\right)d \varphi - a dt\right)^2 \, ,
\label{metric}
\end{equation}
where $\Delta \equiv r^2 - 2MG_Nr + a^2$ and $\rho^2 \equiv r^2 + a^2 {\rm cos}^2 \theta$. $M$ is the mass of the black hole and $J \equiv aMG_N$ its angular momentum. The range of $a$ is $0 \leq |a| \leq MG_N$.

Solutions to the classical equations of motion of (\ref{lag}) in the background (\ref{metric}) have been found numerically in the Nielsen-Olesen case (with a gauge field added to (\ref{lag}) under which $\psi$ is charged) for cosmic strings that coincide with the black hole spin axis \cite{Gregory:2013xca, Kubiznak:2015hsg}. This proves, on the theoretical level, the existence of solutions that describe Kerr black holes pierced by local cosmic strings. In the limit of zero gauge coupling constant the same argument can be used for a global string piercing a Kerr black hole, as considered here (\ref{strings}). The gravitational backreaction of the string on the metric gives rise to deviations from the Kerr metric \cite{Gregory:2013xca, Kubiznak:2015hsg, Tinchev:2013nba, Cohen:1988sg, Gibbons:1988pe, Gregory:1996dd, Gregory:2002tp}. In this work we shall however not take gravitational backreaction into account because, as we will argue, we shall consider only low-tension strings (\ref{tension}) for which the backreaction effects are negligible.
\newline

Given that possible formation mechanisms of pierced black holes exist and that these objects can be well described theoretically, an important question is how the black holes evolve in time after they were formed. Qualitatively, the time evolution in the formation scenario from above can be described as follows: When the Hubble volume shrinks, more and more of the produced black holes (and strings) enter a Hubble volume. The strings frequently intersect and reconnect. In this way some of the strings that pierce one black hole intersect with other strings that pierce the same black hole and form string loops that are attached to the black hole. These loops oscillate, emit gravitational waves and Goldstone bosons, shrink and finally get swallowed by the black hole. As a result of this process many black holes loose their strings. In \cite{Vilenkin:2018zol} it was estimated how many strings that pierce one (or more) black hole(s) survive. The simplest estimate assumed that the fraction of black holes that does not loose their strings scales with time as the fraction of the total length of the string network that survives. This estimate is justified in the simplified scenario where loops appear in the string network randomly and irrespective of the presence of black holes. It was argued in \cite{Vilenkin:2018zol} that this simplified case in fact gives a lower bound on what really happens. As a result, it was shown that at any time in the evolution history of the universe there is a nonzero density of black holes pierced by cosmic strings \cite{Vilenkin:2018zol}.
\newline

Can we expect such pierced black holes to be seeds of supermassive astrophysical black holes and present in today's universe in the center of a galaxy? Typically cosmic strings in the early universe pierce several black holes \cite{Vilenkin:2018zol}. This leads to an attractive force and in this way to a relative acceleration of the black holes in the early universe. As discussed in \cite{Vilenkin:2018zol}, an astrophysical supermassive black hole at present time can be of such origin only in case the induced velocity is sufficiently small such that the black hole could have played a role in structure formation. For a supermassive black hole of order $10^6$ solar masses, the constraint on today's string-induced black hole velocity $v_0 \leq 100 \frac{{\rm km}}{{\rm s}}$, gives rise to an upper bound on the tension $\mu$ of the cosmic string of \cite{Vilenkin:2018zol}
\begin{equation}
G_N \mu \leq 10^{-16} \, .
\label{tension}
\end{equation}
See also \cite{Lake:2015ppa} for related discussions. Therefore, given the reviewed formation mechanism, there is the possibility that black holes which are pierced by such low-tension cosmic strings are present in the center of galaxies today. The attached cosmic strings could have been dragged into the galaxies by the black holes. Other possible formation mechanisms include the capture of a cosmic string by a pre-existing astrophysical black hole while the string moves through the galaxy \cite{Lonsdale:1988xd, DeVilliers:1997nk, DeVilliers:1998nm, Snajdr:2002aa, Dubath:2006vs}. There have been speculations in the literature that the black hole in the center of our milky way might be pierced by a cosmic string \cite{Morris_2017}.

\section{Photons in the background of a black hole pierced by a cosmic string and black hole's light ring}
\label{section:section3}

In this section we shall consider photons scattered by a Kerr black hole that is pierced by a cosmic string of the form (\ref{strings}). First, we review bound photon orbits in Kerr spacetime. We recall what kind of image photons on orbits that slightly deviate from the bound orbits and that eventually reach a distant observer produce on the screen of the observer. Previous works that cover these topics include \cite{Chandrasekhar:1985kt, Teo:2020sey, Bardeen:1973tla, Luminet:1979nyg, Bardeen:1972fi, Gralla:2019xty, Johnson:2019ljv, Gralla:2019drh}. We then investigate what impact a cosmic string that pierces the black hole has on the polarization of linearly polarized light in that image. In section \ref{section:section4} we shall use this result to determine polarization patterns of the black hole image in various setups.

\subsection{Photon orbits in Kerr background and light ring}

\subsubsection{Null geodesics in Kerr spacetime and bound orbits}

It is well known \cite{Chandrasekhar:1985kt, Teo:2020sey} that one can identify three quantities that are conserved along geodesics in Kerr spacetime (constants of motion): the conjugate momenta that correspond to the two killing vector fields $\partial_\varphi$ and $\partial_t$ of (\ref{metric}), $p_\varphi$ and $p_t$, and Carter's constant $Q \equiv p_\theta^2 + {\rm cos}^2 \theta  \left(a^2 \left(-p_\mu p^\mu -p_t^2\right) + \frac{p_\varphi^2}{{\rm sin}^2 \theta} \right)$ \cite{Carter:1968rr}. Null geodesics are independent of the energy $E \equiv -p_t$ of the photons that travel along these geodesics. Therefore, one can characterize null geodesics by two independent conserved quantities which are conveniently defined as $\xi \equiv \frac{p_\varphi}{E}$ and $\eta \equiv \frac{Q}{E^2}$ (with $p_\mu p^\mu = 0$ inserted in the expression for $Q$ from above) \cite{Chandrasekhar:1985kt, Teo:2020sey}. These two quantities can be thought of as the analogue of the single impact parameter that characterizes null geodesics in the plane. For our purposes, only non-negative values of $\eta$ are relevant (see e.g. \cite{Chandrasekhar:1985kt, Teo:2020sey} for a detailed explanation of why this is the case), we therefore restrict to $\eta \geq 0$ in what follows.

The energy-rescaled four-momentum of a photon that travels along a null geodesic can be written in terms of these quantities as (see e.g. \cite{Chandrasekhar:1985kt})
\begin{equation}
\frac{p_t}{E} = -1, \frac{p_r}{E} = \pm \frac{\sqrt{R}}{\Delta}, \frac{p_\varphi}{E} = \xi, \frac{p_\theta}{E} = \pm \sqrt{\Theta} \, ,
\end{equation}
where
\begin{equation}
R \equiv r^4  + \left(a^2  - \xi^2  - \eta\right)r^2 + 2MG_Nr\left(\eta + \left(\xi - a\right)^2\right) - a^2 \eta \, ,
\end{equation}
\begin{equation}
\Theta \equiv \eta + a^2 {\rm cos}^2 \theta - \xi^2 {\rm cot}^2 \theta \, .
\end{equation}
The signs of $p_r$ and $p_\theta$ set the initial directions of motion in the radial and polar direction.
\newline

In a region around the Kerr black hole there exist null geodesics that form unstable bound orbits. Each bound orbit has a constant radius $r$ that lies in the range $r_- \leq r \leq r_+$ with (see e.g. \cite{Teo:2020sey})
\begin{equation}
r_\pm = 2MG_N\left(1 + {\rm cos} \left(\frac{2}{3} {\rm arccos} \left(\pm \frac{|a|}{MG_N}\right)\right)\right) \, .
\label{radiusrange}
\end{equation}
The bound geodesics with $r=r_-$ and $r=r_+$ lie entirely in the equatorial plane ($p_\theta = 0$), see e.g. \cite{Chandrasekhar:1985kt, Teo:2020sey}. A bound geodesic with a radius $r$ in the interval $r_-<r<r_+$ oscillates between the two different polar angles \cite{Chandrasekhar:1985kt, Teo:2020sey}
\begin{equation}
\theta_\pm = {\rm arccos} \left( \mp \sqrt{\mu_+^2}\right) \, ,
\end{equation}
where
\begin{equation}
\mu_\pm^2 \equiv \frac{\left(a^2 - \eta - \xi^2\right) \pm \sqrt{\left(a^2 - \eta - \xi^2\right)^2 + 4 a^2 \xi}}{2a^2} \, .
\end{equation}
The set of all bound photon orbits is known as the ``photon shell". Unperturbed photons on such orbits circle the black hole forever and neither fall into the black hole nor escape to infinity. For $a \neq 0$, the conserved quantities $\xi$ and $\eta$ on an orbit with radius $r$ are given by (see e.g. \cite{Chandrasekhar:1985kt})
\begin{equation}
\xi (r) = - \frac{r^3 - 3MG_Nr^2 + a^2 r + MG_Na^2}{a(r-MG_N)} \, ,
\label{angular}
\end{equation}
\begin{equation}
\eta (r) = - \frac{r^3(r^3 - 6MG_Nr^2 + 9M^2G_N^2r - 4MG_Na^2)}{a^2(r-MG_N)^2} \, .
\label{eta}
\end{equation}
We follow \cite{Johnson:2019ljv, Gralla:2019drh} and say that a photon has moved \textit{once around the black hole} or has made \textit{one orbit around the black hole} when it reached for the first time the same polar angle from which it started. To move once around the black hole, a photon on an orbit with radius $r_- < r < r_+$ needs a time of (see e.g. \cite{Teo:2020sey} for the geodesic equations to derive the following expression from)
\begin{align} \label{timeshift}
&\tau (r) \equiv 4 \int_0^{\theta_+} \frac{dt}{d\theta}d\theta \\ \nonumber &=4\frac{(r^2 + a^2)^2 - 2 MG_Na \xi r - a^2 \Delta}{a \Delta \sqrt{-\mu_-^2}}K\left(\frac{\mu_+^2}{\mu_-^2}\right)- 4 a \sqrt{-\mu_-^2}\left(K\left(\frac{\mu_+^2}{\mu_-^2}\right)-E \left(\frac{\mu_+^2}{\mu_-^2}\right)\right) \, .
\end{align}
During this time, the photon passes an azimuth angle of \cite{Teo:2020sey}
\begin{equation}
\delta(r) \equiv 4 \int_0^{\theta_+}\frac{d\varphi}{d\theta} d \theta = \frac{4}{\sqrt{-\mu_-^2}}\left(\frac{2MG_Nr-a \xi}{\Delta} K\left(\frac{\mu_+^2}{\mu_-^2}\right) + \frac{\xi}{a} \Pi\left(\mu_+^2, \frac{\mu_+^2}{\mu_-^2}\right)\right) \, .
\label{delphi}
\end{equation}
Here $K$, $E$ and $\Pi$ are the complete elliptic integrals of the first kind, second and third type. Photons that move on orbits that lie entirely in the equatorial plane (orbits with $r=r_-$ and $r = r_+$) satisfy $\delta = 2 \pi$ and $\tau = \frac{2 \pi}{\delta(r_+)} \tau(r_+)$ where $\delta(r_+)$ and $\tau(r_+)$ stand for the expressions (\ref{delphi}) and (\ref{timeshift}) with $r_+$ (or $r_-$) inserted.

Typically photons on the bound orbits never return to the same $\varphi$ coordinate when they circle the black hole. Only when $\frac{\delta}{2 \pi}$ is a fractional number, 
\begin{equation}
\frac{\delta}{2 \pi} = \frac{p}{q} \, ,
\label{fractionalnumber}
\end{equation}
the orbits are closed. This includes the orbits that lie entirely in the equatorial plane. In the case of non-rotating black holes ($a = 0$) all bound orbits are closed.

\subsubsection{Light ring}
\label{section:lightring}

Photons that move on geodesics that deviate slightly from the bound geodesics circle the black hole and eventually escape to infinity (or fall into the black hole). The smaller the deviation of a geodesic from the corresponding bound geodesic is, the more times the photons circle the black hole before they reach out to infinity (or fall into the black hole). When the photons reach a distant observer, they produce an image on the screen of the observer that is known as ``photon ring" or ``light ring" \cite{Bardeen:1973tla, Luminet:1979nyg}. In order to understand some basic properties of this image, it is useful to first consider the image-curve on the screen which is produced by the photons that have completed $n \gg 1$ half-orbits around the black hole before reaching the observer. This curve in the limit $n \rightarrow \infty$ is sometimes termed the ``critical curve" \cite{Johnson:2019ljv, Gralla:2019drh}. In Cartesian coordinates $(\alpha, \beta)$ on the screen of a distant observer ($r_o \rightarrow \infty$), which are chosen such that the $\beta$-axis is the projected black hole spin axis, the critical curve takes the form (see e.g. \cite{Bardeen:1973tla, Chandrasekhar:1985kt})
\begin{equation}
(\alpha_c, \beta_c) = \left(-\frac{\xi(r)}{{\rm sin} \theta_o},  \pm \sqrt{\eta(r) + a^2 {\rm cos}^2 \theta_o - \xi^2(r) {\rm cot}^2\theta_o}\right) \, .
\label{coordinatesscreen}
\end{equation}
Here $\theta_o \neq 0$ is the inclination of the observer from the black hole spin axis. Polar coordinates $(\rho, \varphi_\rho)$ on the screen are related to the Cartesian coordinates via
\begin{equation}
(\rho, \varphi_\rho) = \left(\sqrt{\alpha^2 + \beta^2}, {\rm arctan}\left(\frac{-\beta}{\alpha}\right)\right) \, .
\label{polar}
\end{equation}
In what follows we also comment on the $\theta_o = 0$ case.
\newline

As can be seen from (\ref{coordinatesscreen}) and (\ref{polar}), both $\varphi_\rho$ and $\rho$ are in general $r$-dependent ($r$ is the radius at which the photons have circled the black hole). This has several implications: First, for $a \neq 0$ and $\theta_o \neq 0$, the $r$-dependence of $\rho$ implies a shape of the photon ring which is not a circle.\footnote{See Fig. \ref{fig:1}, \ref{fig:2} for some examples. For most parameter choices the shape is very close to circular. Only for large $\theta_o$ and/or large $a$ the photon ring becomes flattened on one side.} Only in the cases $a = 0$ and/or $\theta_o = 0$ the photon ring is perfectly circular. Second, different angular coordinates $\varphi_\rho$ probe different radii $r$ around the black hole or, in other words, photons at different angular points on the screen have circled the black hole at different radii before reaching the observer. Third, there is one degeneracy in what we have just pointed out: The same given radius $r$ of photons that reach the observer maps not only to one but (for $a \neq 0$ and $\theta_o \neq 0$) to exactly two different angular points $\varphi_\rho$ on the screen. This gives rise to a reflection symmetry of the photon ring about the $\alpha$-axis. For $a = 0$, all photons come from one radius in the photon shell (the only existing one in that case, $r_+ = r_- = 3 M G_N$). For $\theta_o = 0$ only one radius is probed as well (as we will reiterate in what follows).
\newline

Only observers in the equatorial plane ($\theta_o = \pi/2$) can see photons in the ring on their screens that come from orbits of all radii in the photon shell, $r_- \leq r \leq r_+$ (see e.g. \cite{Johnson:2019ljv}). The smaller $\theta_o$, the fewer different radii are probed. This is because only photons that circle the black hole at a particular limited range of radii $r$ in the photon shell can reach a distant observer at an inclination $\theta_o \neq \frac{\pi}{2}$. The range of the parameters $r$ that are probed at a given $\theta_o$ can be determined from requiring that $\beta_c$ is real. For observers at the pole ($\theta_o = 0$), all photons on the ring come from one and the same orbit (with one radius).
\newline

A distant observer does not see the critical curve but an image of an infinite sequence of exponentially demagnified nested subrings that approach the critical curve \cite{Gralla:2019xty, Johnson:2019ljv, Gralla:2019drh}. In order to understand how this image emerges and to characterize the structure of the image, it is useful to use three parameters \cite{Gralla:2019drh}: The parameters $\tau$ and $\delta$ as defined in (\ref{timeshift}) and (\ref{delphi}) and the Lyapunov exponent $\gamma$ that describes the instability of the bound photon orbits around the black hole. All three parameters depend on the radius $r$ of the corresponding bound orbit.\footnote{Note that our conventions for $\tau$ and $\delta$ differ from the conventions used in  \cite{Gralla:2019drh} by a factor of 2.}
\newline

For a bound orbit of radius $r$, the Lyapunov coefficient $\gamma$ is given by\footnote{There are various different definitions of $\gamma$ used in the literature. We follow the definition that is used in \cite{Johnson:2019ljv, Gralla:2019drh}. An alternative definition that is widely used in the literature can be found in \cite{Cardoso:2008bp}.} \cite{Johnson:2019ljv, Gralla:2019drh}
\begin{equation}
\gamma = \frac{4}{a} \sqrt{r^2 - \frac{MG_Nr \Delta}{\left(r-MG_N\right)^2}} \frac{1}{\sqrt{-\mu_-^2}} K\left(\frac{\mu_+^2}{\mu_-^2}\right) \, .
\end{equation}
A geodesic that initially has an infinitesimal radial deviation $\Delta r_0$ from a bound geodesic deviates after $n$ half-orbits by
\begin{equation}
\Delta r(n) \approx e^{\gamma n} \Delta r_0 \, .
\label{delta}
\end{equation}
This implies that photons which have completed a different number of half-orbits around the black hole before they reach the observer, arrive in different ranges of radial positions $\rho$ on the observer's screen. The consequence is a characteristic substructure of the photon ring: The photon ring is a ring that consists of an infinite sequence of nested subrings. The photons in the $n^{\rm th}$ subring have completed $n$ half-orbits around the black hole before reaching the observer. The rings become exponentially narrower when they approach the critical curve in the limit $n \rightarrow \infty$. Each subring is an image of photons from all sources in the universe that got lensed by the black hole and reach the observer. In what follows we shall often distinguish between two different classes of subrings: Even subrings (subrings with even $n$) and odd subrings (subrings with odd $n$).
\newline

Let us now consider isotropically emitting light sources. Photons that are emitted at the same location in the sky by such sources can, depending on the emission direction, move on many different geodesics with different  ``impact parameters" $\eta$ (\ref{eta}) and $\xi$ (\ref{angular}) and thus circle the black hole at various different radii $r$. Whether or not photons can reach the screen of a distant observer after having completed $n$ half-orbits around the black hole at a given radius $r$ depends on the values of the impact parameters and on the location of the observer (see e.g. \cite{Johnson:2019ljv}). For any radius $r$ which can be probed by a given observer, photons emitted at the same location can generally reach the observer only after having completed one concrete number of half-orbits. Only when photon orbits are closed (\ref{fractionalnumber}), photons from the same location that orbit the black hole at a given $r$ can reach the same observer after having completed various different numbers of half-orbits. (In this case they appear in various different subrings at the same angular coordinate $\varphi_\rho$.) In general, photons that are emitted at the same location in the sky can however reach the same observer after having completed different numbers of half-orbits when they circle the black hole at different radii $r$. Therefore, photons that are emitted at the same location generally appear in different subrings at different angular coordinates $\varphi_\rho$ (as explained above). Vice versa, photons in different subrings at the same angular coordinates $\varphi_\rho$ are in general emitted at different locations in the sky. The parameter $\delta(r)$ (\ref{delphi}) is the difference between the azimuth angles of the locations in the sky from which photons at an angular coordinate $\varphi_\rho(r)$ in the $n^{\rm th}$ and $(n+2)^{\rm th}$ subrings originate \cite{Gralla:2019drh}. It is completely determined by the black hole parameters and does not depend on the particular light source distribution (\ref{delphi}). The light in the $n^{\rm th}$ subring that orbited the black hole at a radius $r$ is delayed by a time $\tau(r)$ (\ref{timeshift}) when compared to the light in the $(n+2)^{\rm th}$ subring \cite{Gralla:2019drh}. Subrings with larger $n$ are composed of photons which left their source(s) at an earlier time.

\subsection{Impact of the cosmic string on the photons}

\begin{wrapfigure}[22]{r}{6cm}
\begin{center}
\includegraphics[scale=0.15]{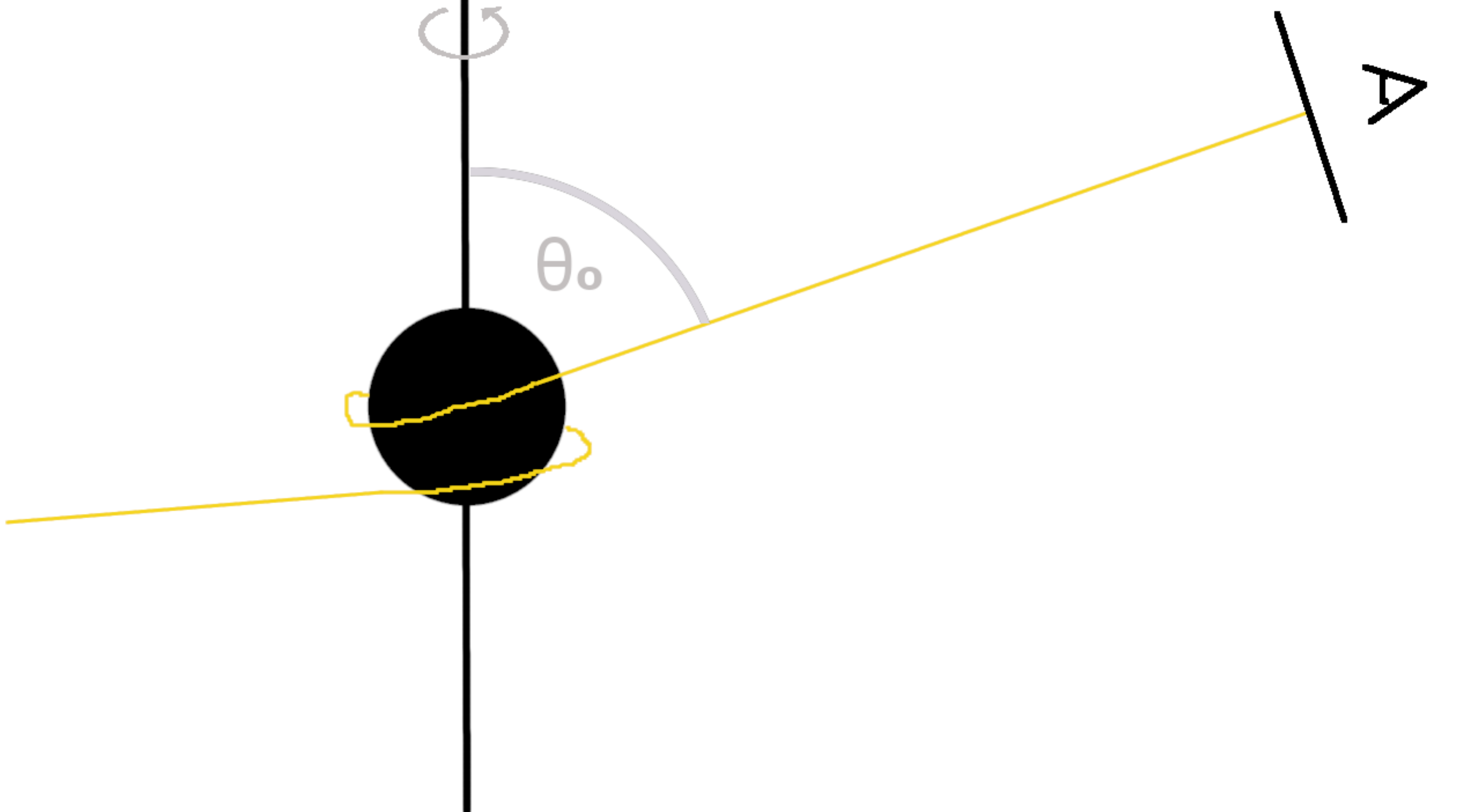}
\end{center}
\caption{\small Illustration of the setup with one ray of light (yellow) that is emitted by a source far away from the black hole: Light circles the rotating black hole that is pierced by a cosmic string which coincides with the black hole spin axis. It then reaches the screen of a distant observer. The angle $\theta_o$ is the inclination of the observer from the black hole spin axis.}
\label{fig:0}
\end{wrapfigure}

Let us now consider a cosmic string of the form (\ref{strings}) and photons that couple to the Goldstone boson $\phi$ via (\ref{coupling}). At the level of the geometric optics approximation, the coupling (\ref{coupling}) gives rise to birefringence when photons move on a worldline along which $\phi$ changes its value: Left-handed/right-handed polarized photons acquire a phase of the form \cite{Harari:1992ea, Fedderke:2019ajk, Schwarz:2020jjh}
\begin{equation}
\epsilon_{L,R} \longrightarrow e^{\pm i C\frac{\Delta \phi}{v}}\epsilon_{L,R} \, ,
\label{epsilonrot}
\end{equation}
where $\epsilon_{L,R} \equiv \frac{1}{\sqrt{2}}\left(\epsilon_1 \pm i \epsilon_2\right)$ is the polarization vector of left-handed (right-handed) polarized photons, $C$ is the constant from (\ref{coupling}) and $\pm$ stands for the left-handed (right-handed) polarization. The phase only depends on the values of $\phi$ at the initial and end points of the photon's worldline. It is independent of the spacetime metric \cite{Schwarz:2020jjh}. For linearly polarized photons (\ref{epsilonrot}) implies a polarization rotation of 
\begin{equation}
\Delta \Phi = C\frac{\Delta \phi}{v} \, .
\end{equation}
In order to illustrate the geometric optics effect in the case of a black hole background, let us, as above, consider a Kerr black hole that is pierced by a single cosmic string in such a way that the string coincides with the black hole spin axis. As discussed above, in the black hole spacetime, there are null geodesics that complete a certain number of half-orbits around the black hole (and the string) before they reach a distant observer (see Figure \ref{fig:0} for an illustration with a source that is located far away from the black hole). The string generates a change in the polarization of the photons when they move on these geodesics around the black hole: Since the field profile of $\phi$ around the string that pierces the rotating black hole is given by $\pm v \varphi$ (\ref{strings}), a linearly polarized photon acquires a polarization rotation of
\begin{equation}
\Delta \Phi_{2n} =  \pm Cn \delta(r)
\label{polrot}
\end{equation}
once it makes $2n$ half-orbits around the string. Here $\pm$ is set by the orientation of the string and $\delta$ is defined in (\ref{delphi}). The phase (\ref{polrot}) is a topological phase, similar as known from the Aharonov-Bohm effect \cite{Aharonov:1959fk}. It does not depend on the concrete worldline of the photon. In the case of a non-rotating black hole ($a=0$), $\delta(r) \rightarrow 2 \pi$ is independent of $r$. For rotating black holes, $\delta(r)$ differs from $2 \pi$ and depends on the radius of the orbiting photons. The underling physical reason for this dependence is that the photons get dragged by a rotating black hole in a way that depends on the radius of the orbit.
\newline

This effect is independent of the vacuum expectation value $v$ of the spontaneously broken symmetry that gave rise to the cosmic string (\ref{lag}). When the constant $C$ is given by (\ref{anomaly}), for any given $n$ and given black hole parameters that set $\delta(r)$, the size of the effect depends only on the anomaly coefficient $\mathcal{A}$ (\ref{anomaly}) and the electromagnetic fine structure constant $\alpha = \frac{1}{137}$. As we shall argue in section \ref{section:section4}, in the black hole setup this effect thus provides us with a promising way to measure (or to constrain) the anomaly coefficient $\mathcal{A}$. A similar way to measure (or constrain) $\mathcal{A}$ via the same effect was recently proposed in different setups \cite{Agrawal:2019lkr} (see also \cite{Jain:2021shf}).\footnote{Various setups without cosmic strings that have been discussed along these lines in the literature (see e.g. \cite{Harari:1992ea, Lue:1998mq, Pospelov:2008gg, Kamionkowski:2008fp, Ivanov:2018byi, Fujita:2018zaj, Fedderke:2019ajk, Blas:2019qqp, Chen:2019fsq}) give rise to much smaller polarization rotations because, in contrast to our setup, in those cases the generated phase depends on the vacuum expectation value $v$.} The value of $\mathcal{A}$ that we can expect from theoretical considerations depends on the theoretical model that gave rise to the anomaly; see e.g. \cite{Agrawal:2019lkr} which includes a short summary of what values of $\mathcal{A}$ we can expect in various models where the anomaly is obtained by integrating out fermions in the UV (see also \cite{Srednicki:1985xd} for a discussion of this kind for minimal GUT models). Knowledge of $\mathcal{A}$ could help us to learn a lot about the UV model that is realized in nature. For $\mathcal{A} = 1$ (obtained in the case of just a single charged fermion of charge one \cite{Agrawal:2019lkr}) and $\delta(r) \approx 2 \pi$ (a good estimate for $n=1$ and slowly rotating black holes), we can expect a polarization rotation of $\mathcal{O}\left(\alpha\right)=\mathcal{O}\left(1\%\right)$. This can become much bigger when the black hole is pierced by more than one cosmic string or when $n \gg 1$.
\newline

In what follows (section \ref{section:section4}) we shall investigate in detail what impact the generated change in the polarization direction has on the polarization of the linear polarized photons in the photon ring on the screen of a distant observer. Since, as pointed out above, different angular coordinates $\varphi_\rho$ of the photon ring generally consist of photons that orbited the black hole at different radii $r$, the polarization rotation (a function of $r$, (\ref{polrot})) depends on the angular coordinate of the photon ring. We shall work out signatures in various setups.

\section{Polarimetric signatures of the photon ring}
\label{section:section4}

We shall first review some aspects of the polarimetric signatures of a black hole photon ring that are generated by the geodesic motion of the photons and the corresponding parallel transport of the polarization vector (see \cite{Himwich:2020msm} for a detailed recent study). We shall then investigate what impact a cosmic string of the form (\ref{strings}) has on these signatures when it pierces the black hole. We focus on the linearly polarized component of the light in the ring that can potentially be measured with future very large baseline interferometers \cite{Himwich:2020msm}. We shall both discuss signatures for time-averaged black hole images and time dependent effects. We do not take into account additional possible polarization effects coming from plasma birefringence (see e.g. \cite{Moscibrodzka:2017gdx, Jimenez-Rosales:2018mpc, Tsunetoe:2020pyz, Ricarte:2020llx, Tsunetoe:2020nws, Jimenez-Rosales:2021ytz}), from other absorption or scattering processes or from potential string-induced plasma effects. A detailed study on the potential impact of such effects on our results requires a further detailed analysis that is beyond the scope of this work.\footnote{As mentioned in \cite{Himwich:2020msm}, there are some good indications that, in the case of black holes with optically thin environments, these ``environmental effects" on the propagating light are either weak or can be well separated from the signal: One can expect that plasma effects can be removed from a signal in case one has data from different frequency components of the light. (This is because the plasma effects are typically frequency dependent whereas the effects that we study are achromatic.) Other effects due to absorption and scattering are expected to be relatively weak in an optically thin environment.}

\subsection{Photon polarization}

Let us first consider linear polarized photons that move with four-momentum $p_\mu$ on null geodesics in the Kerr geometry. The polarization vector $\epsilon_\mu$ of a photon (which obeys $\epsilon^\mu p_\mu = 0$) gets parallel transported along the photon's path. The projection of the parallel transported and unit-normalized polarization vector onto the screen of a distant observer is given by (see e.g. \cite{Chandrasekhar:1985kt})
\begin{equation}
(\epsilon_\alpha, \epsilon_\beta) = \frac{(\beta K_2 - \nu K_1, \beta K_1 + \nu K_2)}{\sqrt{(K_1^2 + K_2^2)(\beta^2 + \nu^2)}} \, .
\label{polarisation}
\end{equation}
Here ($\alpha$, $\beta$) are Cartesian coordinates on the observer's screen and $\nu \equiv -\left(\alpha + a {\rm sin} \theta_o\right)$. $K_1$ and $K_2$ are defined in terms of the initial polarization $\epsilon_\mu$ and the momentum $p_\mu$ of the photon as
\begin{align}
K &\equiv K_1 + i K_2 = {\Big {(}}\left(p^t \epsilon^r - p^r \epsilon^t \right)+a {\rm sin}^2 \theta \left(p^r \epsilon^\varphi - p^\varphi \epsilon^r\right) \nonumber \\
&- i \left(\left(\left(r^2 + a^2\right)\left(p^\varphi \epsilon^\theta - p^\theta \epsilon^\varphi \right) - a \left(p^t \epsilon^\theta - p^\theta \epsilon^t \right)\right){\rm sin} \theta \right){\Big {)}}
\left(r - ia{\rm cos} \theta \right) \, .
\end{align}
In terms of the Stokes parameters $Q$ and $U$ that parametrize the linear polarized part of light, the complex polarization $P$ can be written as
\begin{equation}
P \equiv Q + i U = k I e^{2i \sigma} \, ,
\end{equation}
where $I$ is the intensity, $k$ the degree of polarization and $\sigma$ is defined via
\begin{equation}
{\rm tan} \sigma = - \frac{\epsilon_\alpha}{\epsilon_\beta} \, .
\end{equation}
Using the definition of $\sigma$, (\ref{polarisation}) can also be written as
\begin{equation}
(\epsilon_\alpha, \epsilon_\beta) = (- {\rm sin} \sigma, {\rm cos} \sigma) \, .
\label{polarisation2}
\end{equation}

To see how a cosmic string that pierces the black hole and that couples to the photons via (\ref{coupling}) affects the polarization of a photon on the observer's screen (\ref{polarisation2}), let us define two unit vectors:
\begin{equation}
e_1 = (- {\rm sin} \sigma, {\rm cos} \sigma) \, ,
\end{equation}
\begin{equation}
e_2 = ({\rm cos} \sigma, {\rm sin} \sigma) \, .
\end{equation}
As discussed before, in the absence of a cosmic string the polarization $(\epsilon_\alpha, \epsilon_\beta)$ is given by $e_1$. In case a string is present, once a linear polarized photon makes $2n$ half-orbits around the string, its polarization on the observer's screen gets shifted (\ref{polrot}) and becomes
\begin{equation}
(\epsilon^{2n}_\alpha, \epsilon^{2n}_\beta) = {\rm cos} (\Delta \Phi_{2n}) e_1 - {\rm sin} (\Delta \Phi_k) e_2 = (-{\rm sin}(\sigma + \Delta \Phi_{2n}), {\rm cos} (\sigma + \Delta \Phi_{2n})) \, .
\end{equation}
In terms of the polarization $(\epsilon_\alpha, \epsilon_\beta)$ that would be seen on the screen if the string were absent, this can be written as
\begin{equation}
(\epsilon^{2n}_\alpha, \epsilon^{2n}_\beta) = (\epsilon_\alpha {\rm cos}(\Delta \Phi_{2n}) - \epsilon_\beta {\rm sin}(\Delta \Phi_{2n}), \epsilon_\beta {\rm cos}(\Delta \Phi_{2n}) + \epsilon_\alpha {\rm sin}(\Delta \Phi_{2n})) \, .
\end{equation}
Some algebra gives, that the phase shift $\Delta \Phi_{2n}$ can be obtained from these polarizations as
\begin{equation}
{\rm sin}\left(\Delta \Phi_{2n}\right) = \epsilon^{2n}_\beta \epsilon_\alpha - \epsilon_\beta \epsilon^{2n}_\alpha \, .
\label{deltaphi}
\end{equation}

\subsection{Time-averaged signatures}

Interesting polarimetric signatures can be obtained from taking time-averaged images of a black hole photon ring. Time-averaging has several advantages, for example, local, random fluctuations get averaged out. In this subsection, we shall focus on signatures that appear in the case of light source distributions that are such that, after time-averaging, the entire photon ring gets populated with photons emitted from the sources. We assume a given (partial) initial source polarization of the photons that does not depend on the locations of the light source(s). Our results need to be adjusted in cases where the initial linear polarization of the photons depends on the locations of their source(s).
\newline

At first, one might want to consider string-induced polarization effects of the photons in individual subrings of the photon ring. In general, such effects are generated because photons at different angular coordinates $\varphi_\rho$ in a given subring have generally passed different azimuth angles on their way from the source to the observer (see the discussion in the last paragraph in section \ref{section:lightring}). In this way they have acquired different string-induced polarization rotations. These effects however generally depend on the locations of the light source(s) and the observer (see e.g. \cite{Gralla:2019drh} and the discussion in the last paragraph of section \ref{section:lightring}). Therefore, the resulting polarization patterns in individual subrings are functions of the particular light source distribution and in general have to be determined for each given distribution individually. Given the light source distribution and the location of the observer, one can determine these patterns by integrating the geodesic equation (see e.g. \cite{Gralla:2019drh, Gralla:2019ceu} for an analytic approach). We won't perform such an analysis in this work.
\newline

Instead of considering polarizations of photons in individual subrings of the photon ring, we shall consider the class of even subrings (or the class of odd subrings) and focus on relative polarizations of the photons from two given even/odd subrings. In this way universal string-induced polarization effects that do not depend on the locations of the light sources and the observer emerge (see the discussion in the last paragraph in section \ref{section:lightring}). In what follows we shall study such effects for the case of photons from the $n^{\rm th}$ and $(n+2)^{\rm th}$ subring (for any $n$). In certain special cases of observer locations and source distributions, a similar analysis could be done for relative polarizations between one even and one odd subring \cite{Gralla:2019drh}.
\newline

\begin{figure}

\includegraphics[scale=0.242]{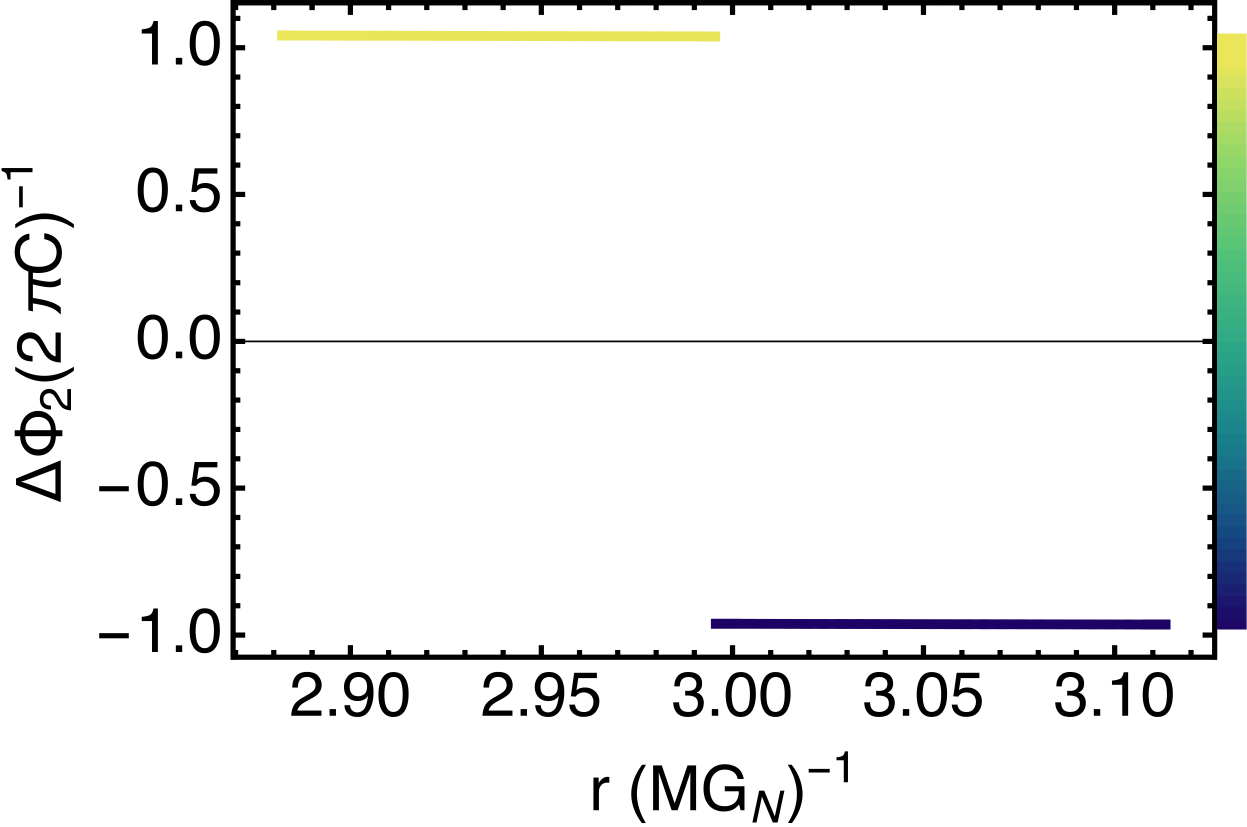} \hspace{0.07em}
\includegraphics[scale=0.242]{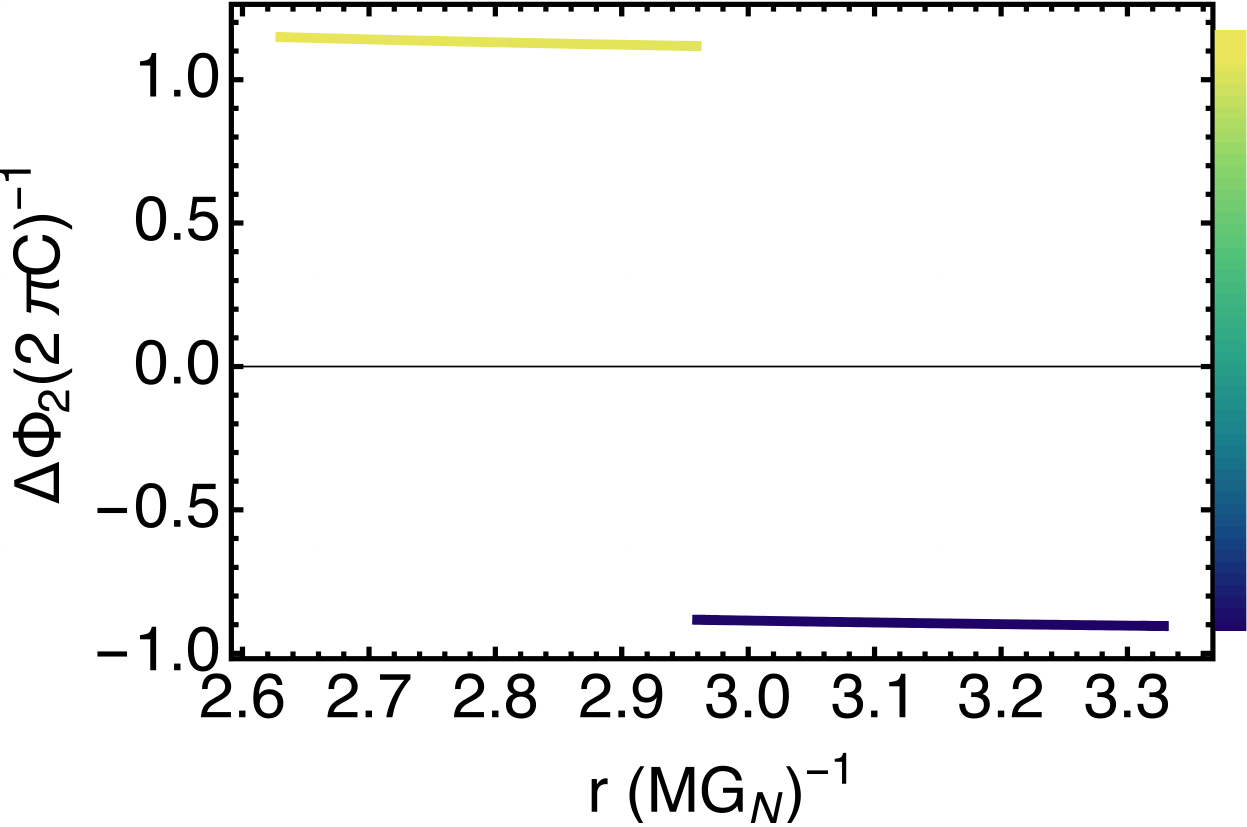} \hspace{0.07em}
\includegraphics[scale=0.242]{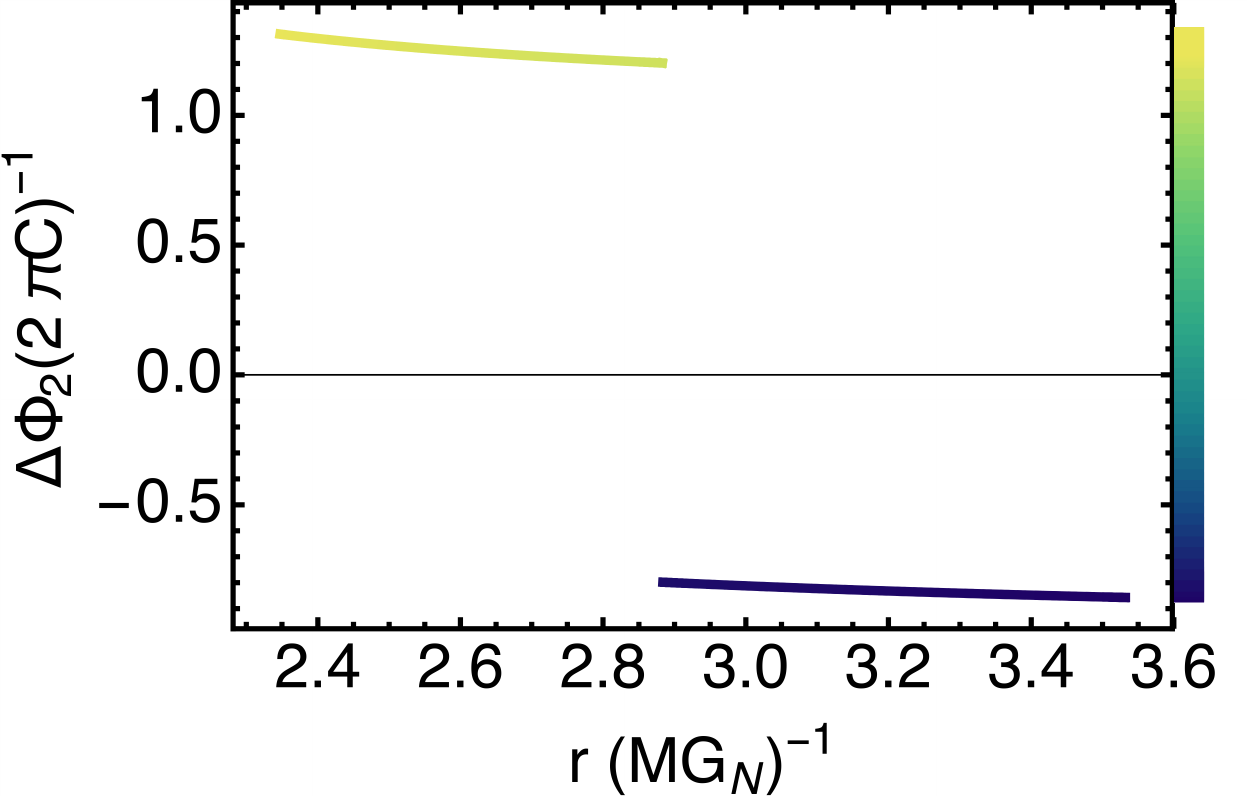} \hspace{0.07em}
\includegraphics[scale=0.242]{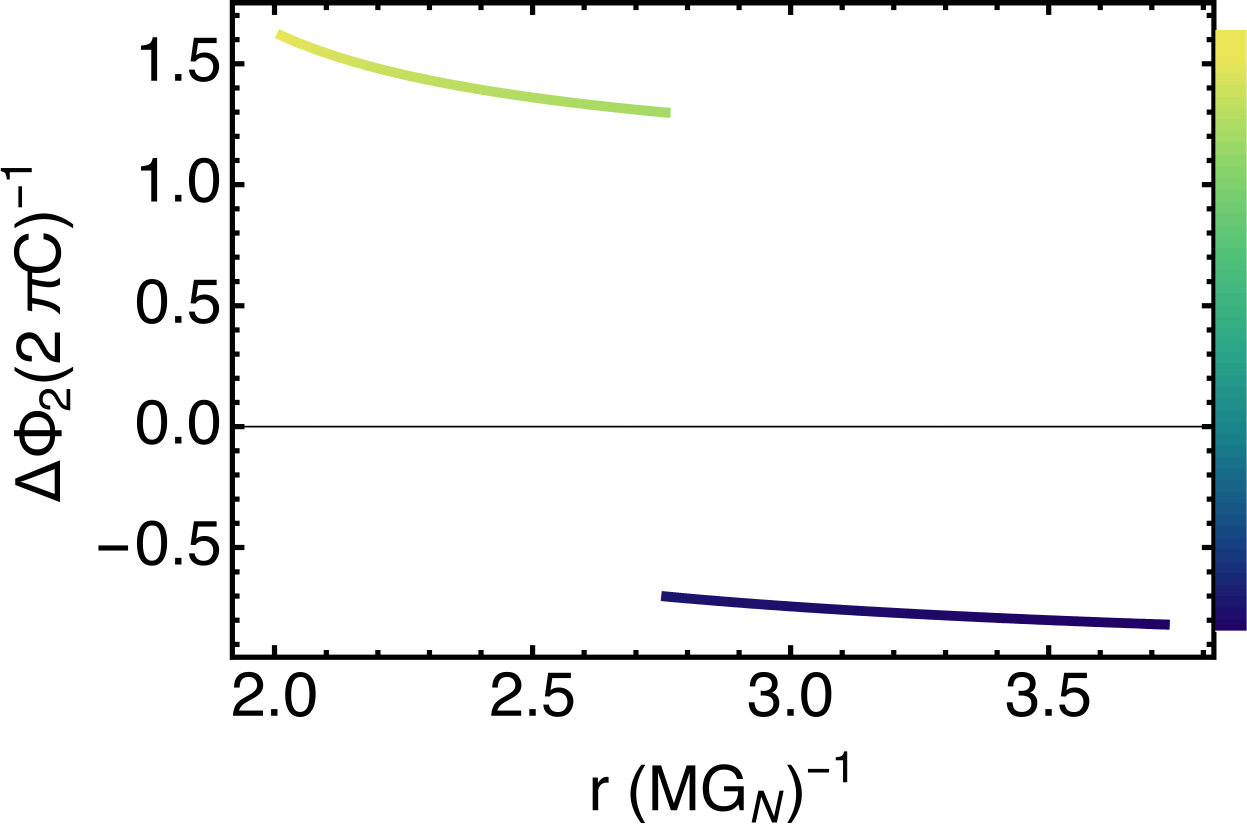} \hspace{0.07em}
\includegraphics[scale=0.242]{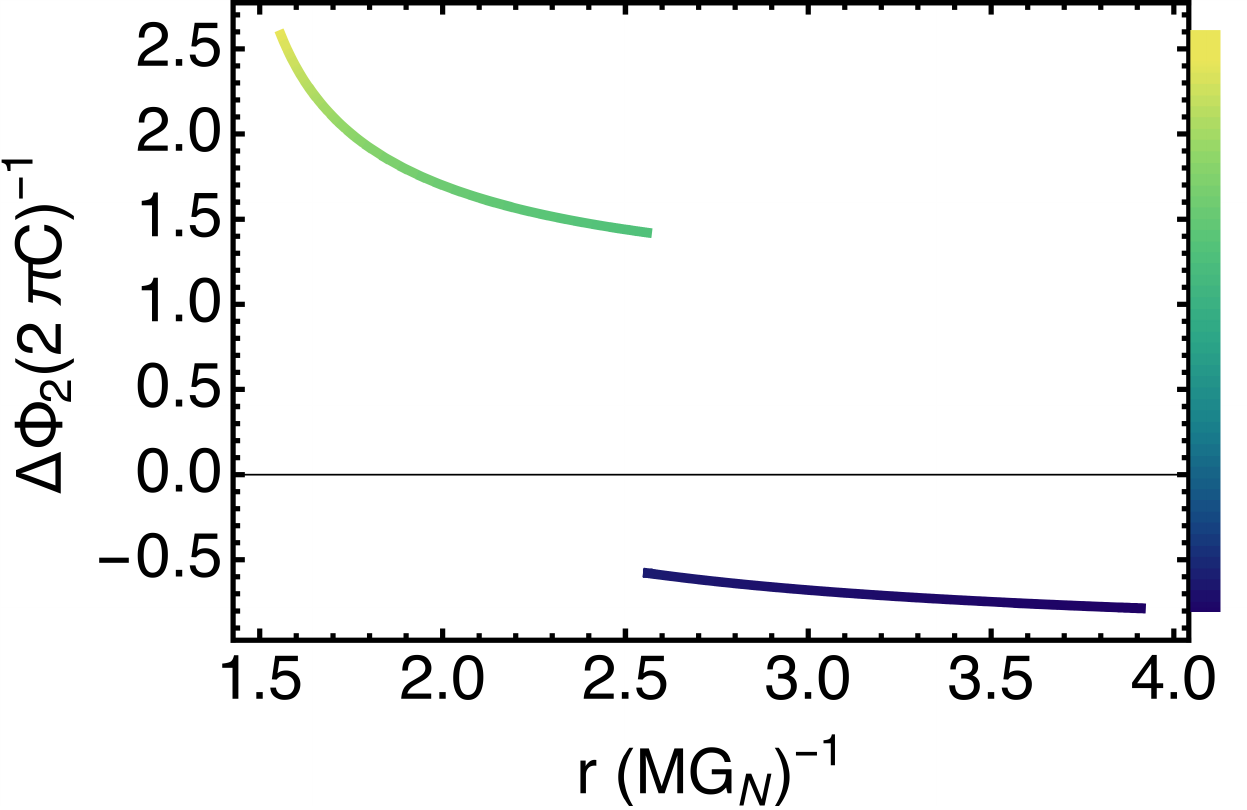}
\newline

\includegraphics[scale=0.241]{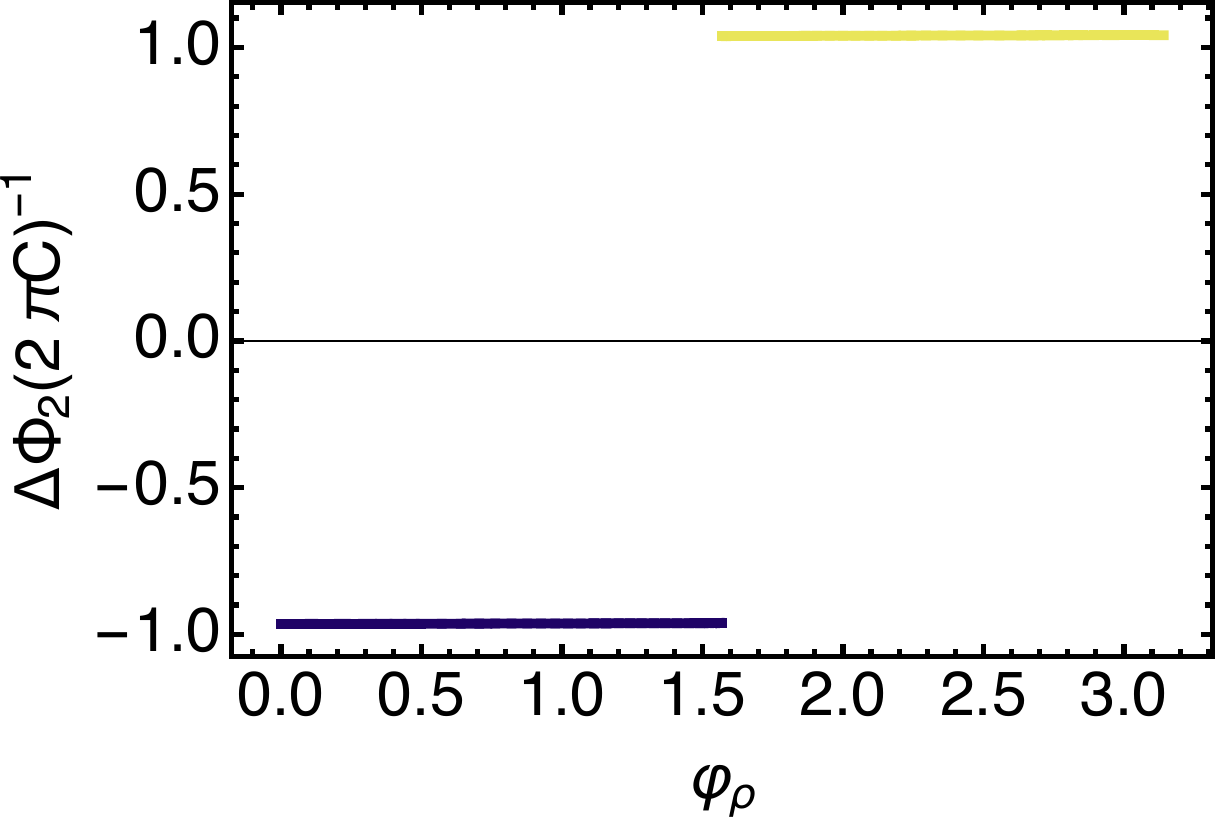} \hspace{0.05em}
\includegraphics[scale=0.241]{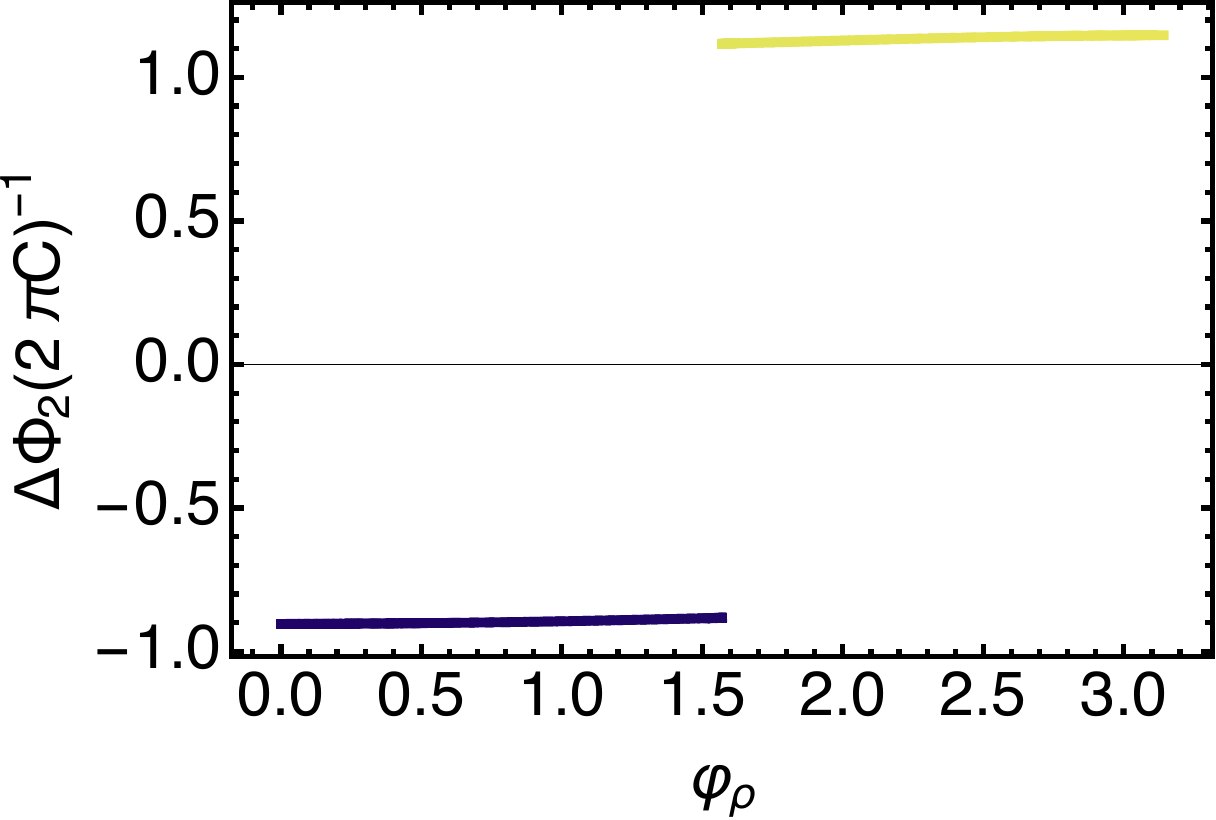} \hspace{0.05em}
\includegraphics[scale=0.241]{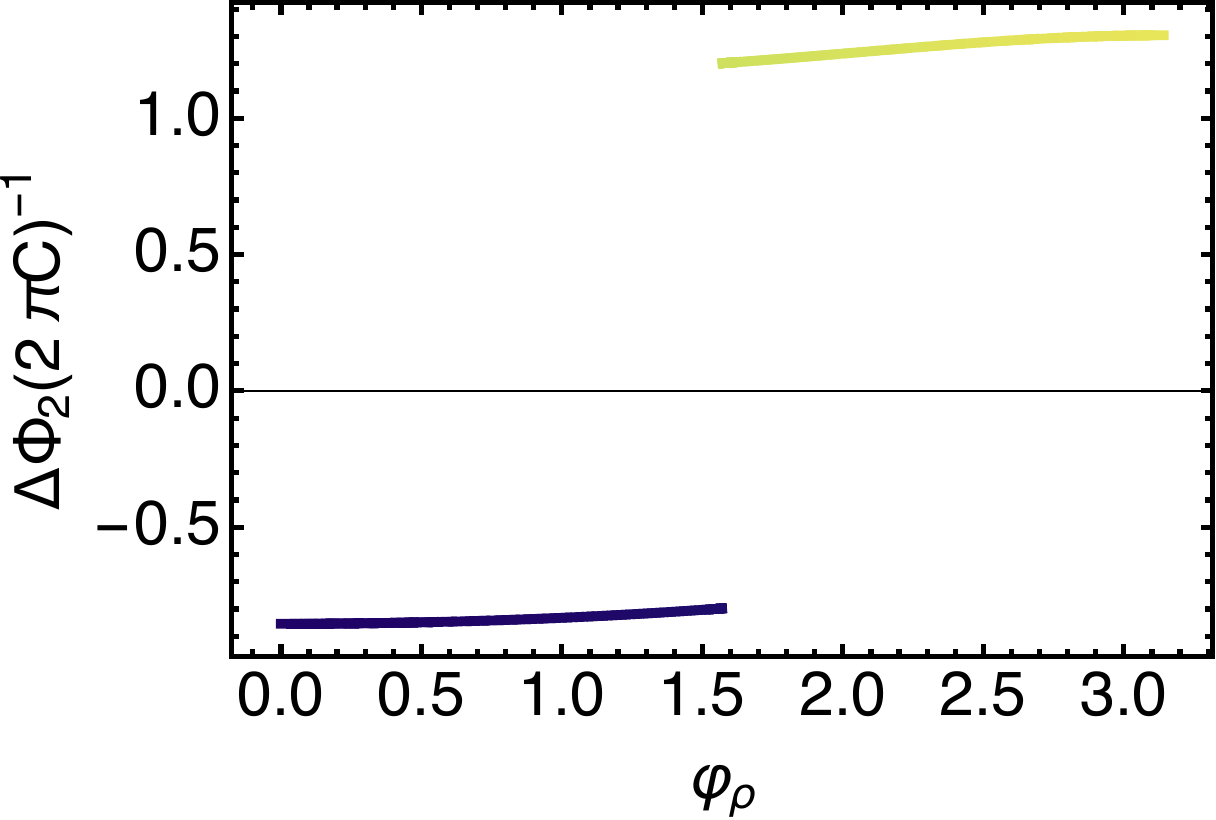} \hspace{0.05em}
\includegraphics[scale=0.241]{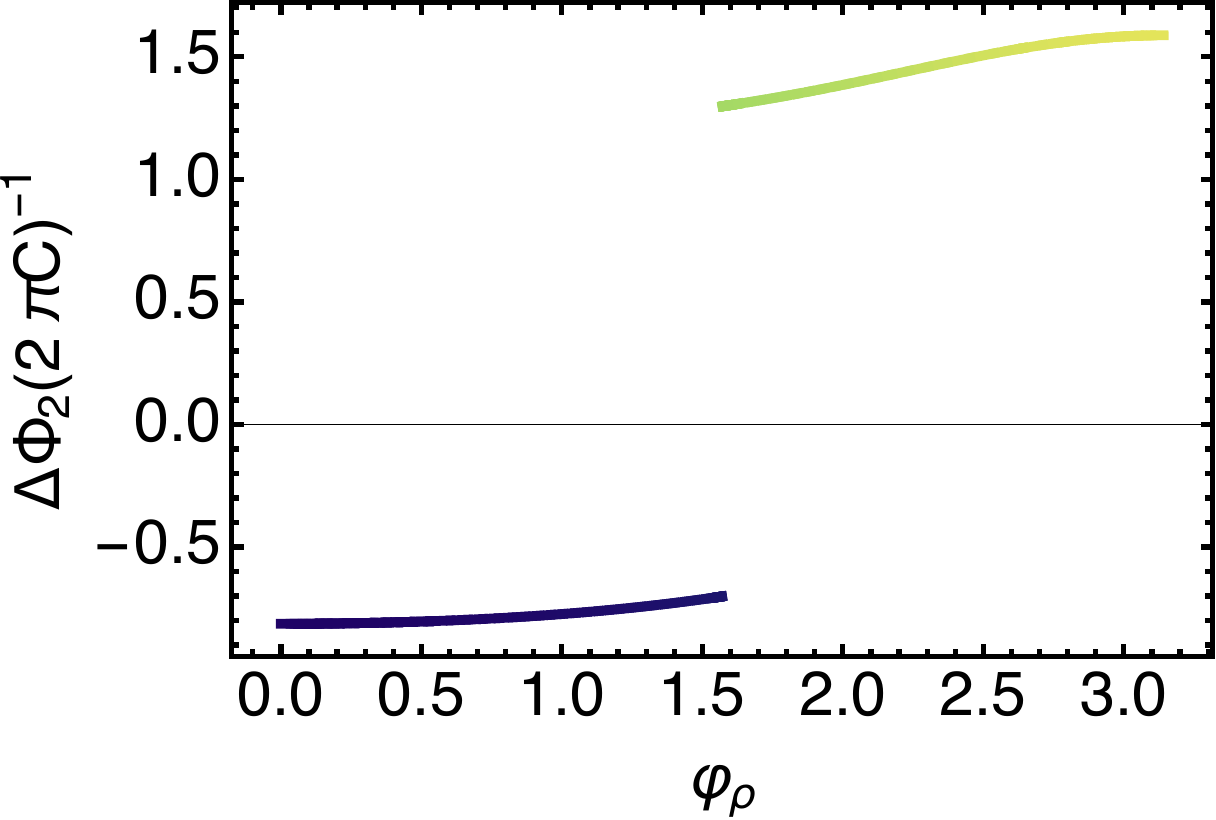} \hspace{0.05em}
\includegraphics[scale=0.241]{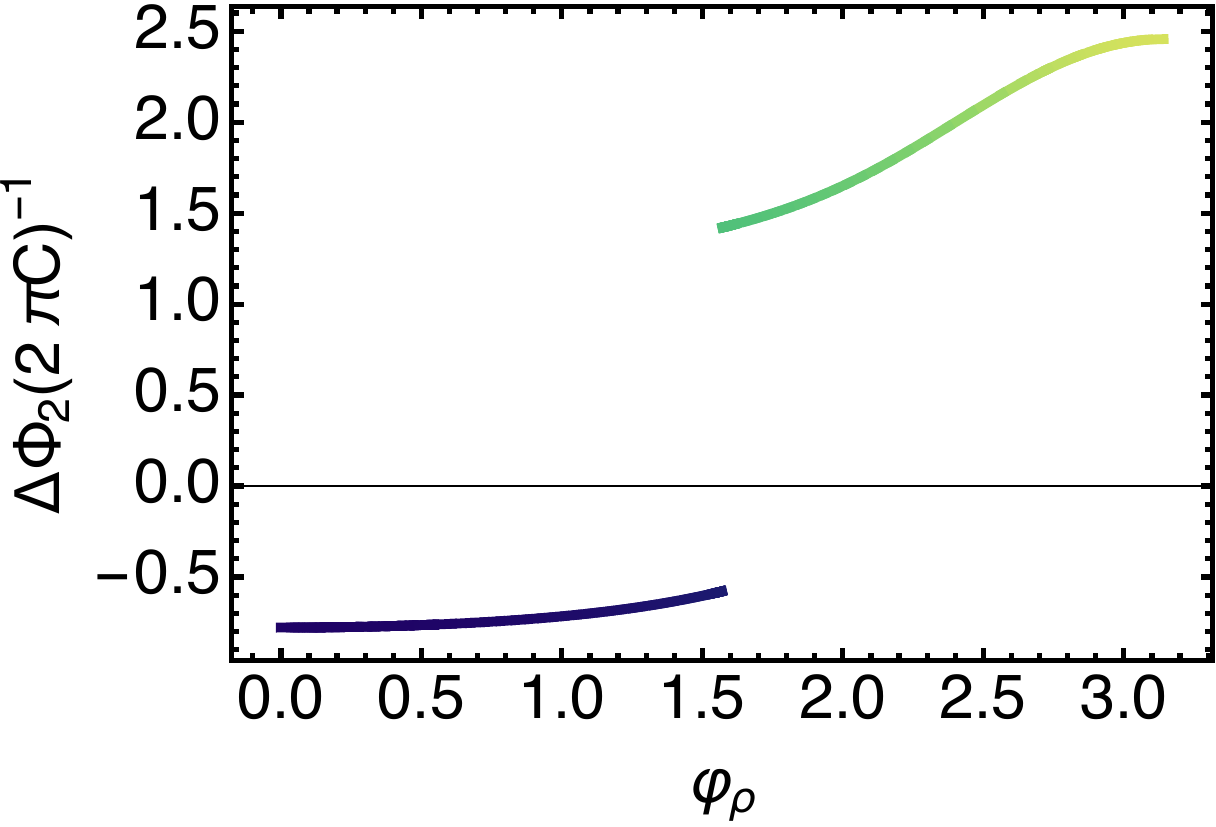}
\newline

\includegraphics[scale=0.245]{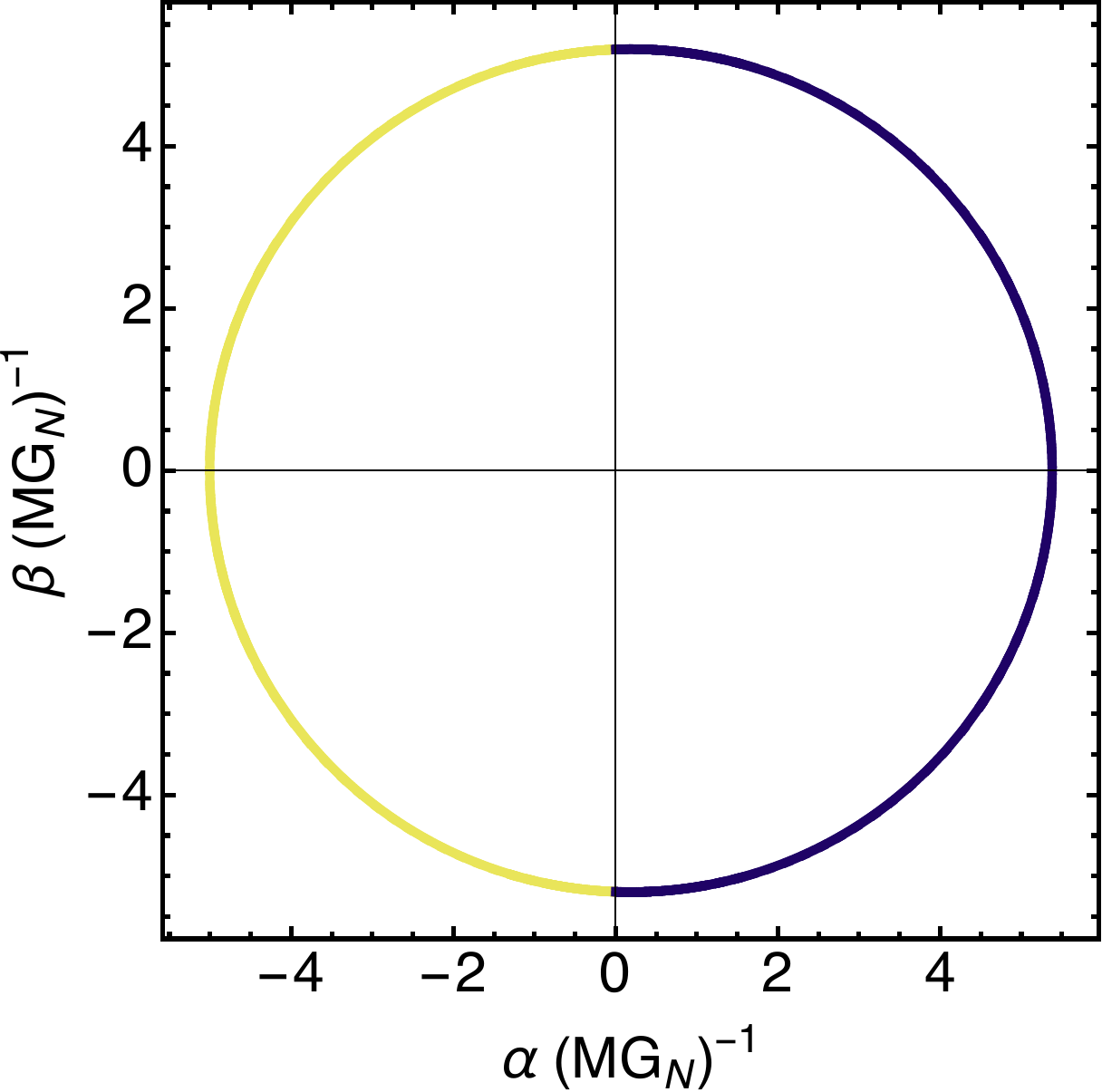} \hspace{0.07em}
\includegraphics[scale=0.245]{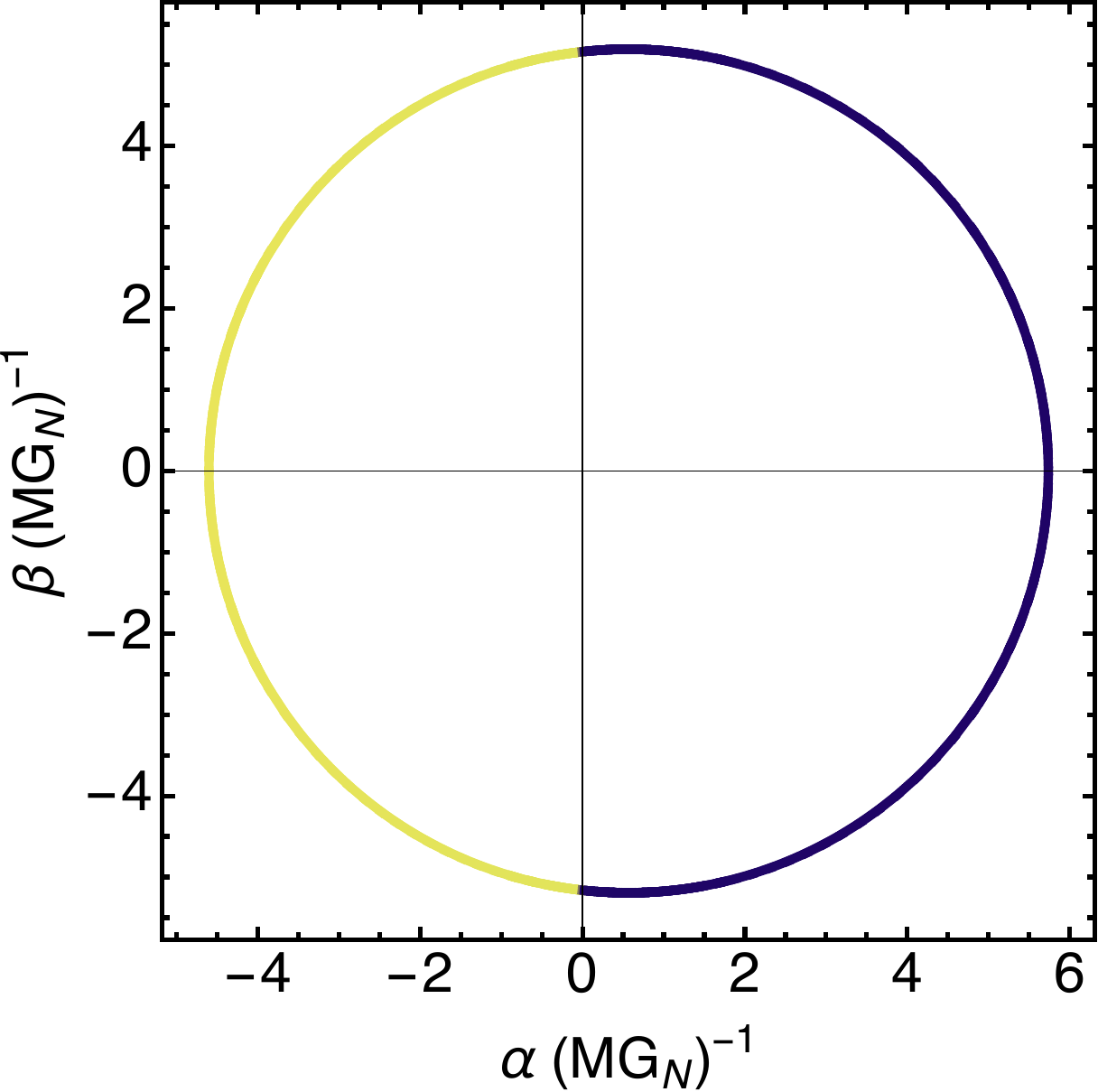} \hspace{0.07em}
\includegraphics[scale=0.245]{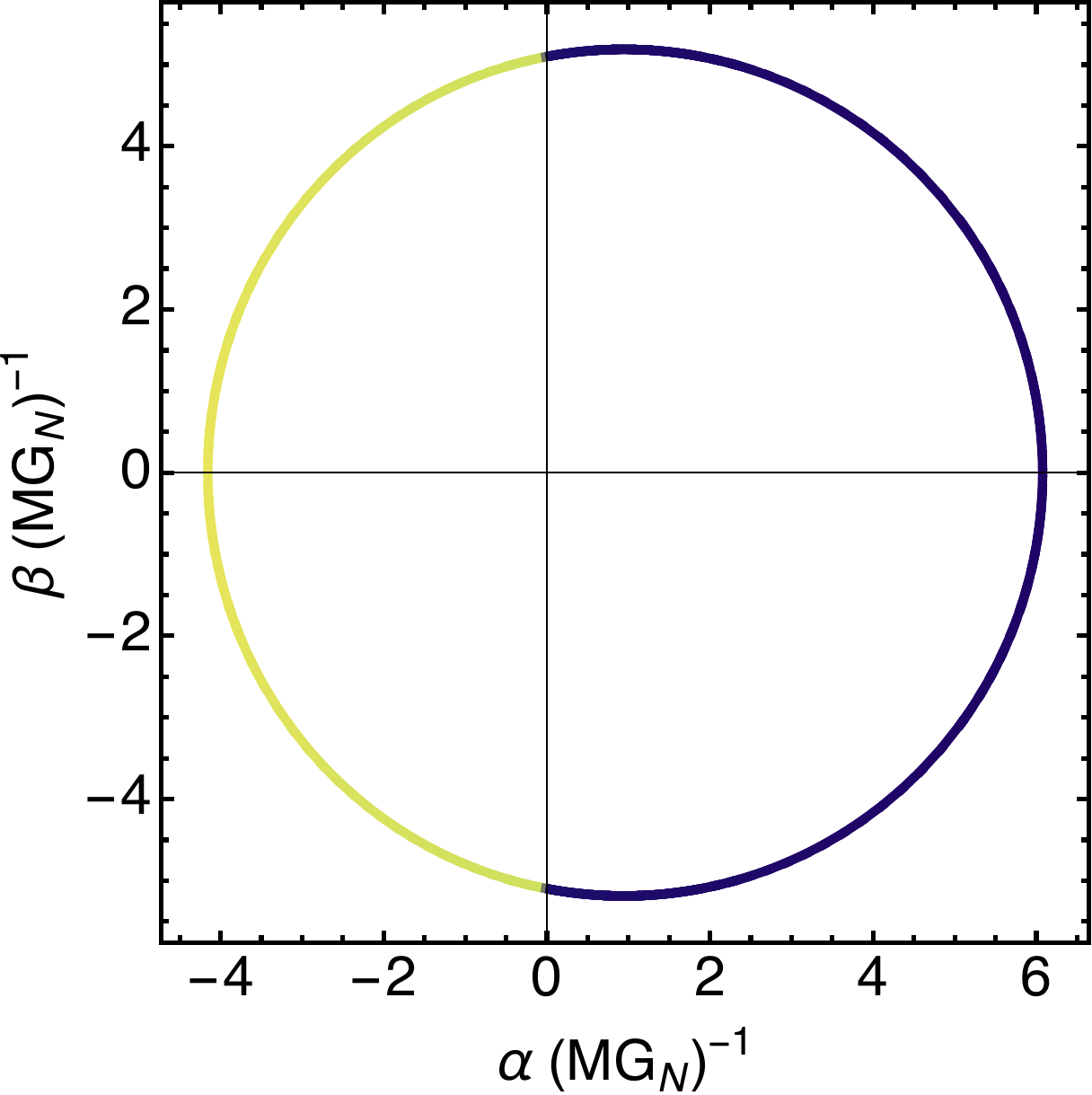} \hspace{0.07em}
\includegraphics[scale=0.245]{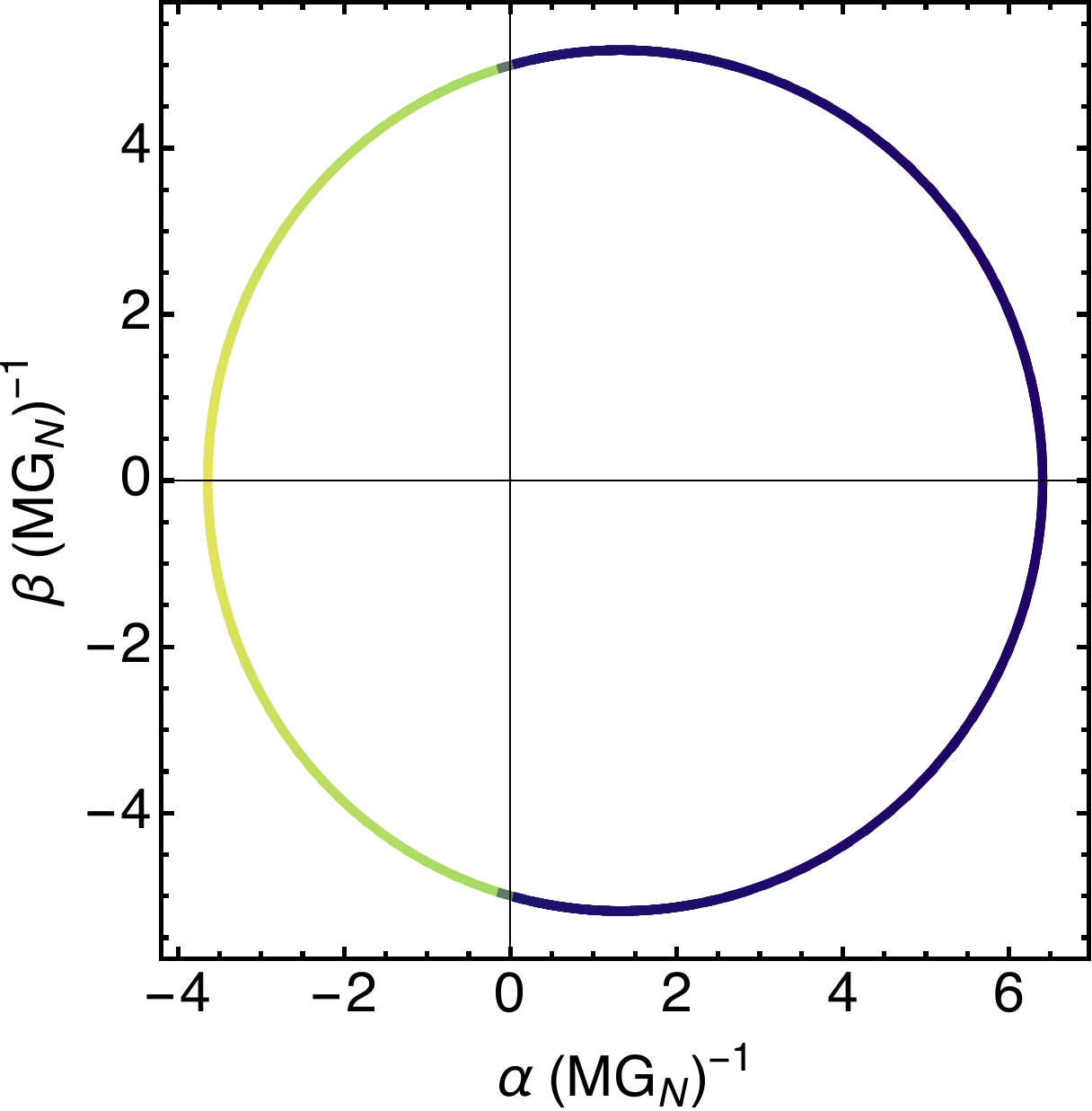} \hspace{0.07em}
\includegraphics[scale=0.245]{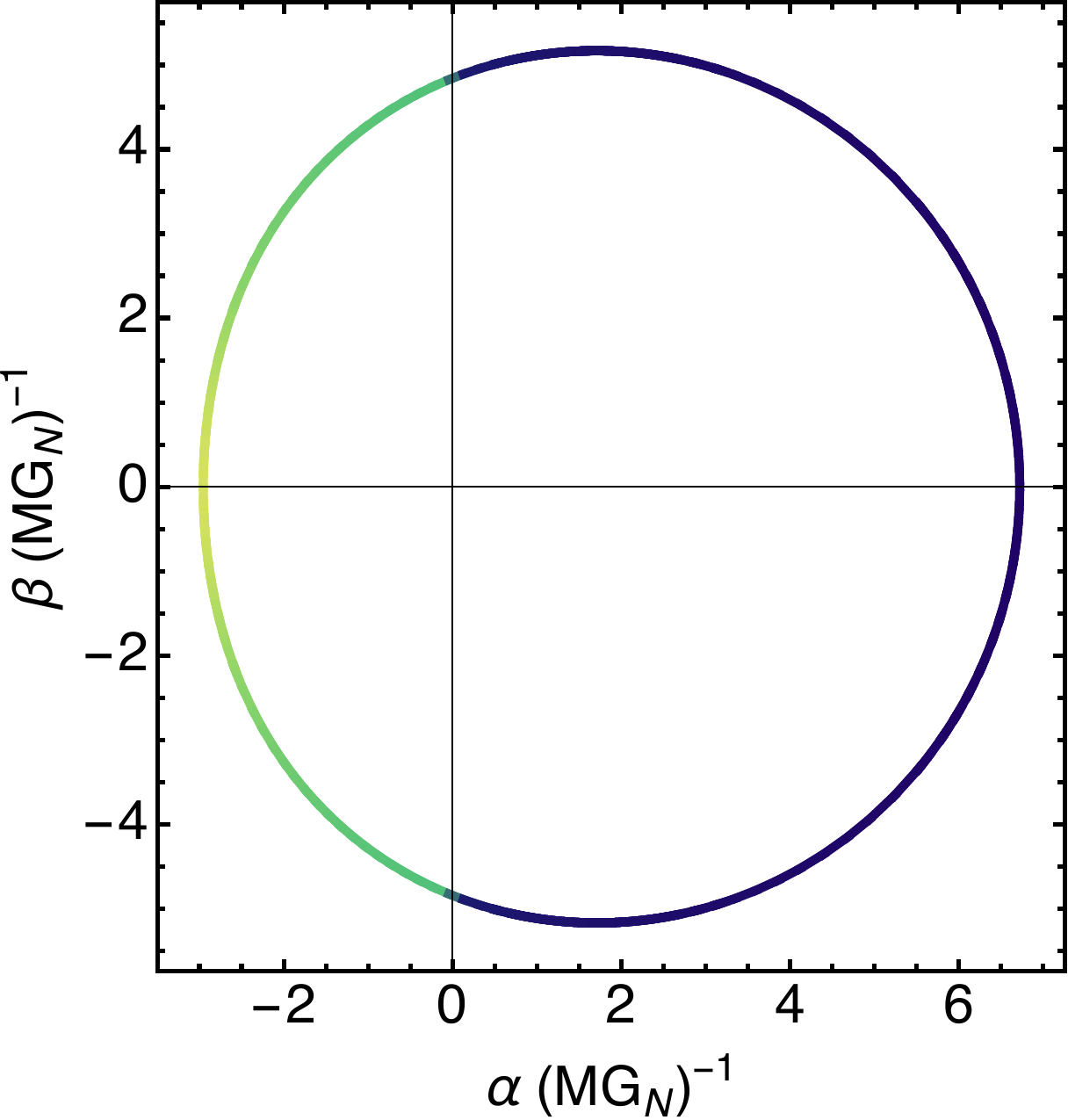}

\caption{\small The relative polarizations of linear polarized light $\Delta \Phi_2$ from the $n^{\rm th}$ and $(n+2)^{\rm th}$ subrings of the photon ring of a black hole that is pierced by a cosmic string. All plots are given for a cosmic string that coincides with the black hole spin axis and has an orientation which corresponds to the $+$ sign in (\ref{polrot}). Plots are provided for various black hole spin parameters $a$ for the given observer inclination $\theta_o = 70^{\circ}$. From left to right: $a (MG_N)^{-1}= 0.1$, $a(MG_N)^{-1}=0.3$, $a(MG_N)^{-1}=0.5$, $a(MG_N)^{-1}=0.7$, $a(MG_N)^{-1}=0.9$. For each chosen $a$, there are three plots: The first plot (top) shows $C^{-1} \Delta \Phi_2$ as a function of the radius $r$ at which photons can circle the black hole. The plotted range of $r$ is $r_-(a) \leq r \leq r_+(a)$. The constant $C$ is defined in (\ref{coupling}), $r_-$ and $r_+$ are given in (\ref{radiusrange}). As discussed in the main text, photons from orbits of a given $r$ that reach a distant observer generally arrive at two different angular coordinates $\varphi_\rho$ on the screen of the observer. The second (middle) and third (bottom) plots show $C^{-1}\Delta \Phi_2$ as a function of the angular coordinate $\varphi_\rho$. In the second plots $C^{-1} \Delta \Phi_2$ is plotted against $\varphi_\rho$ in the range $0 \leq \varphi_\rho \leq \pi$, omitting the intervall $\pi \leq \varphi_\rho \leq 2\pi$ that is symmetric to $0 \leq \varphi_\rho \leq \pi$ since it is comprised of photons from the same $r$ (as just mentioned). The third plots show the shape of the photon ring indicating different values of $C^{-1} \Delta \Phi_2$ by different colors (the Cartesian coordinates $\alpha$ and $\beta$ on the screen are related to $\varphi_\rho$ via (\ref{polar})). For convenience, these colors are also used in the first and second plots. The color profile is indicated in the first plots on the right. Although difficult to see in these plots (because $\theta_o = 70^{\circ}$ is too close to $\theta_o = 90^{\circ}$), not all values of $C^{-1} \Delta \Phi_2$ that appear in the first plot show up in the second and third plots. (This is because observers with $\theta_o \neq 90^{\circ}$ can only probe photons that come from a subset of orbits of the radii $r$ that are plotted in the first plot.) This effect becomes more apparent in Figure \ref{fig:2} in the plots with $\theta_o \ll 90^{\circ}$. Note that, for each chosen $a$, we have normalized the color profile differently, as can be seen on the right in the first plots. Therefore, the same color in plots for different $a$ not necessarily represent the same value of $C^{-1} \Delta \Phi_2$.}
\label{fig:1}
\end{figure}

\begin{figure}

\includegraphics[scale=0.242]{plot17007} \hspace{0.07em}
\includegraphics[scale=0.242]{plot17007} \hspace{0.07em}
\includegraphics[scale=0.242]{plot17007} \hspace{0.07em}
\includegraphics[scale=0.242]{plot17007} \hspace{0.07em}
\includegraphics[scale=0.242]{plot17007}
\newline

\includegraphics[scale=0.241]{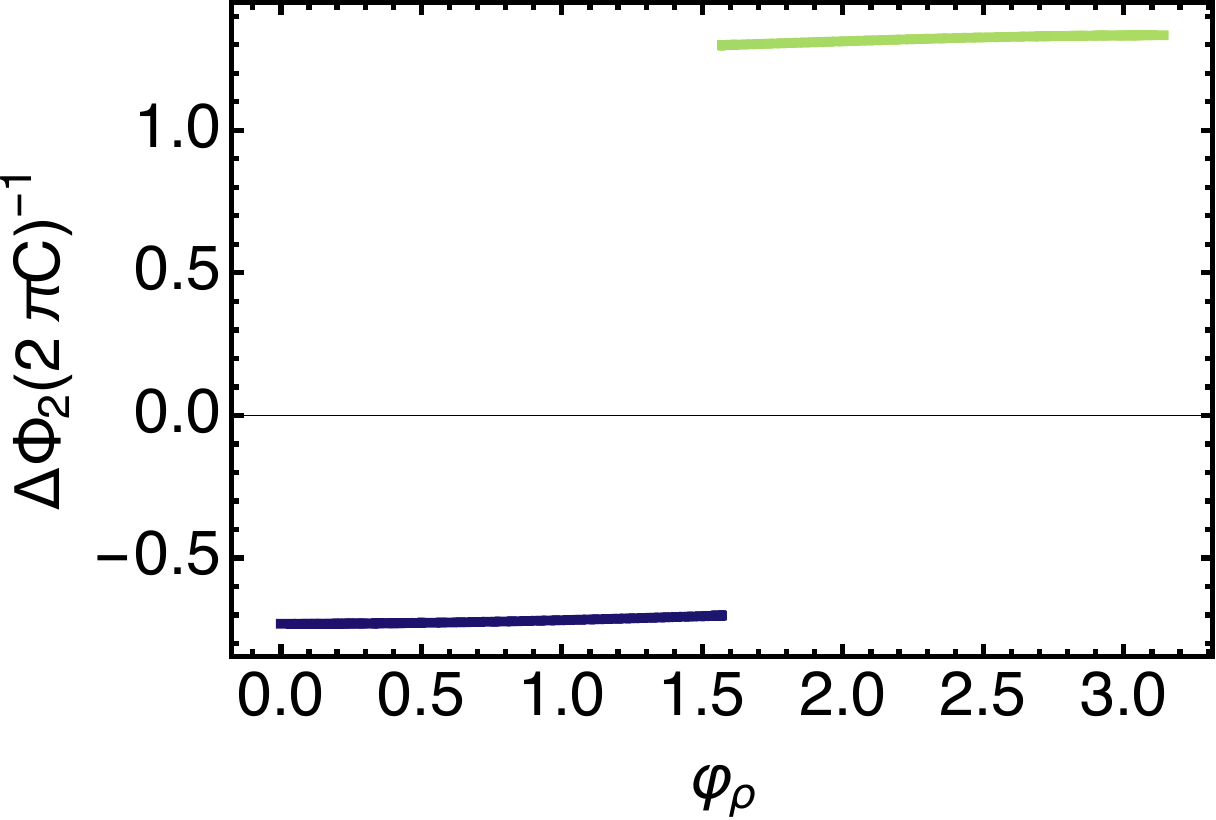} \hspace{0.05em}
\includegraphics[scale=0.241]{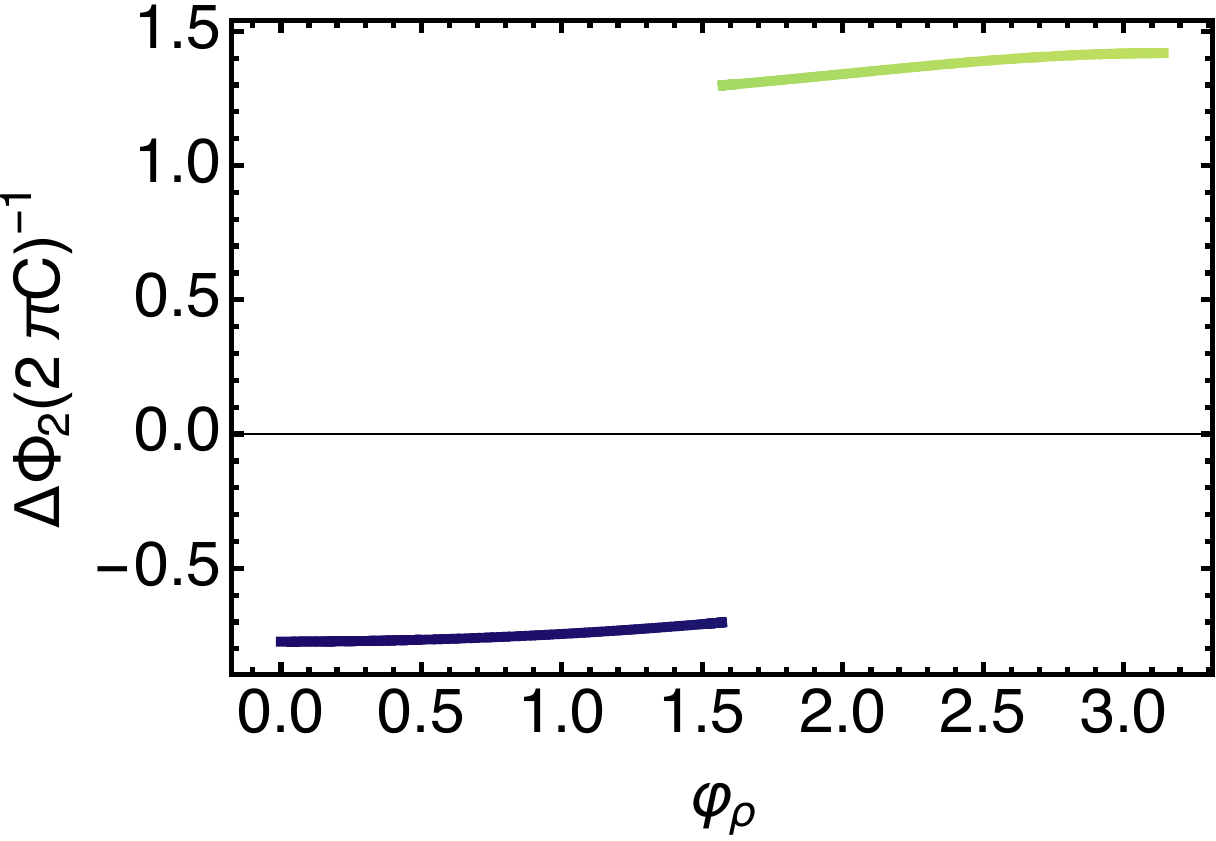} \hspace{0.05em}
\includegraphics[scale=0.241]{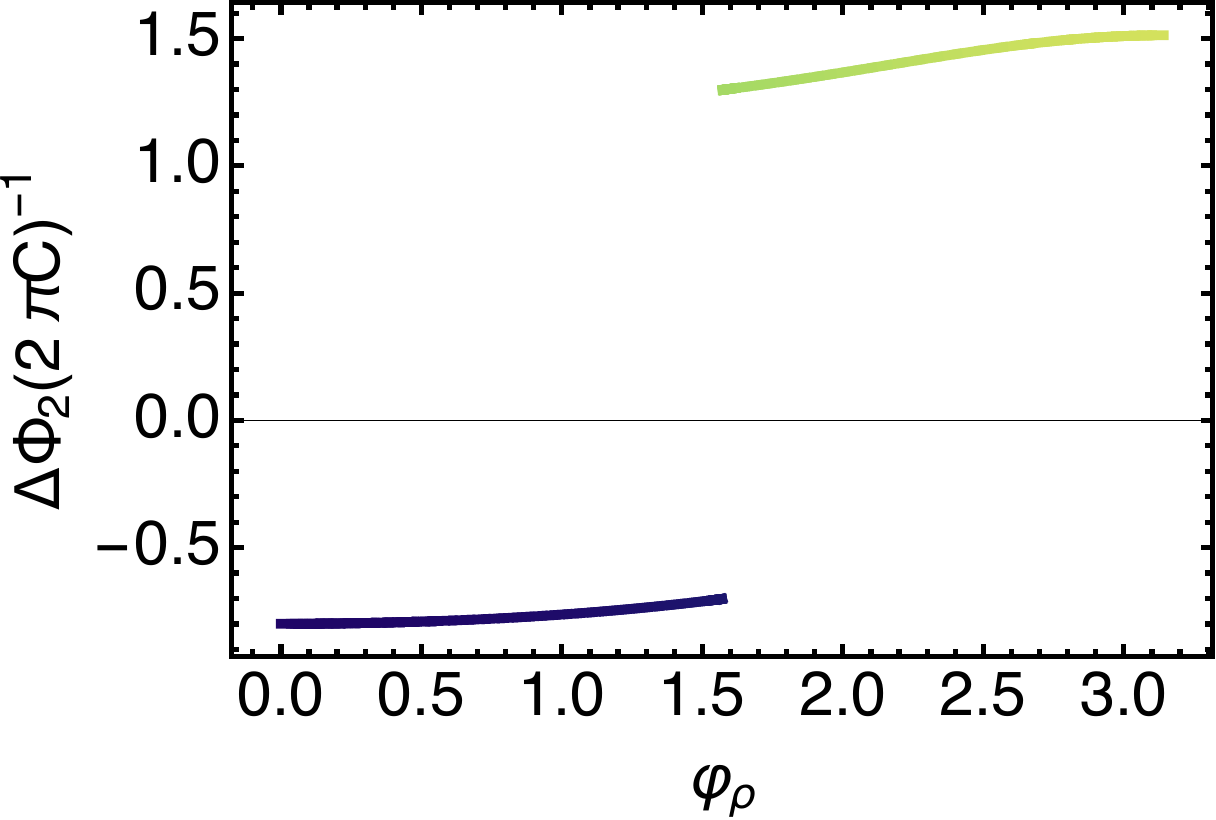} \hspace{0.05em}
\includegraphics[scale=0.241]{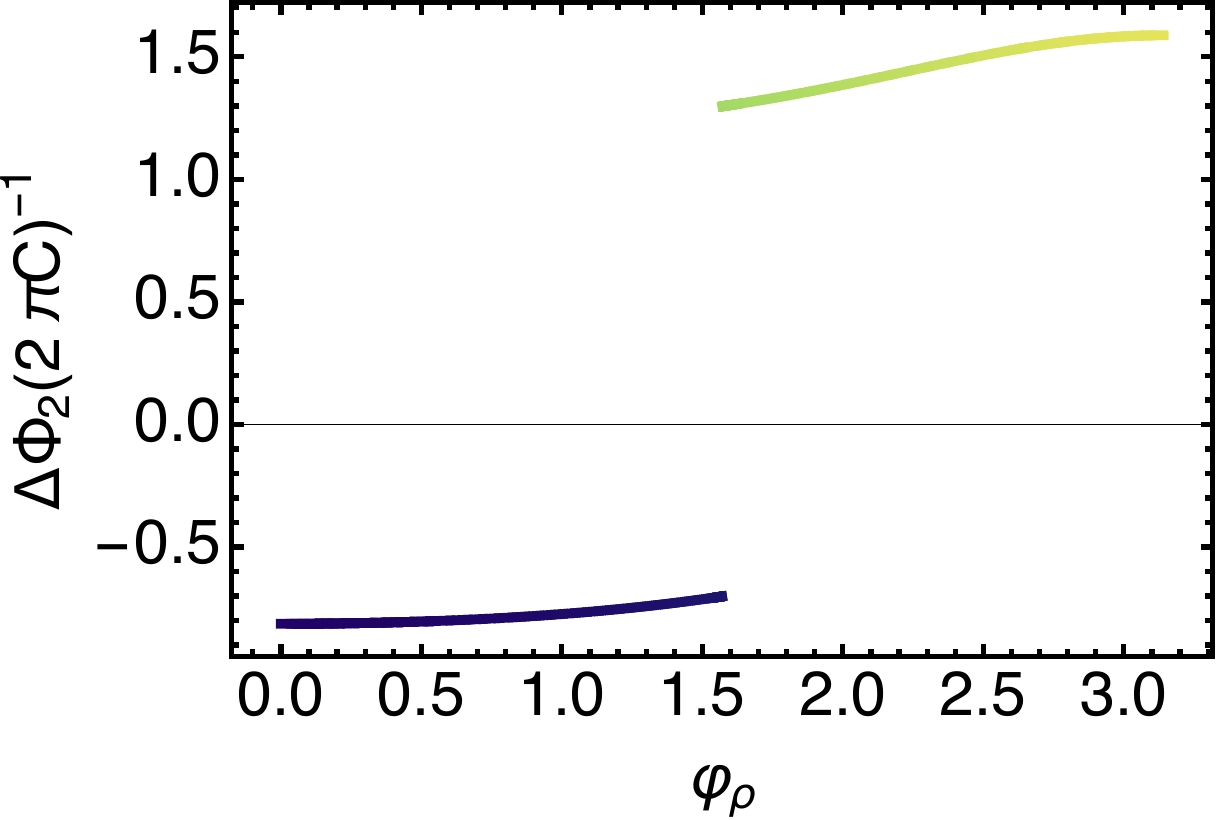} \hspace{0.05em}
\includegraphics[scale=0.241]{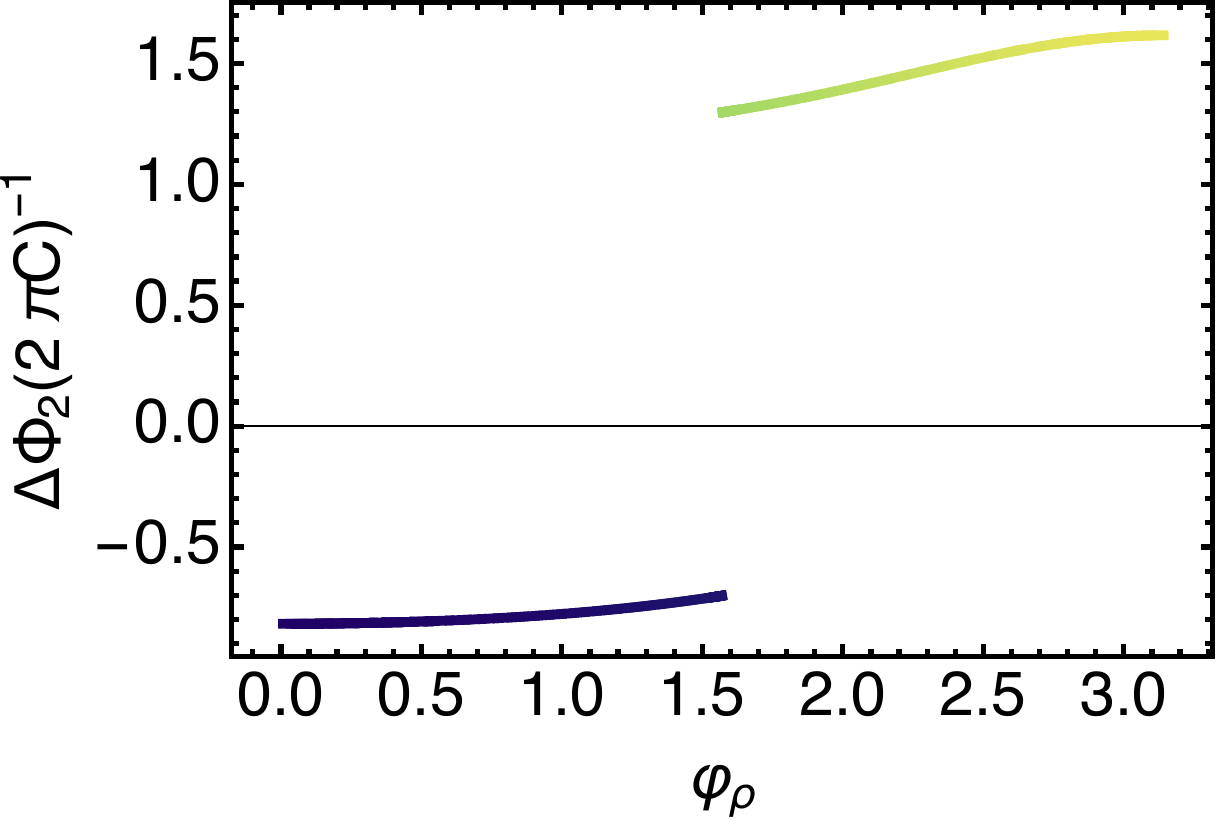}
\newline

\includegraphics[scale=0.245]{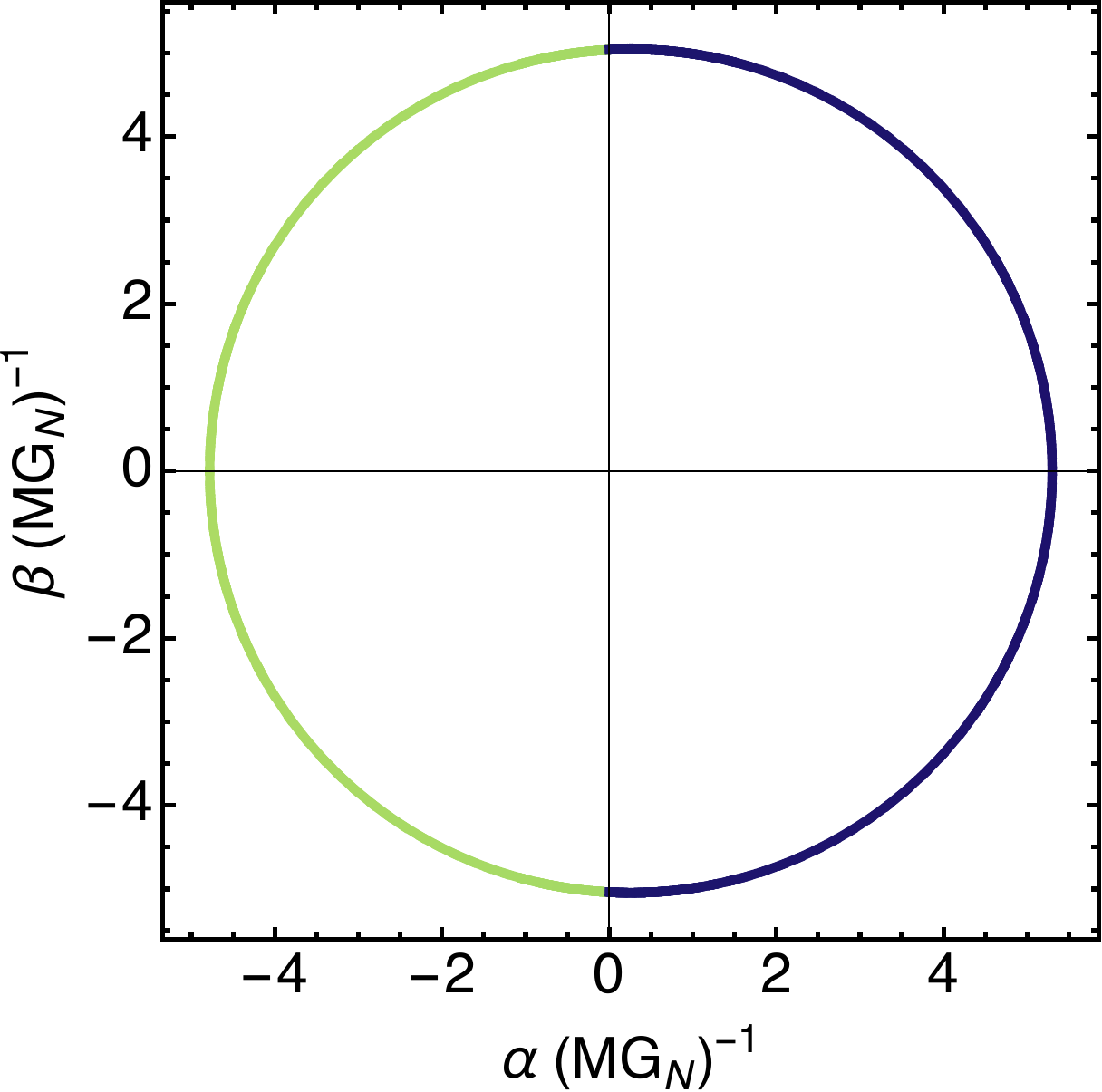} \hspace{0.07em}
\includegraphics[scale=0.245]{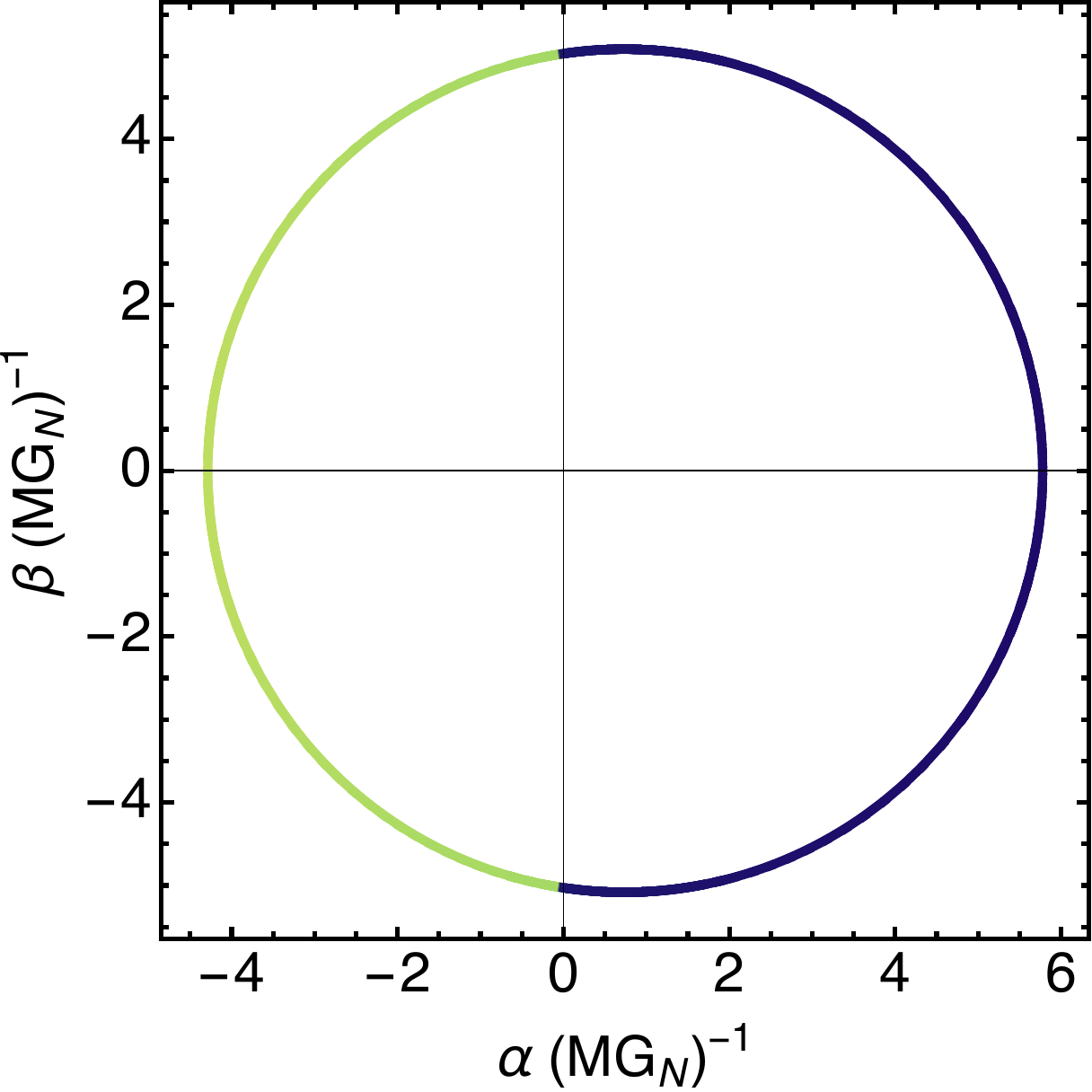} \hspace{0.07em}
\includegraphics[scale=0.245]{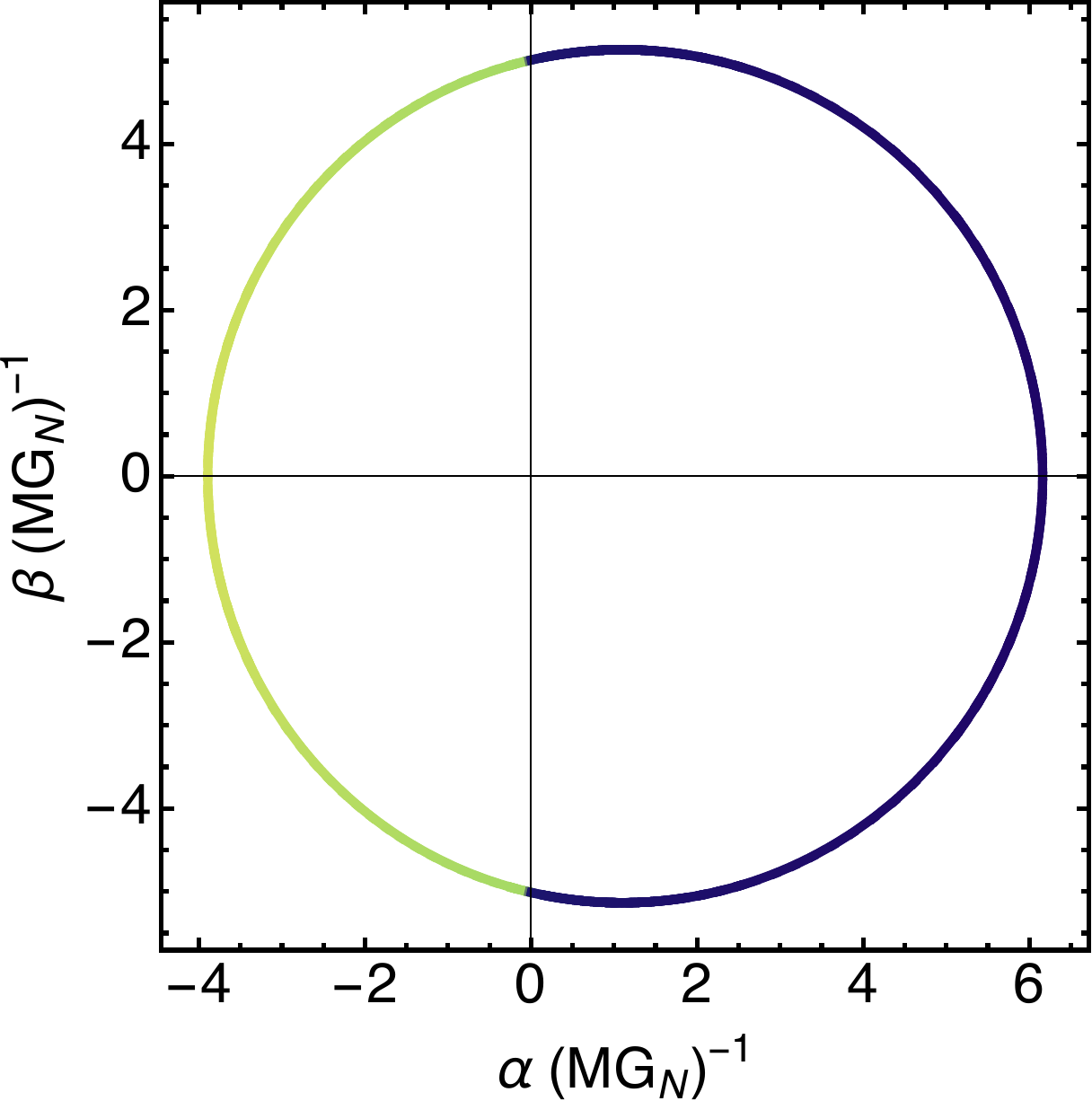} \hspace{0.07em}
\includegraphics[scale=0.245]{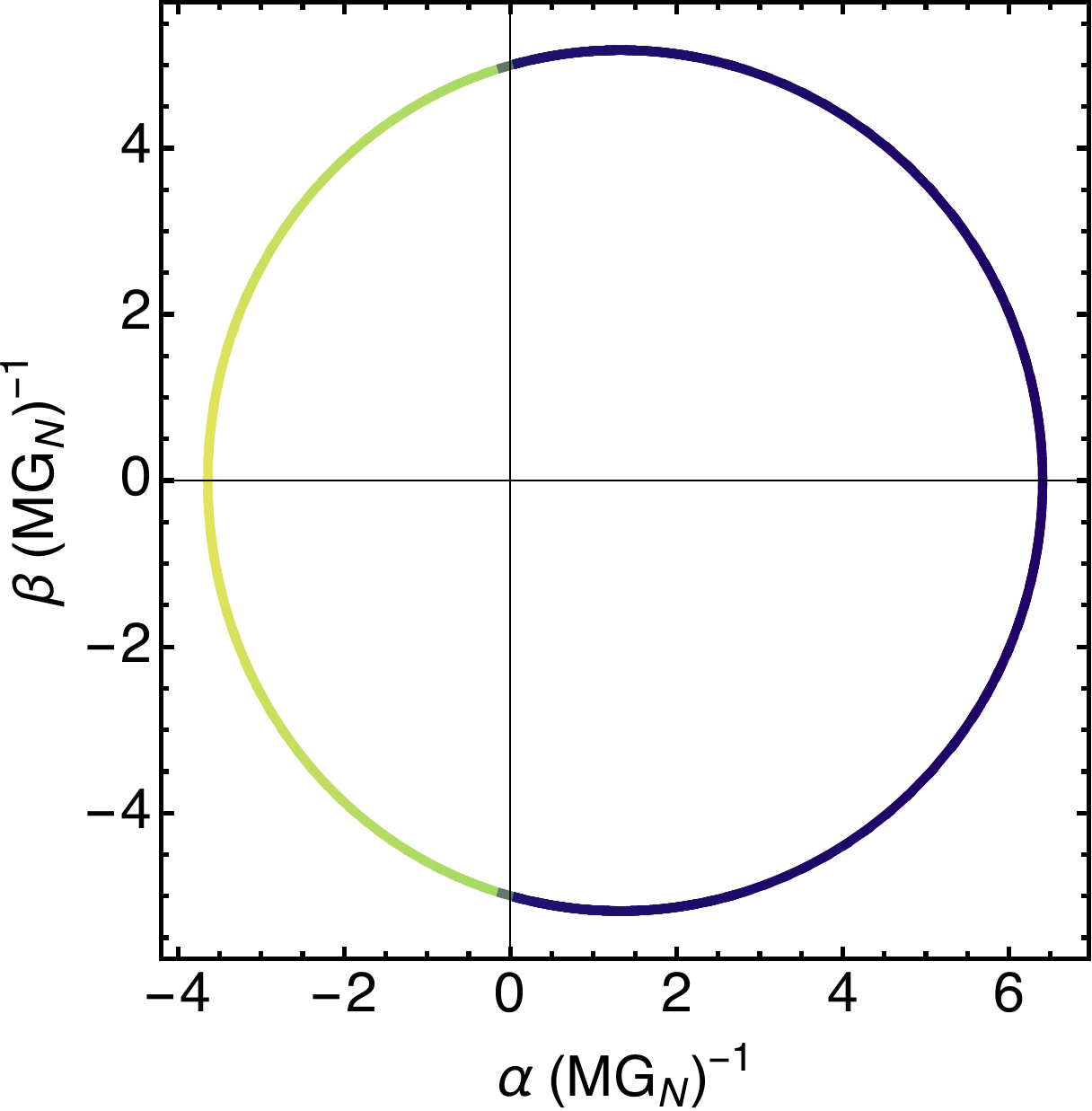} \hspace{0.07em}
\includegraphics[scale=0.245]{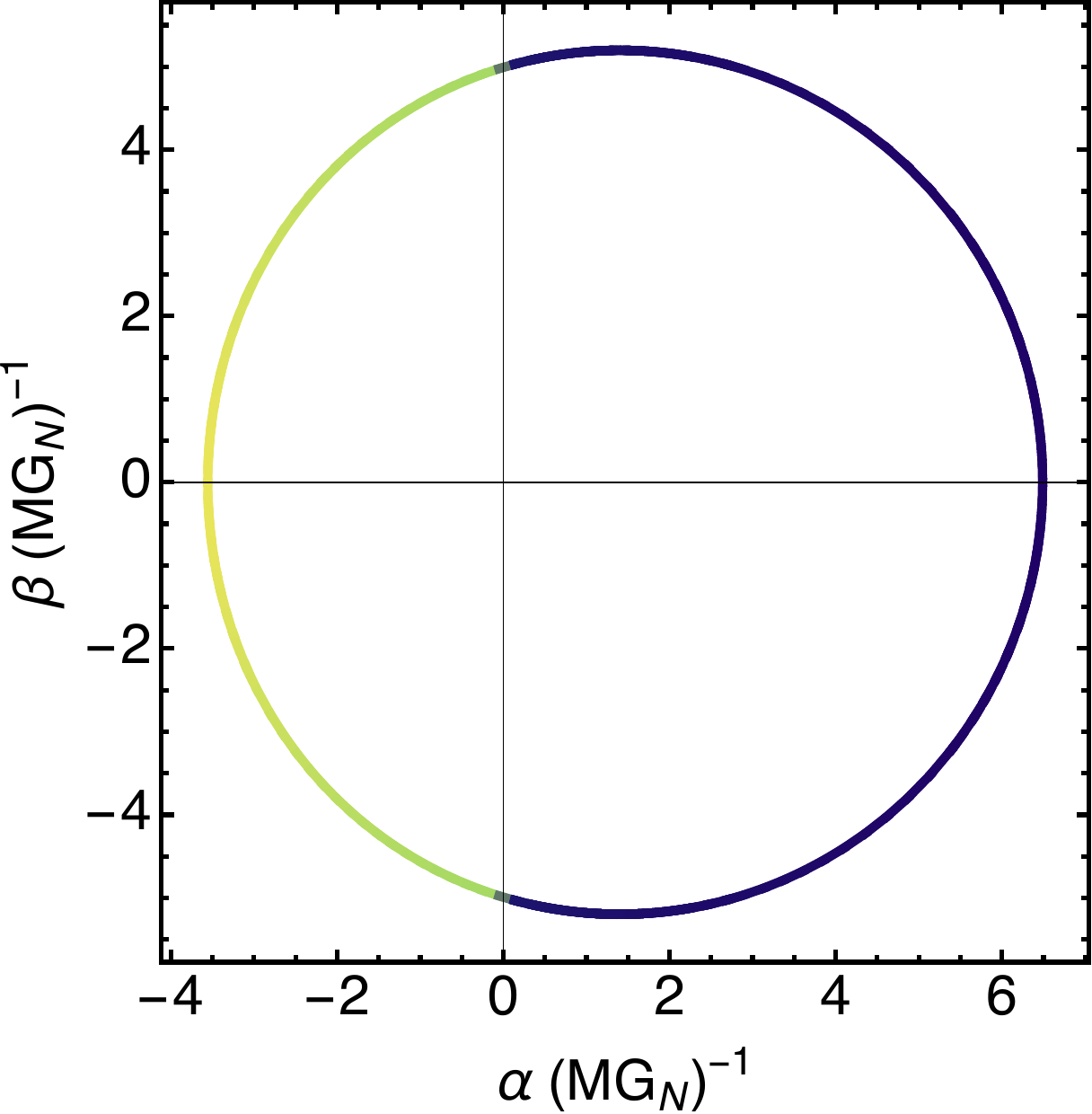}

\caption{\small The same kinds of plots as in Figure (\ref{fig:1}), now for a fixed black hole spin parameter $a (MG_N)^{-1} = 0.7$ and for various observer inclinations $\theta_o$. From left to right: $\theta_o = 10^{\circ}$, $\theta_o=30^{\circ}$, $\theta_o=50^{\circ}$, $\theta_o=70^{\circ}$, $\theta_o=90^{\circ}$. See caption of Figure (\ref{fig:1}) for explanations.}
\label{fig:2}
\end{figure}

Two photons that are located at the same angular coordinate $\varphi_\rho$ on the screen of a distant observer but are in different (but ``closeby") even/odd subrings have moved on orbits that are exponentially close to each other (\ref{delta}) and have the same parallel transported polarizations \cite{Himwich:2020msm}. In case a cosmic string of the form (\ref{strings}) pierces the black hole, the polarization of a linearly polarized photon in the $n^{\rm th}$ subring at the angular coordinate $\varphi_\rho$ differs however from the polarization of the photon in the $(n+2)^{\rm th}$ subring at the same $\varphi_\rho$ by $\Delta \Phi_2 (\varphi_\rho)$ (as discussed in section \ref{section:section3}). Knowing the polarizations of photons at the same $\varphi_\rho$ in the $n^{\rm th}$ and $(n+2)^{\rm th}$ subrings allows us to determine $\Delta \Phi_2$ via (\ref{deltaphi}):
\begin{equation}
{\rm sin} \left(\Delta \Phi_2\right) = \epsilon^{n+2}_\beta \epsilon^{n}_\alpha - \epsilon^{n}_\beta \epsilon^{n+2}_\alpha \, .
\label{deltapol}
\end{equation}
From any given angular coordinate $\varphi_\rho$ on the photon ring (that corresponds to a distinct radius $r$ of photon orbits around the black hole), the anomaly coefficient $\mathcal{A}$ (\ref{anomaly}) can be obtained via (\ref{deltapol}) with (\ref{polrot}), (\ref{delphi}) and (\ref{anomaly}) when the black hole parameters are known. In Figures \ref{fig:1} and \ref{fig:2} we plot the relative polarizations $C^{-1} \Delta \Phi_2$ (\ref{polrot}) between photons from the $n^{\rm th}$ and $(n+2)^{\rm th}$ subrings of the photon ring for several black hole angular momenta and observer inclinations $\theta_o$. Our plots show $C^{-1} \Delta \Phi_2$ for the string orientation that corresponds to the $+$ sign in (\ref{polrot}). For each choice of angular momentum and observer inclination we provide three plots: First, we plot $C^{-1} \Delta \Phi_2$ as a function of $r$ in the range $r_- \leq r \leq r_+$. This shows the values of the relative polarizations for photons that circle the black hole at a radius $r$. Second, we plot $C^{-1} \Delta \Phi_2$ as a function of the angular coordinate $\varphi_\rho$ on the observer's screen. As already mentioned in section \ref{section:section3}, for $\theta_o \neq \frac{\pi}{2}$ an observer can only see photons from a subset of this $r$-range on the screen. Third, we provide the same plot again but now present the shape of the photon ring and indicate different $C^{-1} \Delta \Phi_2$ by different colors. (For convenience, the same colors are also used in the first and second Figures.)
\newline

As one can see from the second/third plots, an obvious signature is given by the discontinuity of the values of $C^{-1} \Delta \Phi_2$ in the photon ring at $\alpha = 0$. The underlying reason for this signature is that the corresponding photons have circled the black hole in opposite directions. A less obvious signature is given by the variations of $C^{-1} \Delta \Phi_2$ as a function of $\varphi_\rho$ for photons that have circled the black hole in the same direction. These variations are present because photons on different orbits are dragged by the rotating black holes in a different way and therefore obtain different polarization rotations when they circle the black hole that is pierced by a cosmic string.

\subsection{Time dependent signatures}

A complementary approach to study properties of the black hole photon ring is to look for time dependent signatures (see e.g.\cite{Moriyama:2015zfa, Moriyama:2019mhz, Hadar:2020fda, Chesler:2020gtw, Wong:2020ziu} for some works in this direction). Such signatures can come from time dependent compact light-emitting regions close to the black hole. This can for example be a gas cloud that is falling into the black hole (see e.g. \cite{Moriyama:2019mhz}) or matter that is rotating around the black hole (see e.g. \cite{Wong:2020ziu}). In order to study time dependent signatures in a simple setup, we shall consider the toy-example of bright point-like spots that are located in the photon shell and that emit linearly polarized light in all directions (for some recent studies along these lines without a string piercing the black hole see \cite{Wong:2020ziu}).
\newline

\begin{figure}

\includegraphics[scale=0.31]{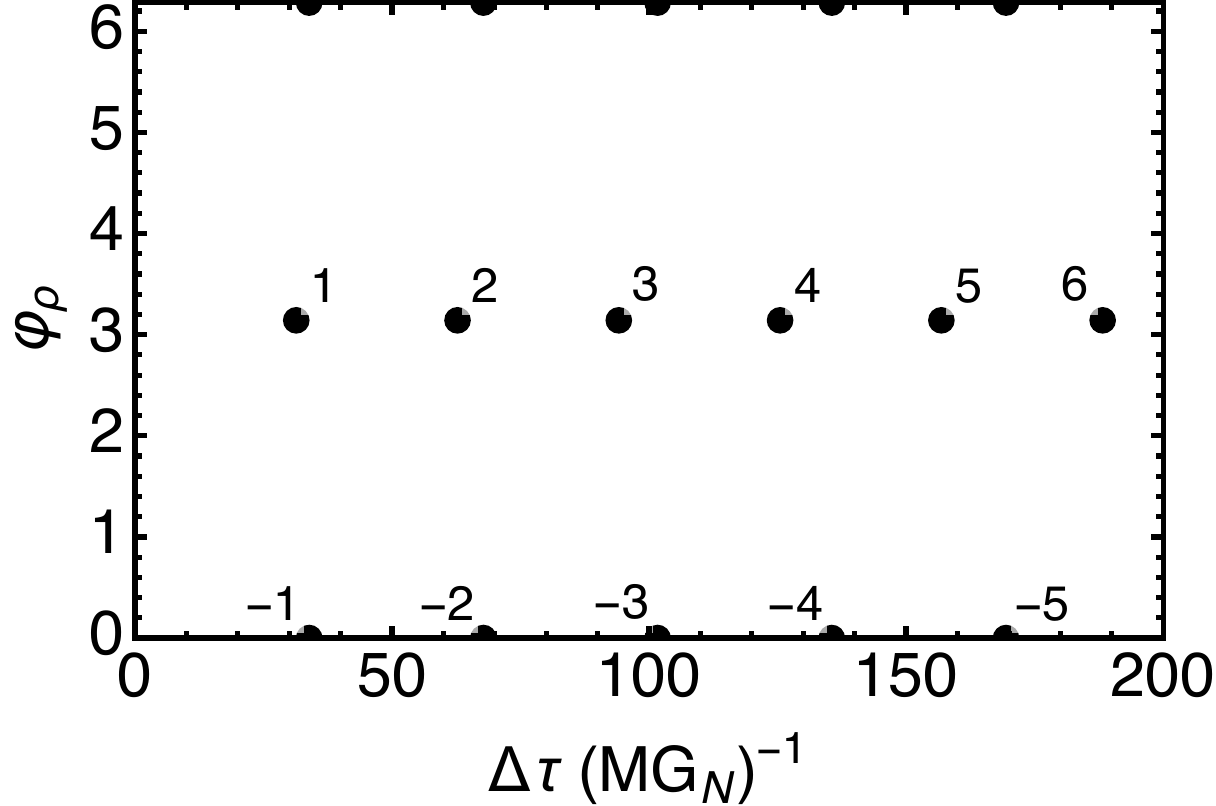} \hspace{0.07em}
\includegraphics[scale=0.31]{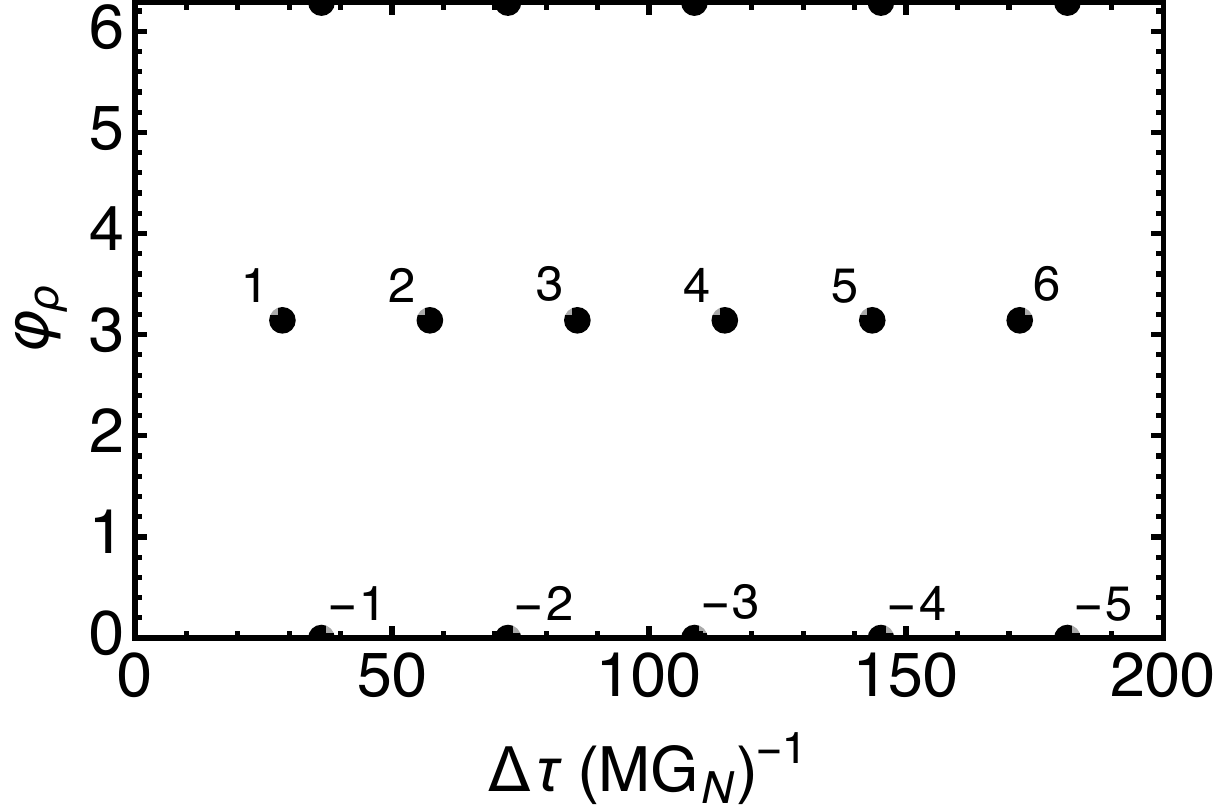} \hspace{0.07em}
\includegraphics[scale=0.31]{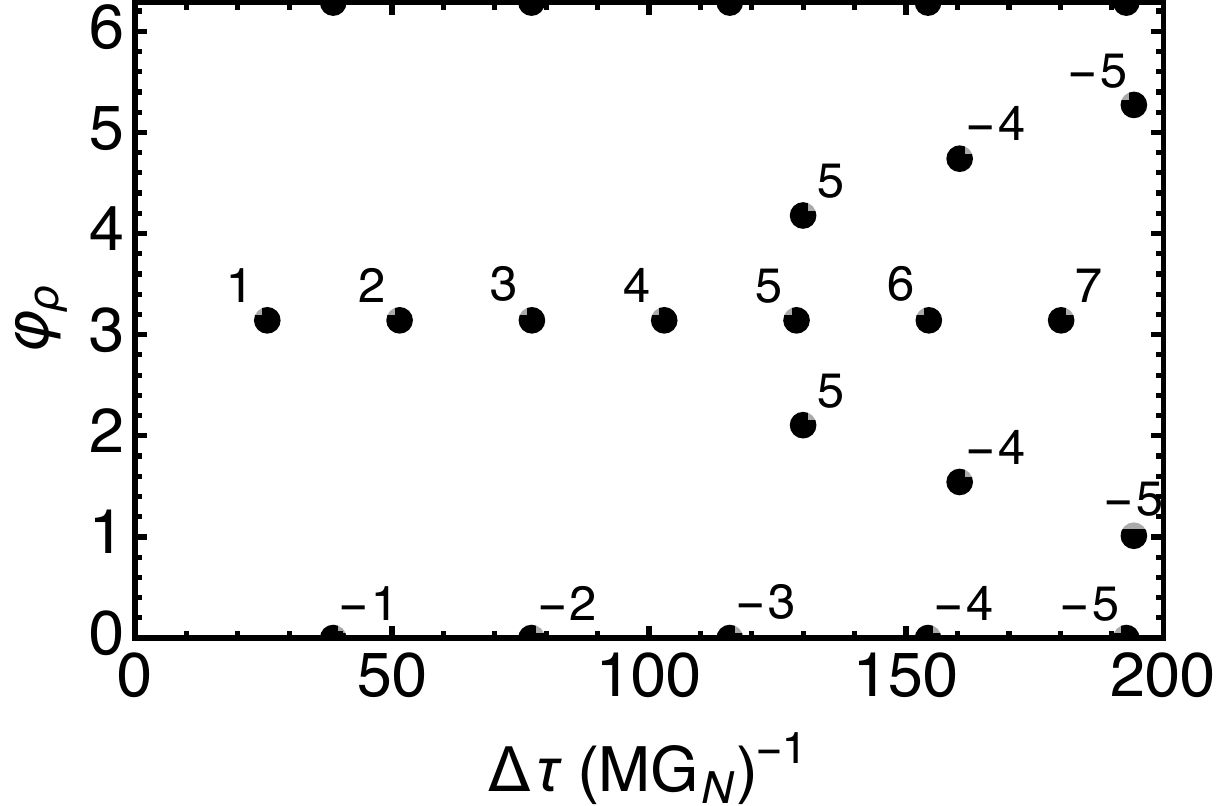} \hspace{0.07em}
\includegraphics[scale=0.31]{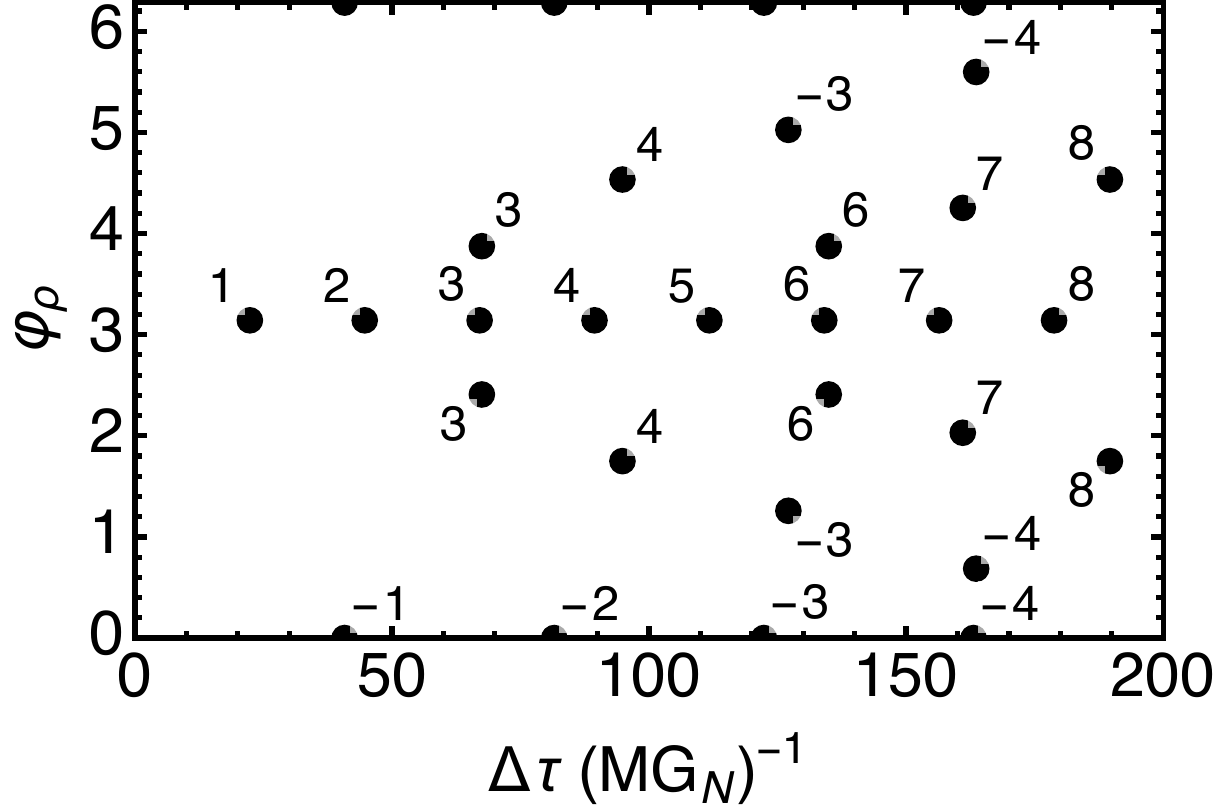} \hspace{0.07em}

\caption{\small Possible echos generated by bright point-like spots that are located in the photon shell of a black hole that is pierced by a cosmic string. The plots are given for a cosmic string that coincides with the black hole spin axis and has an orientation that corresponds to the $+$ sign in (\ref{polrot2}). In these plots the observer inclination $\theta_o$ is fixed to be $\theta_o = 90^{\circ}$ and the black hole spin parameter $a$ varies. From left to right: $a(MG_N)^{-1}= 0.1$, $a(MG_N)^{-1}=0.3$, $a(MG_N)^{-1}=0.5$, $a(MG_N)^{-1}=0.7$. The angular location $\varphi_\rho$ of the echos on the screen of an observer is plotted against the time interval $\Delta \tau$ after which the echos appear. The labels of the dots are the generated phases $\Delta \Phi \left(2\pi C \right)^{-1}$ that rotate the polarization direction of linear polarized photons of that particular echo. These are all integer numbers because the echos come from photons of the same point-like source that circle the black hole on closed orbits which implies that $\Delta \varphi$ is an integer-multiple of $2 \pi$ (see (\ref{phibar}), (\ref{fractionalnumber})).}

\label{fig:3}

\end{figure}

\begin{figure}

\includegraphics[scale=0.31]{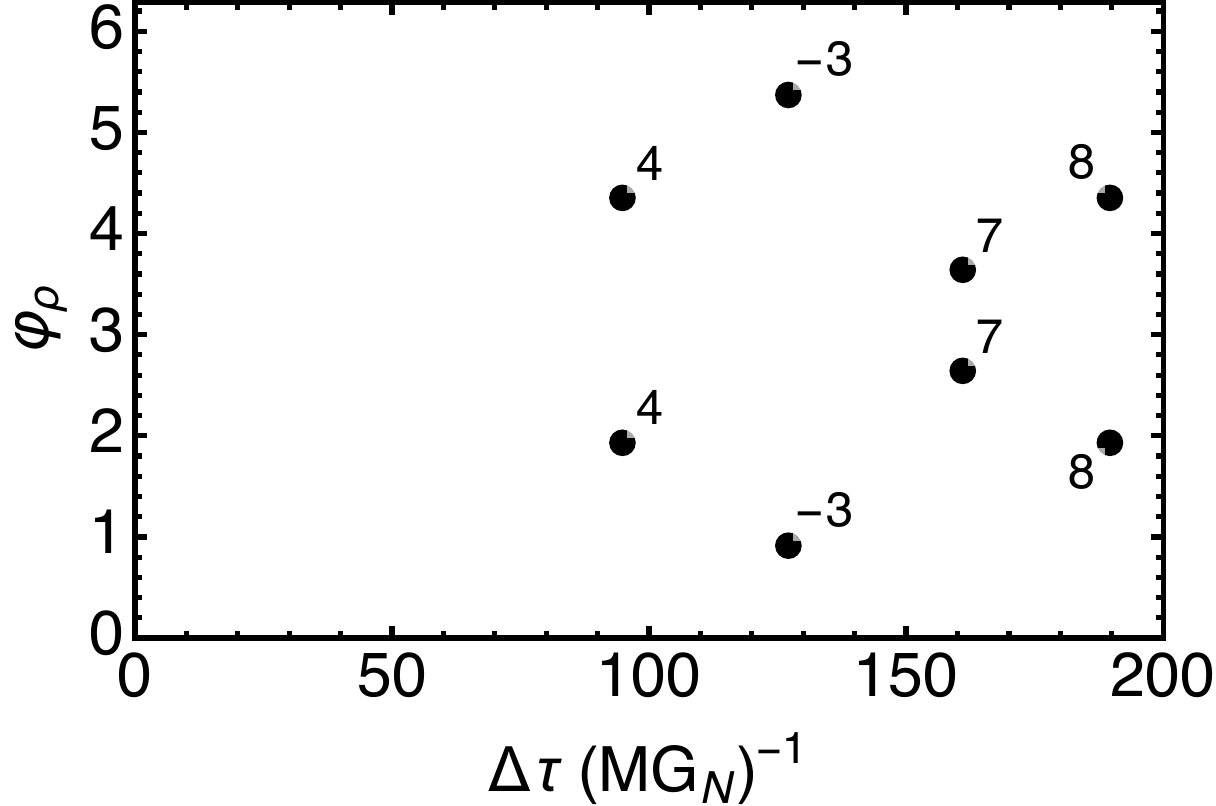} \hspace{0.07em}
\includegraphics[scale=0.31]{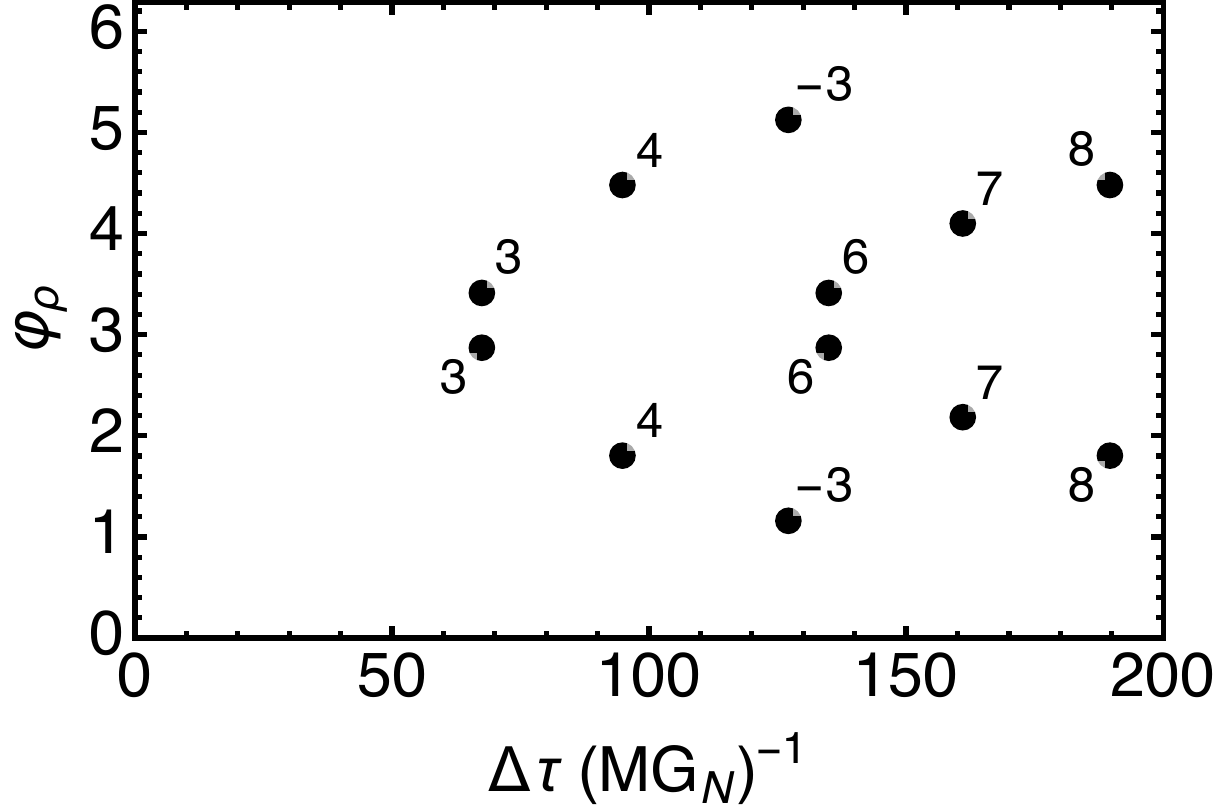} \hspace{0.07em}
\includegraphics[scale=0.31]{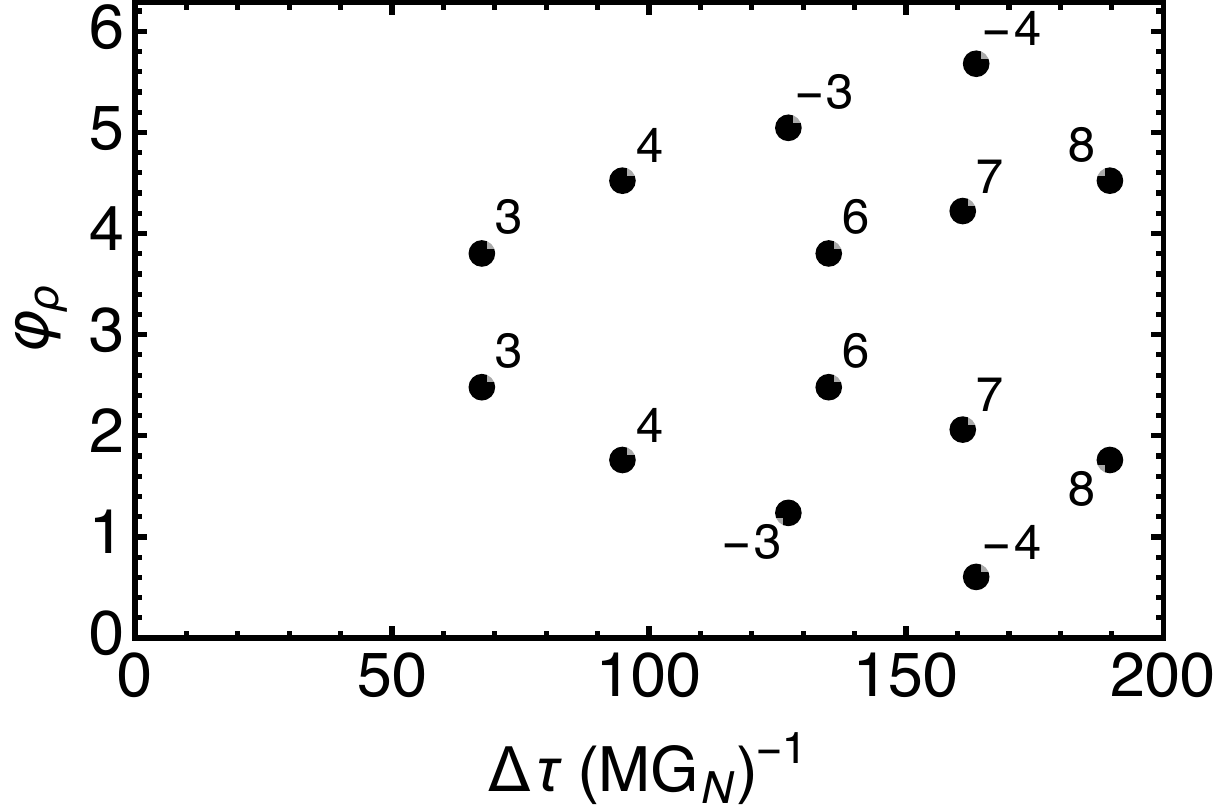} \hspace{0.07em}
\includegraphics[scale=0.31]{plot0790} \hspace{0.07em}

\caption{\small Possible echos generated by bright point-like spots that are located in the photon shell of a black hole that is pierced by a cosmic string. The plots are given for a cosmic string that coincides with the black hole spin axis and has an orientation that corresponds to the $+$ sign in (\ref{polrot2}). In these plots the black hole spin parameter $a$ is fixed to be $a(MG_N)^{-1}=0.7$ and the observer inclination $\theta_o$ varies. From left to right: $\theta_o = 30^{\circ}$, $\theta_o = 50^{\circ}$, $\theta_o = 70^{\circ}$, $\theta_o = 90^{\circ}$. The angular location $\varphi_\rho$ of the echos on the screen of an observer is plotted against the time interval $\Delta \tau$ after which the echos appear. The labels of the dots are the generated phases $\Delta \Phi \left(2\pi C \right)^{-1}$ that rotate the polarization direction of linear polarized photons of that particular echo. These are all integer numbers because the echos come from photons of the same point-like source that circle the black hole on closed orbits which implies that $\Delta \varphi$ is an integer-multiple of $2 \pi$ (see (\ref{phibar}), (\ref{fractionalnumber})).}

\label{fig:4}
\end{figure}

Photons that are emitted by a bright spot that is located at a radius $r$ in the photon shell move on null geodesics at radius $r$ around the black hole. When their orbits slightly deviate from the bound orbits, they leave the photon shell and, depending on the location of the distant observer, can reach the screen of the observer. In case the photon orbits are closed (\ref{fractionalnumber}), the geodesics return to the position of the bright spot (in the simplified case when the spot is not moving). In such situations light from a spot that is emitted at a particular point of time can reach the observer on multiple different paths: After emitted by the point-like spot isotropically in all directions, some photons travel ``directly" to the observer whereas other photons circle once (or several times) around the black hole on the bound orbit before they pass the spot again and reach the observer on the same path as the light that has reached the observer directly (see also the discussion in the last paragraph in section \ref{section:lightring}). In this way an observer can see a signal and successive echos of light from the same source at a given angular coordinate $\varphi_\rho$. The successive echoes typically have smaller and smaller flux \cite{Wong:2020ziu}. In case of closed orbits that do not lie entirely in the equatorial plane, the echos appear after time intervals of
\begin{equation}
\Delta \tau = q \tau \, ,
\end{equation}
where $q$ and $\tau$ are given by (\ref{fractionalnumber}) and (\ref{timeshift}). During this time photons pass an azimuth angle of
\begin{equation}
\Delta \varphi = q \delta \, ,
\label{phibar}
\end{equation}
with $\delta$ as given in (\ref{delphi}). In the case of closed orbits that lie in the equatorial plane (these are orbits located at $r_-$ and $r_+$ in the photon shell) $\Delta \tau = \tau \frac{2 \pi}{q \delta}$ and $\Delta \varphi = 2 \pi$. For linear polarized photons, a phase shift of
\begin{equation}
\Delta \Phi = \pm C \Delta \varphi
\label{polrot2}
\end{equation}
is generated (see (\ref{polrot})). Here $\pm$ is set by the orientation of the cosmic string. Each bound orbit that exists (in total infinitely many) can give rise to such echos if a bright spot (or another light source) is located in the photon shell at the same radius at which photons in that orbit circle the black hole. Only the echos from orbits with short enough path length such that the flux is still big enough to be seen can however be observable. In the presence of a cosmic string piercing the black hole, the polarization direction of linear polarized light is different for different echos: The polarization direction of linear polarized light from the $k^{\rm th}$ echo (that is light that made $k$ orbits around the black hole) is rotated by (\ref{polrot2}). In Figures \ref{fig:3} and \ref{fig:4} we plot the brightest echos and the corresponding polarization directions for several black hole spin parameters $a$ and observer inclinations $\theta_o$.\footnote{Extended sources instead of point-like spots can give rise to more intricate images, see e.g. \cite{Wong:2020ziu} for such a discussion for black holes without cosmic strings.} (Here we plot all possible brightest echos, in practice only the echos that correspond to bound orbits at which a bright spot is located are excited.) As in the case of time-averaged signatures, the observed polarization of light in such echos can be used to determine the anomaly coefficient $\mathcal{A}$ by using (\ref{polrot2}), (\ref{phibar}), (\ref{delphi}) and (\ref{anomaly}).

\section{Conclusion and outlook}
\label{section:section5}

In this work we have considered hyper-light global axion-type cosmic strings that arise in certain particle physics models due to a spontaneously broken global symmetry. We have studied black holes that are pierced by such strings and, in various setups, have investigated polarimetric signatures of the photon rings of pierced black holes. These signatures are generated by two effects: First, by the parallel transport of the polarization vector due to the geodesic motion of the photons. Second, by the change of the polarization direction of linear polarized photons due to the coupling $\phi F \tilde{F}$ between the Goldstone boson $\phi$ (axion) and the photon. When deriving the latter effect, we have worked in the geometric optics approximation. A study of potential higher order effects is left for future work. Polarimetric signatures of different origin, generated for example by plasma interactions, are studied elsewhere in the literature (e.g. \cite{Moscibrodzka:2017gdx, Jimenez-Rosales:2018mpc, Tsunetoe:2020pyz, Ricarte:2020llx, Tsunetoe:2020nws, Jimenez-Rosales:2021ytz}). A further detailed analysis of potential plasma effects which takes into account the existence of the cosmic string that pierces the black hole is also left for future work.
\newline

We discussed the question whether or not we can expect black holes that are pierced by axion-type cosmic strings to be present in the center of galaxies. We have pointed out that there exist several possible formation mechanisms and briefly reviewed one particular example. If existent, it seems however unlikely that these objects are widely spread in the universe because the number of axion-type cosmic strings by itself (if existent at all) is expected to be quite small within one Hubble volume. Though we pointed out that there are speculations in the literature arguing that the black hole in the center of our milky way might be pierced by a cosmic string \cite{Morris_2017}. If correct, this could be a good target for future experiments using very large baseline interferometers with a detector placed in space (or on the moon) \cite{Johnson:2019ljv, Himwich:2020msm}.
\newline

In theories with charged fermions that have Yukawa interactions with the Goldstone boson $\phi$, the value of the coefficient $C$ in the coupling $\frac{C}{2}\frac{\phi}{v} F \tilde{F}$ is set by the coefficient $\mathcal{A}$ of the mixed anomaly of electromagnetism and the (spontaneously broken) global symmetry that gave rise to the cosmic string. As we have pointed out, the polarization rotation of linear polarized photons that circle the black hole does only depend on $C$ and is independent of the vacuum expectation value $v$ of the spontaneously broken symmetry that gave rise to the cosmic string. The detection of the polarimetric signatures that we have studied in this work therefore provides us with a possible way of measuring the anomaly coefficient $\mathcal{A}$. Knowing the value of $\mathcal{A}$ could teach us a lot about the UV theory that is realized in nature and gave rise to the anomaly. A value of $\mathcal{A} = 1$ would give rise to a difference in the polarization directions of linear polarized photons from the $n^{\rm th}$ and $(n+2)^{\rm th}$ subrings of $\mathcal{O} \left(1 \%\right)$. This value becomes bigger when the black hole is pierced by more than one string or when photons from two ``more separated" subrings are taken. In case a black hole is pierced by a long cosmic string, there are different possibilities to measure $\mathcal{A}$, for example by observing linear polarized light from objects that are gravitationally lensed by the galaxy \cite{Agrawal:2019lkr}. In case the black hole is pierced by a (small) string loop, other polarization signatures than the ones in the photon ring are however difficult to imagine.
\newline

A global axion-type cosmic string that pierces a black hole can itself give rise to interesting signatures that are unrelated to the presence of the black hole. In particular, in the case of theories with charged fermions that have Yukawa interactions with $\phi$, the cosmic string becomes superconducting which can lead to many potentially observable signatures. Such signatures have already been studied in the literature (see e.g. \cite{Chudnovsky:1986hc, Gruzinov:2016hqs, Agrawal:2020euj}) and can be considered as complementary to the polarimetric signatures of the black hole photon ring studied here.

\section*{Acknowledgement}
This work was supported by the Deutsche Forschungsgemeinschaft (DFG, German Research Foundation) under the individual grant GU 2052/1-1.


\begin{thebibliography}{9}

\bibitem{Chandrasekhar:1985kt}
S.~Chandrasekhar,
``The mathematical theory of black holes,''
Oxford, UK: Clarendon,
ISBN: 9780198503705.

\bibitem{Teo:2020sey}
E.~Teo,
``Spherical photon orbits around a Kerr black hole,''
Gen. Rel. Grav. \textbf{35}, 1909-1926 (2003)
doi:10.1023/A:1026286607562

\bibitem{Bardeen:1973tla}
J.~M.~Bardeen
``Timelike and null geodesics in the Kerr metric,''
Les Houches Summer School of Theoretical Physics: Black Holes,
1973.

\bibitem{Luminet:1979nyg}
J.~P.~Luminet,
``Image of a spherical black hole with thin accretion disk,''
Astron. Astrophys. \textbf{75}, 228-235 (1979)

\bibitem{Akiyama:2019cqa}
K.~Akiyama \textit{et al.} [Event Horizon Telescope],
``First M87 Event Horizon Telescope Results. I. The Shadow of the Supermassive Black Hole,''
Astrophys. J. Lett. \textbf{875}, L1 (2019)
doi:10.3847/2041-8213/ab0ec7
[arXiv:1906.11238 [astro-ph.GA]].

\bibitem{Akiyama:2019brx}
K.~Akiyama \textit{et al.} [Event Horizon Telescope],
``First M87 Event Horizon Telescope Results. II. Array and Instrumentation,''
Astrophys. J. Lett. \textbf{875}, no.1, L2 (2019)
doi:10.3847/2041-8213/ab0c96
[arXiv:1906.11239 [astro-ph.IM]].

\bibitem{Akiyama:2019sww}
K.~Akiyama \textit{et al.} [Event Horizon Telescope],
``First M87 Event Horizon Telescope Results. III. Data Processing and Calibration,''
Astrophys. J. Lett. \textbf{875}, no.1, L3 (2019)
doi:10.3847/2041-8213/ab0c57
[arXiv:1906.11240 [astro-ph.GA]].

\bibitem{Akiyama:2019bqs}
K.~Akiyama \textit{et al.} [Event Horizon Telescope],
``First M87 Event Horizon Telescope Results. IV. Imaging the Central Supermassive Black Hole,''
Astrophys. J. Lett. \textbf{875}, no.1, L4 (2019)
doi:10.3847/2041-8213/ab0e85
[arXiv:1906.11241 [astro-ph.GA]].

\bibitem{Akiyama:2019fyp}
K.~Akiyama \textit{et al.} [Event Horizon Telescope],
``First M87 Event Horizon Telescope Results. V. Physical Origin of the Asymmetric Ring,''
Astrophys. J. Lett. \textbf{875}, no.1, L5 (2019)
doi:10.3847/2041-8213/ab0f43
[arXiv:1906.11242 [astro-ph.GA]].

\bibitem{Akiyama:2019eap}
K.~Akiyama \textit{et al.} [Event Horizon Telescope],
``First M87 Event Horizon Telescope Results. VI. The Shadow and Mass of the Central Black Hole,''
Astrophys. J. Lett. \textbf{875}, no.1, L6 (2019)
doi:10.3847/2041-8213/ab1141
[arXiv:1906.11243 [astro-ph.GA]].

\bibitem{Gralla:2019xty}
S.~E.~Gralla, D.~E.~Holz and R.~M.~Wald,
``Black Hole Shadows, Photon Rings, and Lensing Rings,''
Phys. Rev. D \textbf{100}, no.2, 024018 (2019)
doi:10.1103/PhysRevD.100.024018
[arXiv:1906.00873 [astro-ph.HE]].

\bibitem{Johnson:2019ljv}
M.~D.~Johnson, A.~Lupsasca, A.~Strominger, G.~N.~Wong, S.~Hadar, D.~Kapec, R.~Narayan, A.~Chael, C.~F.~Gammie and P.~Galison, \textit{et al.}
``Universal interferometric signatures of a black hole\textquoteright{}s photon ring,''
Sci. Adv. \textbf{6} (2020) no.12, eaaz1310
doi:10.1126/sciadv.aaz1310
[arXiv:1907.04329 [astro-ph.IM]].

\bibitem{Gralla:2019drh}
S.~E.~Gralla and A.~Lupsasca,
``Lensing by Kerr Black Holes,''
Phys. Rev. D \textbf{101}, no.4, 044031 (2020)
doi:10.1103/PhysRevD.101.044031
[arXiv:1910.12873 [gr-qc]].

\bibitem{Himwich:2020msm}
E.~Himwich, M.~D.~Johnson, A.~Lupsasca and A.~Strominger,
``Universal polarimetric signatures of the black hole photon ring,''
Phys. Rev. D \textbf{101} (2020) no.8, 084020
doi:10.1103/PhysRevD.101.084020
[arXiv:2001.08750 [gr-qc]].

\bibitem{Bambi:2008jg}
C.~Bambi and K.~Freese,
``Apparent shape of super-spinning black holes,''
Phys. Rev. D \textbf{79}, 043002 (2009)
doi:10.1103/PhysRevD.79.043002
[arXiv:0812.1328 [astro-ph]].

\bibitem{Johannsen:2010ru}
T.~Johannsen and D.~Psaltis,
``Testing the No-Hair Theorem with Observations in the Electromagnetic Spectrum: II. Black-Hole Images,''
Astrophys. J. \textbf{718}, 446-454 (2010)
doi:10.1088/0004-637X/718/1/446
[arXiv:1005.1931 [astro-ph.HE]].

\bibitem{Loeb:2013lfa}
A.~E.~Broderick, T.~Johannsen, A.~Loeb and D.~Psaltis,
``Testing the No-Hair Theorem with Event Horizon Telescope Observations of Sagittarius A*,''
Astrophys. J. \textbf{784}, 7 (2014)
doi:10.1088/0004-637X/784/1/7
[arXiv:1311.5564 [astro-ph.HE]].

\bibitem{Psaltis:2014mca}
D.~Psaltis, F.~Ozel, C.~K.~Chan and D.~P.~Marrone,
``A General Relativistic Null Hypothesis Test with Event Horizon Telescope Observations of the black-hole shadow in Sgr A*,''
Astrophys. J. \textbf{814}, no.2, 115 (2015)
doi:10.1088/0004-637X/814/2/115
[arXiv:1411.1454 [astro-ph.HE]].

\bibitem{Cunha:2015yba}
P.~V.~P.~Cunha, C.~A.~R.~Herdeiro, E.~Radu and H.~F.~Runarsson,
``Shadows of Kerr black holes with scalar hair,''
Phys. Rev. Lett. \textbf{115}, no.21, 211102 (2015)
doi:10.1103/PhysRevLett.115.211102
[arXiv:1509.00021 [gr-qc]].

\bibitem{Johannsen:2015hib}
T.~Johannsen, A.~E.~Broderick, P.~M.~Plewa, S.~Chatzopoulos, S.~S.~Doeleman, F.~Eisenhauer, V.~L.~Fish, R.~Genzel, O.~Gerhard and M.~D.~Johnson,
``Testing General Relativity with the Shadow Size of Sgr A*,''
Phys. Rev. Lett. \textbf{116}, no.3, 031101 (2016)
doi:10.1103/PhysRevLett.116.031101
[arXiv:1512.02640 [astro-ph.GA]].

\bibitem{Johannsen:2016vqy}
T.~Johannsen, C.~Wang, A.~E.~Broderick, S.~S.~Doeleman, V.~L.~Fish, A.~Loeb and D.~Psaltis,
``Testing General Relativity with Accretion-Flow Imaging of Sgr A*,''
Phys. Rev. Lett. \textbf{117}, no.9, 091101 (2016)
doi:10.1103/PhysRevLett.117.091101
[arXiv:1608.03593 [astro-ph.HE]].

\bibitem{Johannsen:2016uoh}
T.~Johannsen,
``Testing the No-Hair Theorem with Observations of Black Holes in the Electromagnetic Spectrum,''
Class. Quant. Grav. \textbf{33}, no.12, 124001 (2016)
doi:10.1088/0264-9381/33/12/124001
[arXiv:1602.07694 [astro-ph.HE]].

\bibitem{Lacroix:2016qpq}
T.~Lacroix, M.~Karami, A.~E.~Broderick, J.~Silk and C.~Boehm,
``Unique probe of dark matter in the core of M87 with the Event Horizon Telescope,''
Phys. Rev. D \textbf{96}, no.6, 063008 (2017)
doi:10.1103/PhysRevD.96.063008
[arXiv:1611.01961 [astro-ph.GA]].

\bibitem{Cunha:2018gql}
P.~V.~P.~Cunha, C.~A.~R.~Herdeiro and M.~J.~Rodriguez,
``Does the black hole shadow probe the event horizon geometry?,''
Phys. Rev. D \textbf{97}, no.8, 084020 (2018)
doi:10.1103/PhysRevD.97.084020
[arXiv:1802.02675 [gr-qc]].

\bibitem{Psaltis:2018xkc}
D.~Psaltis,
``Testing General Relativity with the Event Horizon Telescope,''
Gen. Rel. Grav. \textbf{51}, no.10, 137 (2019)
doi:10.1007/s10714-019-2611-5
[arXiv:1806.09740 [astro-ph.HE]].

\bibitem{Bambi:2019tjh}
C.~Bambi, K.~Freese, S.~Vagnozzi and L.~Visinelli,
``Testing the rotational nature of the supermassive object M87* from the circularity and size of its first image,''
Phys. Rev. D \textbf{100}, no.4, 044057 (2019)
doi:10.1103/PhysRevD.100.044057
[arXiv:1904.12983 [gr-qc]].

\bibitem{Banerjee:2019xds}
I.~Banerjee, S.~Sau and S.~SenGupta,
``Implications of axionic hair on the shadow of M87*,''
Phys. Rev. D \textbf{101}, no.10, 104057 (2020)
doi:10.1103/PhysRevD.101.104057
[arXiv:1911.05385 [gr-qc]].

\bibitem{Vagnozzi:2019apd}
S.~Vagnozzi and L.~Visinelli,
``Hunting for extra dimensions in the shadow of M87*,''
Phys. Rev. D \textbf{100}, no.2, 024020 (2019)
doi:10.1103/PhysRevD.100.024020
[arXiv:1905.12421 [gr-qc]].

\bibitem{Chen:2019fsq}
Y.~Chen, J.~Shu, X.~Xue, Q.~Yuan and Y.~Zhao,
``Probing Axions with Event Horizon Telescope Polarimetric Measurements,''
Phys. Rev. Lett. \textbf{124}, no.6, 061102 (2020)
doi:10.1103/PhysRevLett.124.061102
[arXiv:1905.02213 [hep-ph]].

\bibitem{Gralla:2020srx}
S.~E.~Gralla, A.~Lupsasca and D.~P.~Marrone,
``The shape of the black hole photon ring: A precise test of strong-field general relativity,''
Phys. Rev. D \textbf{102}, no.12, 124004 (2020)
doi:10.1103/PhysRevD.102.124004
[arXiv:2008.03879 [gr-qc]].

\bibitem{Volkel:2020xlc}
S.~H.~V\"olkel, E.~Barausse, N.~Franchini and A.~E.~Broderick,
``EHT tests of the strong-field regime of General Relativity,''
[arXiv:2011.06812 [gr-qc]].

\bibitem{Psaltis:2020ctj}
D.~Psaltis, C.~Talbot, E.~Payne and I.~Mandel,
``Probing the Black Hole Metric. I. Black Hole Shadows and Binary Black-Hole Inspirals,''
[arXiv:2012.02117 [gr-qc]].

\bibitem{Glampedakis:2021oie}
K.~Glampedakis and G.~Pappas,
``Can supermassive black hole shadows test the Kerr metric?,''
[arXiv:2102.13573 [gr-qc]].

\bibitem{Tinchev:2013nba}
V.~K.~Tinchev and S.~S.~Yazadjiev,
``Possible imprints of cosmic strings in the shadows of galactic black holes,''
Int. J. Mod. Phys. D \textbf{23}, 1450060 (2014)
doi:10.1142/S0218271814500606
[arXiv:1311.1353 [gr-qc]].

\bibitem{Hindmarsh:1994re}
M.~B.~Hindmarsh and T.~W.~B.~Kibble,
``Cosmic strings,''
Rept. Prog. Phys. \textbf{58}, 477-562 (1995)
doi:10.1088/0034-4885/58/5/001
[arXiv:hep-ph/9411342 [hep-ph]].

\bibitem{Vilenkin:2000jqa}
A.~Vilenkin and E.~P.~S.~Shellard,
``Cosmic Strings and Other Topological Defects,''
Cambridge University Press,
2000.

\bibitem{Nielsen:1973cs}
H.~B.~Nielsen and P.~Olesen,
``Vortex Line Models for Dual Strings,''
Nucl. Phys. B \textbf{61}, 45-61 (1973)
doi:10.1016/0550-3213(73)90350-7

\bibitem{Peccei:1977hh}
R.~D.~Peccei and H.~R.~Quinn,
``CP Conservation in the Presence of Instantons,''
Phys. Rev. Lett. \textbf{38}, 1440-1443 (1977)
doi:10.1103/PhysRevLett.38.1440

\bibitem{Weinberg:1977ma}
S.~Weinberg,
``A New Light Boson?,''
Phys. Rev. Lett. \textbf{40}, 223-226 (1978)
doi:10.1103/PhysRevLett.40.223

\bibitem{Wilczek:1977pj}
F.~Wilczek,
``Problem of Strong  $P$  and  $T$  Invariance in the Presence of Instantons,''
Phys. Rev. Lett. \textbf{40}, 279-282 (1978)
doi:10.1103/PhysRevLett.40.279

\bibitem{Arvanitaki:2009fg}
A.~Arvanitaki, S.~Dimopoulos, S.~Dubovsky, N.~Kaloper and J.~March-Russell,
``String Axiverse,''
Phys. Rev. D \textbf{81}, 123530 (2010)
doi:10.1103/PhysRevD.81.123530
[arXiv:0905.4720 [hep-th]].

\bibitem{Vilenkin:2018zol}
A.~Vilenkin, Y.~Levin and A.~Gruzinov,
``Cosmic strings and primordial black holes,''
JCAP \textbf{11} (2018), 008
doi:10.1088/1475-7516/2018/11/008
[arXiv:1808.00670 [astro-ph.CO]].

\bibitem{Morris_2017}
Morris, Mark R. and Zhao, Jun-Hui and Goss, W. M.,
``A Nonthermal Radio Filament Connected to the Galactic Black Hole?,",
The Astrophysical Journal \textbf{850} (2017) 2, L23
doi:10.3847/2041-8213/aa9985
[arXiv:1711.04190 [astro-ph.GA]].

\bibitem{Agrawal:2019lkr}
P.~Agrawal, A.~Hook and J.~Huang,
``A CMB Millikan experiment with cosmic axiverse strings,''
JHEP \textbf{07}, 138 (2020)
doi:10.1007/JHEP07(2020)138
[arXiv:1912.02823 [astro-ph.CO]].

\bibitem{Vilenkin:1984ea}
A.~Vilenkin,
``Cosmic strings as gravitational lenses,''
Astrophys. J. Lett. \textbf{282}, L51-L53 (1984)
doi:10.1086/184303

\bibitem{Chudnovsky:1986hc}
E.~M.~Chudnovsky, G.~B.~Field, D.~N.~Spergel and A.~Vilenkin,
``Superconducting cosmic strings,''
Phys. Rev. D \textbf{34}, 944-950 (1986)
doi:10.1103/PhysRevD.34.944

\bibitem{Gruzinov:2016hqs}
A.~Gruzinov and A.~Vilenkin,
``Fireballs from Superconducting Cosmic Strings,''
JCAP \textbf{01} (2017), 029
doi:10.1088/1475-7516/2017/01/029
[arXiv:1608.05396 [astro-ph.HE]].

\bibitem{Agrawal:2020euj}
P.~Agrawal, A.~Hook, J.~Huang and G.~Marques-Tavares,
``Axion string signatures II: A cosmological plasma collider,''
[arXiv:2010.15848 [hep-ph]].

\bibitem{Kibble:1976sj}
T.~W.~B.~Kibble,
``Topology of Cosmic Domains and Strings,''
J. Phys. A \textbf{9}, 1387-1398 (1976)
doi:10.1088/0305-4470/9/8/029

\bibitem{Kibble:1980mv}
T.~W.~B.~Kibble,
``Some Implications of a Cosmological Phase Transition,''
Phys. Rept. \textbf{67}, 183 (1980)
doi:10.1016/0370-1573(80)90091-5

\bibitem{Albrecht:1984xv}
A.~Albrecht and N.~Turok,
``Evolution of Cosmic Strings,''
Phys. Rev. Lett. \textbf{54}, 1868-1871 (1985)
doi:10.1103/PhysRevLett.54.1868

\bibitem{Bennett:1985qt}
D.~P.~Bennett,
``The evolution of cosmic strings,''
Phys. Rev. D \textbf{33}, 872 (1986)
[erratum: Phys. Rev. D \textbf{34}, 3932 (1986)]
doi:10.1103/PhysRevD.33.872

\bibitem{Bennett:1986zn}
D.~P.~Bennett,
``Evolution of cosmic strings. 2.,''
Phys. Rev. D \textbf{34}, 3592 (1986)
doi:10.1103/PhysRevD.34.3592

\bibitem{Bennett:1987vf}
D.~P.~Bennett and F.~R.~Bouchet,
``Evidence for a Scaling Solution in Cosmic String Evolution,''
Phys. Rev. Lett. \textbf{60}, 257 (1988)
doi:10.1103/PhysRevLett.60.257

\bibitem{Allen:1990tv}
B.~Allen and E.~P.~S.~Shellard,
``Cosmic string evolution: a numerical simulation,''
Phys. Rev. Lett. \textbf{64}, 119-122 (1990)
doi:10.1103/PhysRevLett.64.119

\bibitem{Vincent:1996rb}
G.~R.~Vincent, M.~Hindmarsh and M.~Sakellariadou,
``Scaling and small scale structure in cosmic string networks,''
Phys. Rev. D \textbf{56}, 637-646 (1997)
doi:10.1103/PhysRevD.56.637
[arXiv:astro-ph/9612135 [astro-ph]].

\bibitem{Yamaguchi:1998gx}
M.~Yamaguchi, M.~Kawasaki and J.~Yokoyama,
``Evolution of axionic strings and spectrum of axions radiated from them,''
Phys. Rev. Lett. \textbf{82}, 4578-4581 (1999)
doi:10.1103/PhysRevLett.82.4578
[arXiv:hep-ph/9811311 [hep-ph]].

\bibitem{Yamaguchi:2002sh}
M.~Yamaguchi and J.~Yokoyama,
``Quantitative evolution of global strings from the Lagrangian view point,''
Phys. Rev. D \textbf{67}, 103514 (2003)
doi:10.1103/PhysRevD.67.103514
[arXiv:hep-ph/0210343 [hep-ph]].

\bibitem{Kawasaki:2018bzv}
M.~Kawasaki, T.~Sekiguchi, M.~Yamaguchi and J.~Yokoyama,
``Long-term dynamics of cosmological axion strings,''
PTEP \textbf{2018}, no.9, 091E01 (2018)
doi:10.1093/ptep/pty098
[arXiv:1806.05566 [hep-ph]].

\bibitem{Vaquero:2018tib}
A.~Vaquero, J.~Redondo and J.~Stadler,
``Early seeds of axion miniclusters,''
JCAP \textbf{04}, 012 (2019)
doi:10.1088/1475-7516/2019/04/012
[arXiv:1809.09241 [astro-ph.CO]].

\bibitem{Gorghetto:2018myk}
M.~Gorghetto, E.~Hardy and G.~Villadoro,
``Axions from Strings: the Attractive Solution,''
JHEP \textbf{07}, 151 (2018)
doi:10.1007/JHEP07(2018)151
[arXiv:1806.04677 [hep-ph]].

\bibitem{Hindmarsh:2019csc}
M.~Hindmarsh, J.~Lizarraga, A.~Lopez-Eiguren and J.~Urrestilla,
``Scaling Density of Axion Strings,''
Phys. Rev. Lett. \textbf{124}, no.2, 021301 (2020)
doi:10.1103/PhysRevLett.124.021301
[arXiv:1908.03522 [astro-ph.CO]].

\bibitem{Hindmarsh:2021vih}
M.~Hindmarsh, J.~Lizarraga, A.~Lopez-Eiguren and J.~Urrestilla,
``Approach to scaling in axion string networks,''
[arXiv:2102.07723 [astro-ph.CO]].

\bibitem{Gorghetto:2020qws}
M.~Gorghetto, E.~Hardy and G.~Villadoro,
``More Axions from Strings,''
SciPost Phys. \textbf{10}, 050 (2021)
doi:10.21468/SciPostPhys.10.2.050
[arXiv:2007.04990 [hep-ph]].

\bibitem{Callan:1984sa}
C.~G.~Callan, Jr. and J.~A.~Harvey,
``Anomalies and Fermion Zero Modes on Strings and Domain Walls,''
Nucl. Phys. B \textbf{250} (1985), 427-436
doi:10.1016/0550-3213(85)90489-4

\bibitem{Lazarides:1984zq}
G.~Lazarides and Q.~Shafi,
``Superconducting Strings in Axion Models,''
Phys. Lett. B \textbf{151}, 123-126 (1985)
doi:10.1016/0370-2693(85)91398-X

\bibitem{Witten:1984eb}
E.~Witten,
``Superconducting Strings,''
Nucl. Phys. B \textbf{249}, 557-592 (1985)
doi:10.1016/0550-3213(85)90022-7

\bibitem{Aryal:1986sz}
M.~Aryal, L.~H.~Ford and A.~Vilenkin,
``Cosmic Strings and Black Holes,''
Phys. Rev. D \textbf{34}, 2263 (1986)
doi:10.1103/PhysRevD.34.2263

\bibitem{Lonsdale:1988xd}
S.~Lonsdale and I.~Moss,
``The Motion of Cosmic Strings Under Gravity,''
Nucl. Phys. B \textbf{298}, 693-700 (1988)
doi:10.1016/0550-3213(88)90003-X

\bibitem{DeVilliers:1997nk}
J.~P.~De Villiers and V.~P.~Frolov,
``Gravitational capture of cosmic strings by a black hole,''
Int. J. Mod. Phys. D \textbf{7}, 957-967 (1998)
doi:10.1142/S0218271898000632
[arXiv:gr-qc/9711045 [gr-qc]].

\bibitem{DeVilliers:1998nm}
J.~P.~De Villiers and V.~P.~Frolov,
``Gravitational scattering of cosmic strings by nonrotating black holes,''
Class. Quant. Grav. \textbf{16}, 2403-2425 (1999)
doi:10.1088/0264-9381/16/7/317
[arXiv:gr-qc/9812016 [gr-qc]].

\bibitem{Snajdr:2002aa}
M.~Snajdr and V.~P.~Frolov,
``Capture and critical scattering of a long cosmic string by a rotating black hole,''
Class. Quant. Grav. \textbf{20}, 1303-1320 (2003)
doi:10.1088/0264-9381/20/7/305
[arXiv:gr-qc/0211018 [gr-qc]].

\bibitem{Dubath:2006vs}
F.~Dubath, M.~Sakellariadou and C.~M.~Viallet,
``Scattering of cosmic strings by black holes: Loop formation,''
Int. J. Mod. Phys. D \textbf{16}, 1311-1325 (2007)
doi:10.1142/S0218271807010778
[arXiv:gr-qc/0609089 [gr-qc]].

\bibitem{Xing:2020ecz}
H.~Xing, Y.~Levin, A.~Gruzinov and A.~Vilenkin,
``Spinning black holes as cosmic string factories,''
[arXiv:2011.00654 [astro-ph.HE]].

\bibitem{Zeldovich:1967lct}
Y.~B.~;.~N.~Zel'dovich, I.~D.,
``The Hypothesis of Cores Retarded during Expansion and the Hot Cosmological Model,''
Soviet Astron. AJ (Engl. Transl. ), \textbf{10}, 602 (1967)

\bibitem{Hawking:1971ei}
S.~Hawking,
``Gravitationally collapsed objects of very low mass,''
Mon. Not. Roy. Astron. Soc. \textbf{152}, 75 (1971)

\bibitem{Carr:1974nx}
B.~J.~Carr and S.~W.~Hawking,
``Black holes in the early Universe,''
Mon. Not. Roy. Astron. Soc. \textbf{168}, 399-415 (1974)

\bibitem{Lake:2015ppa}
M.~J.~Lake and T.~Harko,
``Can Superconducting Cosmic Strings Piercing Seed Black Holes Generate Supermassive Black Holes in the Early Universe?,''
Fortsch. Phys. \textbf{65}, no.10-11, 1600121 (2017)
doi:10.1002/prop.201600121
[arXiv:1505.01584 [astro-ph.CO]].

\bibitem{Achucarro:1995nu}
A.~Achucarro, R.~Gregory and K.~Kuijken,
``Abelian Higgs hair for black holes,''
Phys. Rev. D \textbf{52}, 5729-5742 (1995)
doi:10.1103/PhysRevD.52.5729
[arXiv:gr-qc/9505039 [gr-qc]].

\bibitem{Frolov:1995vp}
V.~P.~Frolov, S.~Hendy and A.~L.~Larsen,
``How to create a 2-D black hole,''
Phys. Rev. D \textbf{54}, 5093-5102 (1996)
doi:10.1103/PhysRevD.54.5093
[arXiv:hep-th/9510231 [hep-th]].

\bibitem{Bonjour:1998rf}
F.~Bonjour, R.~Emparan and R.~Gregory,
``Vortices and extreme black holes: The Question of flux expulsion,''
Phys. Rev. D \textbf{59}, 084022 (1999)
doi:10.1103/PhysRevD.59.084022
[arXiv:gr-qc/9810061 [gr-qc]].

\bibitem{Santos:1999if}
C.~Santos and R.~Gregory,
``Vortices and black holes in dilatonic gravity,''
Phys. Rev. D \textbf{61}, 024006 (2000)
doi:10.1103/PhysRevD.61.024006
[arXiv:gr-qc/9906107 [gr-qc]].

\bibitem{Ghezelbash:2001pq}
A.~M.~Ghezelbash and R.~B.~Mann,
``Abelian Higgs hair for rotating and charged black holes,''
Phys. Rev. D \textbf{65}, 124022 (2002)
doi:10.1103/PhysRevD.65.124022
[arXiv:hep-th/0110001 [hep-th]].

\bibitem{Gregory:2013xca}
R.~Gregory, D.~Kubiznak and D.~Wills,
``Rotating black hole hair,''
JHEP \textbf{06}, 023 (2013)
doi:10.1007/JHEP06(2013)023
[arXiv:1303.0519 [gr-qc]].

\bibitem{Kubiznak:2015hsg}
D.~Kubiznak,
``Rotating black holes pierced by a cosmic string,''
doi:10.1142/9789813226609\_0190
[arXiv:1512.08807 [gr-qc]].

\bibitem{Gregory:2014uca}
R.~Gregory, P.~C.~Gustainis, D.~Kubiz\v{n}\'ak, R.~B.~Mann and D.~Wills,
``Vortex hair on AdS black holes,''
JHEP \textbf{11}, 010 (2014)
doi:10.1007/JHEP11(2014)010
[arXiv:1405.6507 [hep-th]].


\bibitem{Kinoshita:2016lqd}
S.~Kinoshita, T.~Igata and K.~Tanabe,
``Energy extraction from Kerr black holes by rigidly rotating strings,''
Phys. Rev. D \textbf{94}, no.12, 124039 (2016)
doi:10.1103/PhysRevD.94.124039
[arXiv:1610.08006 [gr-qc]].

\bibitem{Igata:2018kry}
T.~Igata, H.~Ishihara, M.~Tsuchiya and C.~M.~Yoo,
``Rigidly Rotating String Sticking in a Kerr Black Hole,''
Phys. Rev. D \textbf{98}, no.6, 064021 (2018)
doi:10.1103/PhysRevD.98.064021
[arXiv:1806.09837 [gr-qc]].

\bibitem{Boyer:1966qh}
R.~H.~Boyer and R.~W.~Lindquist,
``Maximal analytic extension of the Kerr metric,''
J. Math. Phys. \textbf{8}, 265 (1967)
doi:10.1063/1.1705193

\bibitem{Frolov:1998wf}
V.~P.~Frolov and I.~D.~Novikov,
``Black hole physics: Basic concepts and new developments,''\
Kluwer Academic Publishers,
doi:10.1007/978-94-011-5139-9,
1998.

\bibitem{Cohen:1988sg}
A.~G.~Cohen and D.~B.~Kaplan,
``The Exact Metric About Global Cosmic Strings,''
Phys. Lett. B \textbf{215}, 67-72 (1988)
doi:10.1016/0370-2693(88)91072-6

\bibitem{Gibbons:1988pe}
G.~W.~Gibbons, M.~E.~Ortiz and F.~Ruiz Ruiz,
``Existence of Global Strings Coupled to Gravity,''
Phys. Rev. D \textbf{39}, 1546-1551 (1989)
doi:10.1103/PhysRevD.39.1546

\bibitem{Gregory:1996dd}
R.~Gregory,
``Nonsingular global strings,''
Phys. Rev. D \textbf{54}, 4955-4962 (1996)
doi:10.1103/PhysRevD.54.4955
[arXiv:gr-qc/9606002 [gr-qc]].

\bibitem{Gregory:2002tp}
R.~Gregory and C.~Santos,
``Space-time structure of the global vortex,''
Class. Quant. Grav. \textbf{20} (2003), 21-36
doi:10.1088/0264-9381/20/1/302
[arXiv:hep-th/0208037 [hep-th]].

\bibitem{Carter:1968rr}
B.~Carter,
``Global structure of the Kerr family of gravitational fields,''
Phys. Rev. \textbf{174}, 1559-1571 (1968)
doi:10.1103/PhysRev.174.1559

\bibitem{Cardoso:2008bp}
V.~Cardoso, A.~S.~Miranda, E.~Berti, H.~Witek and V.~T.~Zanchin,
``Geodesic stability, Lyapunov exponents and quasinormal modes,''
Phys. Rev. D \textbf{79}, 064016 (2009)
doi:10.1103/PhysRevD.79.064016
[arXiv:0812.1806 [hep-th]].

\bibitem{Bardeen:1972fi}
J.~M.~Bardeen, W.~H.~Press and S.~A.~Teukolsky,
``Rotating black holes: Locally nonrotating frames, energy extraction, and scalar synchrotron radiation,''
Astrophys. J. \textbf{178}, 347 (1972)
doi:10.1086/151796

\bibitem{Harari:1992ea}
D.~Harari and P.~Sikivie,
``Effects of a Nambu-Goldstone boson on the polarization of radio galaxies and the cosmic microwave background,''
Phys. Lett. B \textbf{289}, 67-72 (1992)
doi:10.1016/0370-2693(92)91363-E

\bibitem{Fedderke:2019ajk}
M.~A.~Fedderke, P.~W.~Graham and S.~Rajendran,
``Axion Dark Matter Detection with CMB Polarization,''
Phys. Rev. D \textbf{100}, no.1, 015040 (2019)
doi:10.1103/PhysRevD.100.015040
[arXiv:1903.02666 [astro-ph.CO]].

\bibitem{Schwarz:2020jjh}
D.~J.~Schwarz, J.~Goswami and A.~Basu,
``Geometric optics in the presence of axion-like particles in curved space-time,''
[arXiv:2003.10205 [hep-ph]].

\bibitem{Aharonov:1959fk}
Y.~Aharonov and D.~Bohm,
``Significance of electromagnetic potentials in the quantum theory,''
Phys. Rev. \textbf{115}, 485-491 (1959)
doi:10.1103/PhysRev.115.485

\bibitem{Jain:2021shf}
M.~Jain, A.~J.~Long and M.~A.~Amin,
``CMB birefringence from ultra-light axion string networks,''
[arXiv:2103.10962 [astro-ph.CO]].

\bibitem{Lue:1998mq}
A.~Lue, L.~M.~Wang and M.~Kamionkowski,
``Cosmological signature of new parity violating interactions,''
Phys. Rev. Lett. \textbf{83}, 1506-1509 (1999)
doi:10.1103/PhysRevLett.83.1506
[arXiv:astro-ph/9812088 [astro-ph]].

\bibitem{Pospelov:2008gg}
M.~Pospelov, A.~Ritz, C.~Skordis, A.~Ritz and C.~Skordis,
``Pseudoscalar perturbations and polarization of the cosmic microwave background,''
Phys. Rev. Lett. \textbf{103}, 051302 (2009)
doi:10.1103/PhysRevLett.103.051302
[arXiv:0808.0673 [astro-ph]].

\bibitem{Kamionkowski:2008fp}
M.~Kamionkowski,
``How to De-Rotate the Cosmic Microwave Background Polarization,''
Phys. Rev. Lett. \textbf{102}, 111302 (2009)
doi:10.1103/PhysRevLett.102.111302
[arXiv:0810.1286 [astro-ph]].

\bibitem{Ivanov:2018byi}
M.~M.~Ivanov, Y.~Y.~Kovalev, M.~L.~Lister, A.~G.~Panin, A.~B.~Pushkarev, T.~Savolainen and S.~V.~Troitsky,
``Constraining the photon coupling of ultra-light dark-matter axion-like particles by polarization variations of parsec-scale jets in active galaxies,''
JCAP \textbf{02}, 059 (2019)
doi:10.1088/1475-7516/2019/02/059
[arXiv:1811.10997 [astro-ph.CO]].

\bibitem{Fujita:2018zaj}
T.~Fujita, R.~Tazaki and K.~Toma,
``Hunting Axion Dark Matter with Protoplanetary Disk Polarimetry,''
Phys. Rev. Lett. \textbf{122}, no.19, 191101 (2019)
doi:10.1103/PhysRevLett.122.191101
[arXiv:1811.03525 [astro-ph.CO]].

\bibitem{Blas:2019qqp}
D.~Blas, A.~Caputo, M.~M.~Ivanov and L.~Sberna,
``No chiral light bending by clumps of axion-like particles,''
Phys. Dark Univ. \textbf{27}, 100428 (2020)
doi:10.1016/j.dark.2019.100428
[arXiv:1910.06128 [hep-ph]].

\bibitem{Srednicki:1985xd}
M.~Srednicki,
``Axion Couplings to Matter. 1. CP Conserving Parts,''
Nucl. Phys. B \textbf{260}, 689-700 (1985)
doi:10.1016/0550-3213(85)90054-9

\bibitem{Moscibrodzka:2017gdx}
M.~Moscibrodzka, J.~Dexter, J.~Davelaar and H.~Falcke,
``Faraday rotation in GRMHD simulations of the jet launching zone of M87,''
Mon. Not. Roy. Astron. Soc. \textbf{468}, no.2, 2214-2221 (2017)
doi:10.1093/mnras/stx587
[arXiv:1703.02390 [astro-ph.HE]].

\bibitem{Jimenez-Rosales:2018mpc}
A.~Jim\'enez-Rosales and J.~Dexter,
``The impact of Faraday effects on polarized black hole images of Sagittarius A*,''
Mon. Not. Roy. Astron. Soc. \textbf{478}, no.2, 1875-1883 (2018)
doi:10.1093/mnras/sty1210
[arXiv:1805.02652 [astro-ph.HE]].

\bibitem{Tsunetoe:2020pyz}
Y.~Tsunetoe, S.~Mineshige, K.~Ohsuga, T.~Kawashima and K.~Akiyama,
``Polarization imaging of M 87 jets by general relativistic radiative transfer calculation based on GRMHD simulations,''
Publ. Astron. Soc. Jap. \textbf{72}, no.2, 32 (2020)
doi:10.1093/pasj/psaa008
[arXiv:2002.00954 [astro-ph.HE]].

\bibitem{Ricarte:2020llx}
A.~Ricarte, B.~S.~Prather, G.~N.~Wong, R.~Narayan, C.~Gammie and M.~Johnson,
``Decomposing the Internal Faraday Rotation of Black Hole Accretion Flows,''
Mon. Not. Roy. Astron. Soc. \textbf{498}, no.4, 5468-5488 (2020)
doi:10.1093/mnras/staa2692
[arXiv:2009.02369 [astro-ph.HE]].

\bibitem{Tsunetoe:2020nws}
Y.~Tsunetoe, S.~Mineshige, K.~Ohsuga, T.~Kawashima and K.~Akiyama,
``Polarization images of accretion flow around supermassive black holes: imprints of toroidal field structure,''
[arXiv:2012.05243 [astro-ph.HE]].

\bibitem{Jimenez-Rosales:2021ytz}
A.~Jim\'enez-Rosales, J.~Dexter, S.~M.~Ressler, A.~Tchekhovskoy, M.~Baub\"ock, Y.~Dallilar, P.~T.~De Zeeuw, A.~Drescher, F.~Eisenhauer and S.~Von Fellenberg, \textit{et al.}
``Relative depolarization of the black hole photon ring in GRMHD models of Sgr A* and M87*,''
doi:10.1093/mnras/stab784
[arXiv:2103.06292 [astro-ph.HE]].

\bibitem{Gralla:2019ceu}
S.~E.~Gralla and A.~Lupsasca,
``Null geodesics of the Kerr exterior,''
Phys. Rev. D \textbf{101}, no.4, 044032 (2020)
doi:10.1103/PhysRevD.101.044032
[arXiv:1910.12881 [gr-qc]].

\bibitem{Moriyama:2015zfa}
K.~Moriyama and S.~Mineshige,
``New method for black-hole spin measurement based on flux variation from an infalling gas ring,''
Publ. Astron. Soc. Jap. \textbf{67}, no.6, 106 (2015)
doi:10.1093/pasj/psv074
[arXiv:1508.03334 [astro-ph.HE]].

\bibitem{Moriyama:2019mhz}
K.~Moriyama, S.~Mineshige, M.~Honma and K.~Akiyama,
``Black hole Spin Measurement Based on Time-domain VLBI Observations of Infalling Gas Cloud,''
doi:10.3847/1538-4357/ab505b
[arXiv:1910.10713 [astro-ph.HE]].

\bibitem{Hadar:2020fda}
S.~Hadar, M.~D.~Johnson, A.~Lupsasca and G.~N.~Wong,
``Photon Ring Autocorrelations,''
[arXiv:2010.03683 [gr-qc]].

\bibitem{Chesler:2020gtw}
P.~M.~Chesler, L.~Blackburn, S.~S.~Doeleman, M.~D.~Johnson, J.~M.~Moran, R.~Narayan and M.~Wielgus,
``Light echos and coherent autocorrelations in a black hole spacetime,''
[arXiv:2012.11778 [gr-qc]].

\bibitem{Wong:2020ziu}
G.~N.~Wong,
``Black Hole Glimmer Signatures of Mass, Spin, and Inclination,''
Astrophys. J. \textbf{909}, no.2, 217 (2021)
doi:10.3847/1538-4357/abdd2d
[arXiv:2009.06641 [astro-ph.HE]].



\end{thebibliography}
\end{document}